\documentclass[
	reprint,
	showkeys,
	superscriptaddress,
	aps,
	pra,
	longbibliography,
	nofootinbib
]{revtex4-2}

\usepackage[utf8]{inputenc}

\usepackage{amsmath}
\usepackage{amssymb}
\usepackage{comment}

\usepackage{physics}
\usepackage{slashed}
\usepackage{cancel}
\usepackage{tensor}

\usepackage{dsfont}
\usepackage{upgreek}
\usepackage{bm}
\usepackage{mathtools}

\usepackage[hyperref]{xcolor}
	\definecolor{goethe-blau}{cmyk}{1.0,0.2,0.0,0.4}
	\definecolor{hellgrau}{cmyk}{0.04,0.04,0.05,0.02}
	\definecolor{sandgrau}{cmyk}{0.12,0.09,0.13,0.0}
	\definecolor{dunkelgrau}{cmyk}{0.25,0.25,0.30,0.75}
	\definecolor{emo-rot}{cmyk}{0.04,1.0,0.8,0.07}
	\definecolor{purple}{cmyk}{0.08,1.0,0.3,0.36}
	\definecolor{senfgelb}{cmyk}{0.01,0.25,1.0,0.05}
	\definecolor{gruen}{cmyk}{0.62,0.4,0.87,0.09}
	\definecolor{magenta}{cmyk}{0.08,0.86,0.12,0.12}
	\definecolor{orange}{cmyk}{0.0,0.7,1.0,0.04}
	\definecolor{sonnengelb}{cmyk}{0.0,0.12,0.95,0.0}
	\definecolor{helles-gruen}{cmyk}{0.4,0.17,0.81,0.07}
	\definecolor{lichtblau}{cmyk}{0.8,0.0,0.06,0.04}

\usepackage[
	colorlinks,
	pdfpagelabels,
	breaklinks,
	pdfstartview=FitH,
	bookmarksopen=true,
	bookmarksnumbered=true,
	bookmarksopenlevel=2,
	plainpages=false,
	hypertexnames=false,
	citecolor=emo-rot,
	linkcolor=goethe-blau,
	urlcolor=purple,
	pdftitle={Detecting inhomogeneous chiral condensation from the bosonic two-point function in the (1 + 1)-dimensional Gross-Neveu model in the mean-field approximation}, 
	pdfauthor={Adrian Koenigstein, Laurin Pannullo, Stefan Rechenberger, Martin J. Steil, Marc Winstel},
	pdfkeywords={Gross-Neveu model, phase diagram, mean-field, stability analysis, two-point function, inhomogeneous phases, wave-function renormalization, moat regime}
]{hyperref}

\usepackage{bookmark}
\usepackage[capitalise]{cleveref}
\usepackage{multirow}
\usepackage{xstring}
\usepackage{graphicx}
\usepackage{dcolumn}

\allowdisplaybreaks 

\providecommand{\Rcite}[1]{\begingroup
	\def\tempx{0}\StrCount{#1}{,}[\tempx]\ifnum\tempx > 0 
	Refs.~\else
	Ref.~\fi
	\endgroup
	\cite{#1}}

\newcommand{\infN}{\ensuremath{\Nf \to \infty}}
\newcommand{\pbArrow}{\,\leftrightarrow\,}

\newcommand{\ii}{\ensuremath{\mathrm{i}}}
\newcommand{\D}{\ensuremath{\mathrm{d}}}
\newcommand{\e}{\ensuremath{\mathrm{e}}}
\newcommand{\I}{\ensuremath{\mathds{I}}}
\newcommand{\U}{\ensuremath{\mathrm{U}}}
\newcommand{\SU}{\ensuremath{\mathrm{SU}}}

\DeclareMathOperator{\arcoth}{arcoth}
\DeclareMathOperator{\artanh}{artanh}

\newcommand{\Det}{\ensuremath{\mathrm{Det}}}
\newcommand{\diag}{\ensuremath{\mathrm{diag}}}
\newcommand{\xTwo}{\ensuremath{\tau}}

\newcommand{\xOne}[1]{\ensuremath{#1}}
\newcommand{\ta}[1]{\ensuremath{\xTwo_{#1}}}

\newcommand{\te}{\ensuremath{\ta{x}} }
\newcommand{\ex}{\ensuremath{\xOne{x}}}

\newcommand{\st}[1]{\ensuremath{\mathbf{#1}}}
\newcommand{\xst}{\ensuremath{\st{x}}}

\newcommand{\Hom}[1]{\ensuremath{\bar{#1}}}

\newcommand{\s}{\ensuremath{\sigma}}
\newcommand{\stilde}{\ensuremath{\tilde{\s}}}
\newcommand{\shom}{\ensuremath{\Hom{\s}}}
\newcommand{\smin}{\ensuremath{\Sigma}}
\newcommand{\sminhom}{\ensuremath{\Hom{\smin}}}
\newcommand{\snull}{\ensuremath{\sminhom_0}}

\newcommand{\ds}{\ensuremath{\delta \s}}
\newcommand{\dsx}[1]{\ensuremath{\ds(#1)}}
\newcommand{\dst}{\ensuremath{\delta \stilde}}
\newcommand{\dstx}[1]{\ensuremath{\dst(#1)}}

\newcommand{\qs}{\ensuremath{q_{\smin}}}
\renewcommand{\S}{\mathcal{S}}
\newcommand{\seff}{\mathcal{S}_{\text{eff}}}
\newcommand{\Z}{\ensuremath{\mathbb{Z}}}
\newcommand{\Nf}{\ensuremath{N}}
\newcommand{\gp}{\ensuremath{U}}
\newcommand{\gphom}{\ensuremath{\Hom{\gp}}}
\newcommand{\coupling}{\ensuremath{g^2}}
\newcommand{\rscoupling}{\ensuremath{g^2}}
\newcommand{\grandp}{\ensuremath{\Omega}}
\newcommand{\grandphom}{\ensuremath{\Hom{\grandp}}}

\newcommand{\energy}[1]{\ensuremath{E_{#1}}}

\newcommand{\gch}{\ensuremath{\gamma^\mathrm{ch}}}
\newcommand{\dop}{\ensuremath{D}}
\newcommand{\dophom}{\ensuremath{\Hom{\dop}}}

\newcommand{\gtwo}{\ensuremath{\Gamma^{(2)}}}
\newcommand{\gtwovar}[4]{\ensuremath{\gtwo ( #1, #2, #3, #4)}}
\newcommand{\gtwovardef}{\ensuremath{\gtwovar{\shom}{\mu}{T}{q}}}
\newcommand{\qmin}{\ensuremath{Q}}

\newcommand{\zwave}[3]{\ensuremath{Z ( #1, #2, #3 )}}

\newcommand{\Eqref}[1]{Eq.\ (\ref{#1})}

\DeclareMathOperator{\sumint}{\, \underset{\st{p}}{\mathrlap{\sum}\int} \, \,}

\DeclareUnicodeCharacter{2A0B}{\sumint}

\usepackage[acronym]{glossaries-extra}
\glssetcategoryattribute{acronym}{nohyperfirst}{true}
\setabbreviationstyle[acronym]{long-short}

\newacronym[plural=QFTs,firstplural=quantum field theories]{qft}{QFT}{quantum field theory}
\newacronym{qcd}{QCD}{Quantum Chromodynamics}
\newacronym{frg}{FRG}{Functional Renormalization Group}
\newacronym{cp}{CP}{critical point}
\newacronym{lp}{LP}{Lifshitz point}
\newacronym{cep}{CEP}{critical endpoint}
\newacronym{plp}{PLP}{pseudo Lifshitz point}
\newacronym{njl}{NJL}{Nambu-Jona-Lasinio}
\newacronym{qm}{QM}{Quark-Meson}
\newacronym{gn}{GN}{Gross-Neveu}
\newacronym{hbp}{HBP}{homogeneously broken phase}
\newacronym{ip}{IP}{inhomogeneous phase}
\newacronym{sp}{SP}{symmetric phase}
\newacronym{cdw}{CDW}{chiral density wave}
\newacronym{uv}{UV}{ultra violet}
\newacronym{ir}{IR}{infra red}

\begin{document}

\preprint{APS/123-QED}

\title{
	Detecting inhomogeneous chiral condensation from the bosonic two-point function in the \texorpdfstring{$(1 + 1)$}{(1+1)}-dimensional Gross-Neveu model in the mean-field approximation
}

\thanks{
	The list of authors is sorted by (former) affiliations and surnames. All authors contributed in equal shares to this work.
	S.~Rechenberger initiated this project and produced preliminary results before leaving academia.
}

\author{Adrian Koenigstein}
	\email{koenigstein@th.physik.uni-frankfurt.de}
	\affiliation{
		Institut für Theoretische Physik, Johann Wolfgang Goethe-Universität,
		\\
		Max-von-Laue-Straße 1, D-60438 Frankfurt am Main, Germany.
	}
\author{Laurin Pannullo}
	\email{pannullo@itp.uni-frankfurt.de}
	\affiliation{
		Institut für Theoretische Physik, Johann Wolfgang Goethe-Universität,
		\\
		Max-von-Laue-Straße 1, D-60438 Frankfurt am Main, Germany.
	}
\author{Stefan Rechenberger}
	\affiliation{
		Institut für Theoretische Physik, Johann Wolfgang Goethe-Universität,
		\\
		Max-von-Laue-Straße 1, D-60438 Frankfurt am Main, Germany.
	}
\author{Martin J. Steil}
	\email{msteil@theorie.ikp.physik.tu-darmstadt.de}

	\affiliation{
		Technische Universität Darmstadt, Department of Physics, Institut für Kernphysik, Theoriezentrum,
		\\
		Schlossgartenstraße 2, D-64289 Darmstadt, Germany.
	}

\author{Marc Winstel}
	\email{winstel@itp.uni-frankfurt.de}
		
	\affiliation{
		Institut für Theoretische Physik, Johann Wolfgang Goethe-Universität,
		\\
		Max-von-Laue-Straße 1, D-60438 Frankfurt am Main, Germany.
	}

\date{\today}

\begin{abstract}
	The phase diagram of the $(1 + 1)$-dimensional Gross-Neveu model is reanalyzed for (non-)zero chemical potential and (non-)zero temperature within the mean-field approximation.
	By investigating the momentum dependence of the bosonic two-point function, the well-known second-order phase transition from the $\mathbb{Z}_2$ symmetric phase to the so-called inhomogeneous phase is detected.
	In the latter phase the chiral condensate is periodically varying in space and translational invariance is broken.
	This work is a proof of concept study that confirms that it is possible to correctly localize second-order phase transition lines between phases without condensation and phases of spatially inhomogeneous condensation via a stability analysis of the homogeneous phase. 
	To complement other works relying on this technique, the stability analysis is explained in detail and its limitations and successes are discussed in context of the Gross-Neveu model.
	Additionally, we present explicit results for the bosonic wave-function renormalization in the mean-field approximation, which is extracted analytically from the bosonic two-point function.
	We find regions -- a so-called moat regime -- where the wave function renormalization is negative accompanying the inhomogeneous phase as expected.
\end{abstract} 

\maketitle

\tableofcontents

\section{Introduction}
\label{sec:intro}

	In this work we examine and cross-check the functionality of a technical method in \gls{qft} that can be used to investigate the thermodynamic phase structure in a broad range of systems that exhibit condensation phenomena, namely: the stability analysis of a spatially homogeneous ground state.
	The method and closely related techniques were promoted \cite{Rechenberger:2018talk,Braun:2014fga} and already applied \cite{Thies:2003kk,Nakano:2004cd,Boehmer:2007ea,Boehmer:2008uq,Basar:2009fg,Nickel:2009ke,Abuki:2011pf,deForcrand:2006zz,Wagner:2007he,Tripolt:2017zgc,Winstel:2019zfn,Carignano:2019ivp,Buballa:2018hux,Thies:2019ejd,Buballa:2020nsi,Buballa:2020xaa,Winstel:2021yok} in the context of low-energy (effective) models and theories of systems that describe different kinds of strongly interacting matter.
	
	In general, a lot of the aforementioned systems undergo phase transitions, if some external energy scale like the temperature, chemical potentials or a magnetic field \textit{etc.} is tuned.
	These phase transitions go hand in hand with the breaking and restoration of one or more symmetries and usually the formation of some condensates.
	Understanding this (thermodynamic) behavior of strongly-interacting matter is one of the major challenges in high-energy and condensed matter physics.

	Oftentimes, for the sake of simplicity and feasibility of calculations, it is assumed that condensation is homogeneous in space and that the corresponding condensates do not oscillate in time, which is usually a good first approach.
	However, it is well-known that certain systems show spatially inhomogeneous condensation -- at least in certain approximations and for specific choices of the external parameters, like chemical potentials \textit{etc.} -- meaning that the corresponding condensate oscillates in space and thus breaks translational invariance.
	In short, it has some crystal-like structure that mostly emerges in systems at high densities\footnote{Although the possibility of time-crystals was discussed and is still an object of active research, see, \textit{e.g.}, \Rcite{Wilczek:2012jt,Shapere:2012nq}, we exclude this option or related structures from our analysis for simplicity.
	}.\\
	
	Such crystalline structures are commonly found in solid-state problems, where spin-density waves are known ground states, see \Rcite{Gruner:1994zz} for a review and \Rcite{Bulgac:2008tm,Radzihovsky:2011zz,Gubbels:2012kcu,Roscher:2013cma,Baarsma:2013in} as typical examples.
	Originally such phases were predicted by P.~Fulde, R.~A.~Ferrel, A.~Larkin, and Y.~Ovchinnikov in superconducting materials at large magnetic fields \cite{Fulde:1964zz,Larkin:1964wok}.
	
	In the context of particle physics spatially inhomogeneous condensation was studied first in nuclear matter \cite{Dautry:1979bk} in the form of inhomogeneous pion condensation and later in model studies for quark matter \cite{Kutschera:1989yz,Broniowski:1990dy,Kutschera:1990xk,Deryagin:1992rw}.
	Since then, such thermodynamic phases were found or discussed to be favored over phases of spatially homogeneous condensation for certain thermodynamic state variables in various effective \glspl{qft}, which are in use to model and study various features and aspects of \gls{qcd} \cite{Kojo:2009ha,Kojo:2011cn,Buballa:2014tba,Carignano:2012sx,Carignano:2014jla,Carignano:2018hvn,Carignano:2019ivp,Buballa:2019clw,Buballa:2020xaa}.
	For example, the \gls{njl} model \cite{Nambu:1961tp,Nambu:1961fr,Klevansky:1992qe} or the \gls{qm} model, \textit{e.g.}, \Rcite{GellMann:1960np,Scavenius:2000qd, Schaefer:2006ds,Asakawa:1989bq,Buballa:2003qv}, are used to model the spontaneous breaking of chiral symmetry in \gls{qcd} -- assuming spatially homogeneous condensation/vaporization -- qualitatively very well.
	When allowing for spatially inhomogeneous condensation in these types of models, the \gls{ip} typically covers the first-order boundary between a phase of spatially homogeneous chiral condensation and the approximately\footnote{
		When considering finite, realistic quark-masses chiral symmetry is never fully restored and the approximate symmetry restoration above the critical end point is associated with a crossover transition. Therefore it is common \cite{Buballa:2019clw} to use the terms \gls{cep}, \gls{plp}, and approximate symmetry restoration instead of the terms critical point (CP), Lifshitz point (LP), and symmetry restoration used in this publication in the chiral limit.
		}
		chirally symmetric phase up to critical (end) point -- at least, if these models are studied within the mean-field approximation, where bosonic quantum fluctuations are artificially suppressed, \textit{cf.}\ \Rcite{Nakano:2004cd,Nickel:2009wj,Nickel:2009ke,Abuki:2011pf,Heinz:2013hza,Carignano:2014jla,Heinz:2015lua,Braun:2015fva,Buballa:2018hux,Lakaschus:2020caq}.
	
	A recent \gls{frg} study of full \gls{qcd} found a region in the phase diagram of \gls{qcd} with a negative wave-function renormalization \cite{Fu:2019hdw}.
	While this is only a necessary condition for inhomogeneous chiral condensation, it serves as an indicator for the possibility for such and other related exotic phases in \gls{qcd}.
	Possible experimental signals of inhomogeneities in so-called Lifshitz or moat regimes were recently discussed in \Rcite{Pisarski:2020gkx,Pisarski:2020dnx,Pisarski:2021qof,Rennecke:2021ovl}.
	We will discuss possibility of inhomogeneous phases in moat regimes further throughout this paper, especially in Sections \ref{subsec:phase_diagram_stability_analysis} and \ref{subsec:wavevector}.
	
	However, maybe the most prominent examples for spatially inhomogeneous condensation in relativistic \gls{qft} are observed in $(1 + 1)$-dimensions.
	In the $(1 + 1)$-dimensional \gls{gn} model \cite{Gross:1974jv} spatially oscillating condensates have been proven to be the true absolute ground states in some regions of the phase diagram \cite{Brzoska:2001iq,Thies:2003kk,Thies:2003br,Schnetz:2004vr,Schnetz:2005ih,Schnetz:2005vh,Thies:2005wv,Thies:2006ti}.
	Even the exact spatial modulation of the inhomogeneous condensate was derived analytically in terms of known functions \cite{Schnetz:2004vr,Schnetz:2005ih} deploying techniques of supersymmetric quantum mechanics \cite{Dunne:1997ia,Cooper:2001weo}.
	Also more involved $(1 + 1)$-dimensional models with more complicated symmetry breaking patterns exhibit an \gls{ip} \cite{Schon:2000he,Schon:2000qy,Thies:2019ejd,Thies:2020ofv,Heinz:2015lua}.
	
	For a general review regarding \glspl{ip} in the context of high-energy physics, we refer to \Rcite{Buballa:2014tba}.\\
	
	Notwithstanding all of these findings, it usually remains extremely hard to correctly predict the shapes of these inhomogeneous condensates or to merely guess appropriate ansatz functions for the search of the true absolute ground state of a system.
	Additionally, the direct search for spatially inhomogeneous condensation gets more difficult, when studying models with more elaborated features, improved approximation schemes, and/or more involved numerical schemes.
	
	Thus, in literature one typically relies on certain ansatz functions for the condensates, \textit{e.g.}, by embedding known one-dimensional solutions, such as the kink-antikink solution from the \gls{gn} model \cite{Thies:2003kk,Schnetz:2004vr,Schnetz:2005ih} or the \gls{cdw} solution as found in the chiral \gls{gn} model \cite{Schon:2000he,Schon:2000qy}, in higher-dimensional models, which reduces the functional minimization to a minimization in certain parameters of the ansatz, see, \textit{e.g.}, \Rcite{Nakano:2004cd,Kojo:2009ha,Kojo:2011cn,Carignano:2012sx,Adhikari:2017ydi,Carignano:2018hvn,Steil:2021RGMF} as well as the review \Rcite{Buballa:2014tba}.
	This, however, has the drawback that the application of ansatz functions does not exclude the possibility of other, energetically preferred oscillating solutions which are not captured by the chosen ansatz.
	An alternative possibility is, thus, the discretization of the model via lattice field theory or related methods \cite{deForcrand:2006zz, Wagner:2007he,Winstel:2019zfn,Winstel:2021yok,Heinz:2015lua, Narayanan:2020uqt, Pannullo:2021edr}, which reduces the problem to a high-dimensional optimization problem and requires continuum limits of non-trivial inhomogeneous modulations.
	This is, in general, a very challenging and time consuming numerical task and is usually restricted to severe truncations of the effective action, like the mean-field approximation. \\

	Due to these difficulties, the idea came up, that an indirect search for these exotic states of matter and thermodynamic phases of spatially inhomogeneous condensation might yield a feasible and computationally cheap alternative to direct computation.
	A possible indirect detection is possible with a so-called stability analysis.
	The main idea behind this method is to determine the ground state assuming a spatially homogeneous condensate and to study position dependent perturbations of this state in a second step.
	Hence, one expands the full-quantum effective action in the \gls{ir} in powers of the perturbation.
	By inspecting the two-point function, which is basically the curvature of the action at its homogeneous minimum (the homogeneous ground state), one can classify stable and unstable directions in field space from the sign of the two-point function.
	Thus, one is performing a functional curve sketching and searches for expansion points that are saddle points of the action.
	
	This relatively simple concept allows to examine a sufficient, but not necessary condition for an \gls{ip}, \textit{i.e.}, if the homogeneous ground state is unstable with respect to inhomogeneous perturbations, the ground state must be inhomogeneous.
	In simple terms, stability of the homogeneous condensate can be found, but the true global minimum of the action can still be an inhomogeneous field configuration located outside the range of validity of the stability analysis around the homogeneous ground state -- which is of course a limitation of this approach.
	We discuss this limitation at length in Section \ref{subsec:phase_diagram_stability_analysis}.
	
	Howsoever, the great advantage of the technique is that it is basically applicable to all kinds of models and theories and works independent of the technical method and approximation that is used.
	For example, it was applied in mean-field studies of a broad range of models, but also used in \gls{frg} calculations or in the context of lattice field theory.
	There are multiple studies, see, \textit{e.g.}, \cite{deForcrand:2006zz,Wagner:2007he,Winstel:2019zfn,Nakano:2004cd,Abuki:2011pf,Abuki:2013pla,Tripolt:2017zgc,Buballa:2018hux,Buballa:2020nsi,Buballa:2020xaa, Pannullo:2021edr}, which are based on a stability analysis or directly related approaches.
	
	To the best of our knowledge, there has not been a significant attempt to discuss the limitations and successes of this method in great detail using a fully-understood/solved model, where the exact solution is well-known.
	In addition, the afore mentioned publications were mainly focused on the discussion of physics and phenomenology. 
	Though, a simple low-level technical guide for this method seems to be missing in literature.
	Hence, our goal is to close this gap by providing a simple proof of concept.
	To this end, we revisit the $(1 + 1)$-dimensional \gls{gn} model, as it has been solved analytically in \Rcite{Schnetz:2004vr,Schnetz:2005ih} with an exact solution for the ground state for all temperatures $T$ and chemical potentials $\mu$, and extend earlier ``stability analyses''\footnote{In these works the term ``almost degenerate perturbation theory (ADPT)'' is used instead of ``stability analysis''. However, the approach is quite similar.} within this and closely related models \cite{Thies:2003kk,Boehmer:2008uq,Boehmer:2007ea,Thies:2019ejd}.
	
	A related approach is the so-called \textit{fermion doubler trick} that relies on a simple harmonic ansatz for the bosonic ground state and the minimization of the effective potential with respect to the parameters of the ansatz.
	This technique was already quantitatively benchmarked against the exact solution of the $(1 + 1)$-dimensional \gls{gn} model in \Rcite{Braun:2014fga}.
	It was shown that the lowest term in an expansion of the effective potential obtained with this technique is equivalent to the exact treatment of the bosonic two-point function.
	Therefore, some of the results produced by the approach in this work and the one presented in \Rcite{Braun:2014fga} agree exactly.
	While the fermion doubler trick is also able to produce additional qualitative results that the inspection of the bosonic two-point function alone cannot provide. The stability analysis of the homogeneous ground state, on the other hand, does not rely on a specific ansatz.  \\
	
	At this point we emphasize that this work is explicitly not about groundbreaking new results or a concept that is original to this work.
	This publication is meant primarily as a pedagogical and detailed supplement, and as a complement to existing literature -- especially \Rcite{Thies:2003kk,Basar:2009fg,Braun:2014fga}.

\subsubsection*{Structure}

	Our work is structured as follows: 
	We start our discussion with the theory \cref{sec:theory}.
	There, in \cref{subsec:the_gross-neveu_model}, we introduce the \gls{gn} model as our testing ground and afterwards, in \cref{subsec:phenomenology}, briefly recapitulate its phenomenology in the \infN{} limit, hence its phase diagram at (non-)zero baryon densities and (non-)zero temperature.
	In \cref{sec:gp} the grand canonical potential and renormalization condition are presented and the homogeneous ground state of the model is determined for arbitrary $\mu$ and $T$.
	We close the theory part with \cref{sec:stability} by introducing the formalism for the stability analysis of bosonic two-point function with respect to inhomogeneous perturbations.
	Our numerical results and the proof of concept are shown in \cref{sec:results}.
	In \cref{subsec:momentum_structure} the corresponding two-point function is evaluated and discussed for all relevant cases in the model.
	The results for the detection of inhomogeneous condensation in the phase diagram via the stability analysis are presented and compared to the analytical solution of the model in \cref{subsec:phase_diagram_stability_analysis}.
	In two additional subsections, namely \cref{subsec:wavevector,subsec:wave-function renormalization}, we compare the dominant mode of the exact inhomogeneous condensate with the dominant wave vector from the stability analysis and further comment on the values of the bosonic wave-function renormalization and its implications.
	Finally, in \cref{sec:conclusion_and_outlook} we conclude and give a brief outlook on possible promising applications.

\subsubsection*{Conventions}

	In order to simplify and structure our work, we use the following conventions:
	
	Without loss of generality and for the sake of simplicity we mostly restrict our discussion in the text and the figures to positive (background) field values and condensates ($\s$, $\shom$, $\smin$, $\ldots$) and positive values of the chemical potential $\mu$.
	However, all formulae are manifestly invariant under substitutions $\mu \mapsto - \mu$ \textit{etc.}.
	
	In all figures of this work, dimensionful quantities are measured and plotted as multiples of appropriate powers of the vacuum minimum (fermion mass) $\snull$, which has the dimension of an energy, \textit{e.g.}, we simply use $T$ instead of $T \, [ \snull ]$ or $T/\snull$ to label the axis of our plots.
	In the text and figure captions we abstain from this shorthand notation.

\section{Theory}
\label{sec:theory}

This section is dedicated to the theoretical background and setup of our analysis.

To this end, we start in \cref{subsec:the_gross-neveu_model} by introducing the $(1 + 1)$-dimensional \gls{gn} model as a toy model and prototype \gls{qft}, which serves as our testing ground. 
Afterwards in \cref{subsec:phenomenology} we recapitulate the known phenomenology of the \gls{gn} model in mean-field at (non-)zero chemical potential $\mu$ and (non-)zero temperature $T$ and briefly comment on the phenomenology beyond the \infN\ approximation.
The results of the stability analysis will later be examined against these established (mean-field) results.
In \cref{sec:gp} we introduce the renormalization condition via the gap equation and the effective potential, which are needed in the subsequent calculations for the bosonic two-point function and the wave-function renormalization.
Readers who are familiar with these topics may just skim over these subsections to get familiar with our conventions and notation.

The actual theoretical setup for the stability analysis is presented in \cref{sec:stability}, where we also derive and discuss the formulae for the bosonic two-point function and the bosonic wave-function renormalization. 
\subsection{The Gross-Neveu model}
\label{subsec:the_gross-neveu_model}

	The original \gls{gn} model \cite{Gross:1974jv} is a relativistic quantum field theory describing $\Nf \in \mathbb{N}$ fermion flavors with a four-fermion interaction in the scalar channel \cite{ZinnJustin:1991yn,ZinnJustin:2002ru,Peskin:1995ev}.
	In medium, in $(1 + 1)$ space-time dimensions (${x_1 = \ex \in \mathbb{R}}$ is the spatial coordinate and ${x_2 = \te \in [ 0, \beta )}$ is the periodic Euclidean temporal coordinate, with $\beta \in \mathbb{R}^+$) the classical action of the \gls{gn} model is given by 
		\begin{align}
			& \S [ \bar{\psi}, \psi ] =	\vphantom{\Bigg(\Bigg)}	\label{eq:gn-model}
			\\
			= \, & \int_{- \infty}^{\infty} \D x \int_{0}^{\beta} \D \tau \, \Big[ \bar{\psi} \, ( \slashed{\partial} - \mu \, \gamma^2 )  \, \psi - \tfrac{\coupling}{2 \Nf} \, ( \bar{\psi} \, \psi )^2 \Big] \, .	\vphantom{\Bigg(\Bigg)}	\nonumber
		\end{align}
	Here, ${\psi = \left( \psi_1 , \dots, \psi_{\Nf} \right)}$ contains $\Nf > 1$ two-component Dirac spinors describing massless fermion fields and $\rscoupling \in \mathbb{R}^+$ is the coupling constant.
	Non-zero baryon density is introduced by the chemical potential $\mu \in \mathbb{R}$ and the inverse temperature ${\beta = \frac{1}{T}}$ fixes the extent of the compactified Euclidean temporal direction, such that the space-time manifold is flat and presents as a cylinder.
	The Euclidean Gamma matrices are defined as the irreducible representation of the Clifford algebra 
		\begin{align}
			&	\left\{ \gamma^\mu, \gamma^\nu \right\} = 2 \, \eta_{\mu \nu} \, \I_2 \, ,	&&	\mu, \nu \in \{ 1, 2 \} \, ,
			\label{eq:Cliff}
		\end{align} 
	where ${\eta = \diag\left( 1, 1 \right)}$ is the metric of the two-dimensional cylinder and $\I_2$ is the two-dimensional identity matrix in Dirac space.
	For more details, see, \textit{e.g.}, \Rcite{Stoll:2021ori}.
	
	The action \labelcref{eq:gn-model} is invariant under translations along the spatial and Euclidean temporal direction as well as parity transformations, which form the isometries of the cylinder.
	In the limit of vanishing temperature $T$ (sending the radius of the cylinder to infinity) and vanishing chemical potential $\mu$ (removing the artificial distinction of the Euclidean temporal direction) the model \labelcref{eq:gn-model} recovers the full Euclidean Poincar\'e symmetry of flat Euclidean space-time -- including ``Euclidean boosts'' (rotations) between the spatial and temporal direction.
	A direct consequence of the masslessness and indistinguishability of the $\Nf$ fermions as well as their realization as Dirac spinors is that the model is also invariant under transformations of the group
		\begin{align}
			\U ( \Nf ) \times \Z_2 \, ,
		\end{align}
	where the $\Z_2$ symmetry is called and realizes a \textit{discrete chiral symmetry} generated by $\gch$
		\begin{align}
			&	\bar{\psi} \mapsto - \bar{\psi} \, \gch \, ,	&&	\psi \mapsto \gch \, \psi \, .	\label{eq:disc_sym_ferm}
		\end{align} 
	The operator $\gch$ is defined as the only matrix that anti-commutes with the Euclidean Gamma matrices\footnote{For the irreducible, $2 \times 2$ representation of the $(1 + 1)$-dimensional Clifford algebra one can choose ${\gamma^\nu = \sigma_\nu}$ for ${\nu = 1, 2}$, where $\sigma_\nu$ denotes the Pauli matrices or linear combinations of the Pauli matrices. Consequently, $\gch$ is typically proportional to the remaining third Pauli matrix $\sigma_3$.}.
	The $\U(\Nf)$ symmetry group can be further decomposed into a so-called phase symmetry, elements of $\U ( 1 )$, and a flavor symmetry group, elements of $\SU ( \Nf )$.
	The phase symmetry leads to conservation of the baryon number density $\bar{\psi} \, \gamma^2 \, \psi/\Nf$ that is tuned by the chemical potential $\mu$ for the fermions \cite{Fitzner:2010nv,Dunne:2011wu,Thies:2017fkr,Lenz:2020cuv}.
	On the other hand, the flavor symmetry group leads to the conservation of a vector current.
	
	It can be shown \cite{Gross:1974jv,ZinnJustin:2002ru,Peskin:1995ev} that the (grand-canonical) partition function of the GN model is equivalent to the partition function of a partially bosonized model (up to a physically irrelevant constant)\footnote{This can be shown through the isolation of a shifted Gaussian integral in the bosonic field variable in the partition function of the bosonized model.}, which is described by the following action
		\begin{align}
			& \S [ \bar{\psi}, \psi, \phi ] =	\vphantom{\Bigg(\Bigg)}	\label{eq:thermal_gny_action}
			\\
			= \, & \int_{- \infty}^{\infty} \D x \int_{0}^{\beta} \D \tau \, \Big[ \bar{\psi} \, ( \slashed{\partial} - \mu \, \gamma^2 + h \, \phi ) \, \psi + \tfrac{\Nf h^2}{2 \coupling} \, \phi^2 \Big] \, .	\vphantom{\Bigg(\Bigg)}	\nonumber
		\end{align}
	Here, $\phi$ is an auxiliary/constraint bosonic scalar field, which is real valued and has canonical energy dimension ${[ \phi ] = \text{energy}^0}$ \cite{Harrington:1974tf}.
	In this action, the four-Fermi interaction is replaced by a Yukawa-type interaction with Yukawa coupling $h$.
	To ease notation, the Yukawa coupling, that has canonical energy dimension ${[ h ] = \text{energy}}$, is absorbed into the bosonic field $\s \equiv h \phi$ (the fermion mass), \textit{i.e.}, $\s$ has the dimension of an energy.
	
	Through a specific Ward identity, see, \textit{e.g.}, \Rcite{Pannullo:2019}, the expectation value of the scalar field can be related to the fermionic expectation value $\langle \bar{\psi} \, \psi \rangle$,
		\begin{align}
			\langle \s \rangle = -\tfrac{\coupling}{\Nf} \langle \bar{\psi} \, \psi \rangle \, .	\label{eq:sigm_eqofmotion} 
		\end{align}
	For this expectation value the discrete symmetry transformation \eqref{eq:disc_sym_ferm} is realized as follows,
		\begin{equation}
			\langle \bar{\psi} \, \psi \rangle \mapsto - \langle \bar{\psi} \, \psi \rangle \, .
		\end{equation}
	Since the expectation value of $\s$ is directly proportional to the condensate, a non-vanishing $\langle \s \rangle$ implies a spontaneous breaking of the discrete chiral symmetry and generates a non-zero fermion mass -- a process that cannot be observed within a purely perturbative framework \cite{Gross:1974jv}.
	
	The bosonized action \eqref{eq:thermal_gny_action} is bilinear in the fermion field, such that the fermion fields can be integrated out to obtain an effective purely bosonic action \cite{Wolff:1985av}
		\begin{align}
			& \seff [ \s ] =	\vphantom{\bigg(\bigg)}	\label{eq:S_eff}
			\\
			= \, & \tfrac{1}{2 \rscoupling} \int_{- \infty}^{\infty} \D x \int_{0}^{\beta} \D \tau \, \s^2 - \ln \big( \Det ( \slashed{\partial} - \gamma^2 \, \mu + \s ) \big) \, ,	\vphantom{\bigg(\bigg)}	\nonumber
		\end{align}
	and grand canonical partition function
		\begin{align}
			\mathcal{Z} = \, & \mathcal{N} \int [ \D \sigma ] \, \e^{- \Nf \, \seff [\s]} \, ,	\vphantom{\bigg(\bigg)}	\label{eq:partition_sigma}
		\end{align}
	where $\mathcal{N}$ is a physically irrelevant normalization factor.
	In the limit $\Nf \to \infty$ all bosonic quantum fluctuations in \cref{eq:partition_sigma} are suppressed, such that the so-called mean-field approximation, \textit{i.e.}, the disregard all of bosonic fluctuations, becomes exact \cite{Harrington:1974tf}.
	In this approximation only global minima of $\seff [ \s ]$ contribute to the partition function \cref{eq:partition_sigma}.
	In the following, we assume that there exists one unique global minimum ${\s = \smin}$ of $\seff [ \s ]$\footnote{This is of course a simplifying assumption as there can be many equivalent minima that are connected by symmetry transformations, \textit{e.g.}, a chiral transformation. For most of our analysis, however, it suffices to consider only one of these equivalent minima.
	}.
	
	For studies of the \gls{gn} model beyond the \infN{} approximation, we refer to the discussion at the end of the following subsection. 
\subsection{Phenomenology of the Gross-Neveu model at (non-)zero \texorpdfstring{$\mu$}{mu} and (non-)zero \texorpdfstring{$T$}{T} -- the phase diagram}
\label{subsec:phenomenology}

	The phase diagram of the \gls{gn} model for $\Nf \to \infty$ is well-known, which makes it an ideal testing ground for methods in \glspl{qft}.
	Therefore, we will briefly summarize the established phenomenology of the \gls{gn} in the $\mu$-$T$ plane as benchmark and reference values for the proof of concept of the stability analysis in the following.
	
	For related (and more comprehensive) discussions and original works of the rich large-$\Nf$ phenomenology and physics of the \gls{gn} model we refer to \Rcite{Jacobs:1974ys,Dashen:1974xz,Harrington:1974te,Harrington:1974tf,Dashen:1975xh,Affleck:1981bn,Cohen:1981qz,Cohen:1983nr,Wetzel:1984nw,Shankar:1985zc,Wolff:1985av,Karsch:1986hm,Treml:1989,Rosenstein:1990nm,Gracey:1990sx,Gracey:1990wi,Gracey:1991vy,Pausch:1991ee,Chodos:1993mf,Barducci:1994cb,Schon:2000he,Schon:2000qy,Brzoska:2001iq,Blaizot:2002nh,Thies:2003br,Thies:2003kk,Thies:2003zr,Schnetz:2004vr,Thies:2005vq,Schnetz:2005ih,Thies:2005wv,Schnetz:2005ih,Schnetz:2005vh,deForcrand:2006zz,Thies:2006ti,Karbstein:2006er,Boehmer:2007ea,Karbstein:2007be,Karbstein:2007bg,Wagner:2007he,Basar:2008im,Boehmer:2008uq,Basar:2009fg,Brendel:2009pq,Zinn-Justin:2010,Fitzner:2010nv,Dunne:2011wu,Fitzner:2012gg,Fitzner:2012kb,Dunne:2013rka,Thies:2014ida,Heinz:2015lua,Thies:2017mbl,Ahmed:2018tcs,Bermudez:2018eyh,Narayanan:2020uqt,Roose:2020znu,Quinto:2021lqn,lopes2021excitonic}\footnote{This list of references is not exhaustive.}.\\
	
	Enforcing the ground state (the $\bar{\psi} \psi$-condensate) of the auxiliary field to be constant (homogeneous) in spacetime, thus ${\s(\xst ) = \shom = \mathrm{const.}}$, the so-called \textit{homogeneous phase diagram} of the Gross-Neveu model can be derived semi-analytically in the mean-field approximation.
	The entire phase structure solely depends on a single dimensionful parameter, which sets the scale for all other observables and can be chosen, \textit{e.g.}, by fixing the vacuum condensate (fermion mass) $\snull$ or any other dimensionful observable.
	
\begin{figure}
			\includegraphics[width=\columnwidth]{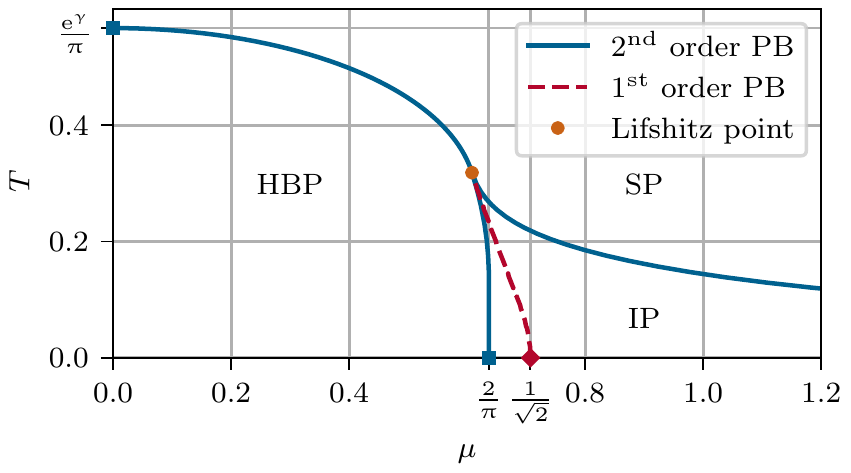}
			\caption{\label{fig:analytical_pd}The phase diagram of the \gls{gn} model in the \infN{} limit. The dashed red curve corresponds to the first-order phase boundary that is obtained if spatially homogeneous condensation is assumed \cite{Wolff:1985av,Schon:2000qy}.
				The solid blue lines correspond to second-order phase transitions, if spatially inhomogeneous condensation is taken into account \cite{Schnetz:2004vr,Schnetz:2005ih,Schnetz:2005vh,Basar:2009fg}.
			}
		\end{figure}

	One then finds a so-called \gls{hbp} with a condensate $\sminhom ( \mu, T ) \neq 0$ (broken $\Z_2$ symmetry) at small $\mu$ and $T$ and a so-called \gls{sp} with ${\sminhom ( \mu, T ) = 0}$ with restored $\Z_2$ symmetry in the rest of the $\mu$-$T$ plane, \textit{cf.}\ \cref{fig:analytical_pd}.
	The phase transition is of second-order at ${( \mu, T_c )/\snull = ( 0, \e^\upgamma/\uppi ) \simeq (0,0.567 )}$\footnote{Here $\upgamma\approx0.577$ denotes the Euler-Mascheroni constant.} \cite{Dashen:1974xz,Jacobs:1974ys,Harrington:1974tf} and ranges from ${\mu = 0}$ to the \gls{cp} at $( \mu_L, T_L )/\snull \simeq ( 0.608, 0.318 )$ \cite{Wolff:1985av}.
	At this point the phase transition becomes first-order and continues to lower temperatures until it finally terminates at ${( \mu_{c, \text{hom}}, T )/\snull = ( 1/\sqrt{2}, 0.0 ) \simeq ( 0.707, 0.0 )}$.\\ 
	
	Relaxing the restrictive assumption of homogeneous condensation and allowing for a spatially varying background field ${\s ( \xst ) =  \s ( x )}$, a modified phase diagram is obtained. Here, one finds an \gls{ip} where the ground state ${\smin ( \mu, T ) = \smin ( \mu, T, x )}$ is an oscillating function in space.
	This phase emerges for temperatures $T < T_L \simeq 0.318$ and moderate chemical potentials $\mu >  \mu_L \simeq 0.608 \, \snull$ and grows in $\mu$-direction for decreasing temperature, \textit{cf.}\ \cref{fig:analytical_pd} and  \Rcite{Thies:2003kk,Thies:2003br,Schnetz:2004vr,Schnetz:2005ih,Schnetz:2005vh,Thies:2006ti}.
	The former homogeneous first-order phase boundary is completely engulfed by the \gls{ip}.
	The novel phase transition between the \gls{ip} and the \gls{hbp} is of second-order and ranges from ${( \mu_{c}, T )/\snull = ( 2/\uppi, 0. 0 ) \simeq ( 0.637, 0.0 )}$ to a so-called \gls{lp} which is located at the position $( \mu_L, T_L )/\snull \simeq ( 0.608, 0.318 )$ of the former \gls{cp}. At the \gls{lp} three phases -- a homogeneously broken, an inhomogeneously broken, and a restored phase -- meet.
	
	At the \gls{hbp}$\pbArrow$\gls{ip} phase boundary, the phase-transition is not linked to the $\Z_2$ symmetry breaking/restoration, but rather to the breaking/restoration of spatial translational invariance, because discrete chiral symmetry is always (periodically) broken by the condensate.
	The other phase boundary from the  \gls{ip} to the \gls{sp} is also of second-order and thus all phase boundaries of the \emph{correct}/\emph{revised} mean-field phase diagram correspond to second-order phase transitions.
	Crossing the \gls{sp}$\pbArrow$\gls{ip} phase boundary the discrete chiral symmetry as well as spatial translational invariance are broken/restored.
	This crucial difference between discrete chiral symmetry and translational invariance breaking/restoration is of great importance for the remainder of this work and the limitations of the stability analysis.
	
\begin{figure}
			\includegraphics[width=\columnwidth]{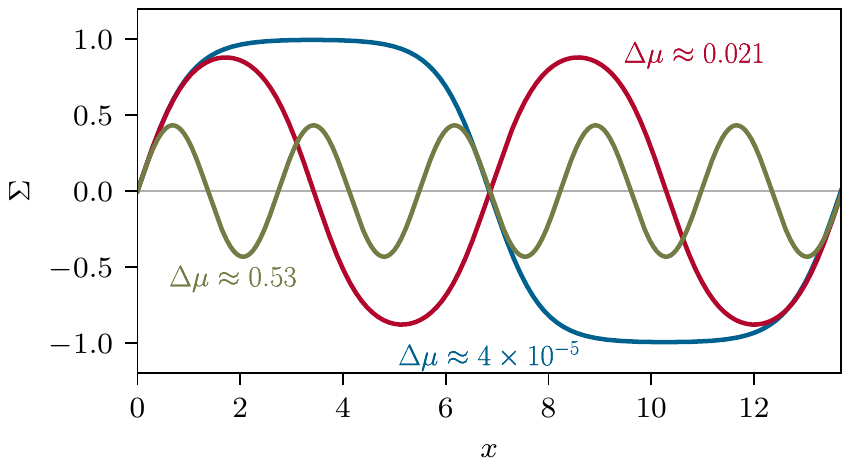}
			\caption{\label{fig:inhomogeneous_condensate}Spatial inhomogeneous chiral condensate $\smin ( \mu, T, \ex )$ at ${T = 0}$ for various chemical potentials $\mu$ with ${{\Delta\mu = ( \mu - \mu_c )/\snull}}$ in the \gls{ip} (where ${\mu_c/\snull = 2/\uppi}$).
				The curves are calculated using expressions of Refs.~\cite{Thies:2003kk,Schnetz:2004vr,Schnetz:2005ih}.
				This figure is inspired by Fig.~2 of Ref.~\cite{Thies:2003kk}.
			}
		\end{figure}

	The spatially inhomogeneous chiral condensate in the \gls{ip} is described by Jacobi elliptic functions\footnote{For definitions, properties, and relations of the involved Jacobi elliptic functions, see, \textit{e.g.}, Chap.~22 of Ref.~\cite{NIST:DLMF}.} and for increasing chemical potential 
		\begin{itemize}
			\item its shape evolves from a kink-antikink shape to a sine-like shape,
			
			\item its amplitude decreases,
			
			\item its frequency increases,
		\end{itemize}
	as shown in \cref{fig:inhomogeneous_condensate} for zero temperature, see also \Rcite{Schnetz:2004vr,Schnetz:2005ih}.
	The general behavior of the condensate is very similar at non-zero temperature, see \Rcite{Schnetz:2004vr} for details.\\
	
	Before the more technical aspects of our work are discussed, we want to point out that the phase structure of the \gls{gn} model was also investigated at imaginary chemical potential \cite{Karbstein:2006er} and non-zero bare fermion mass \cite{Schnetz:2005ih,Schnetz:2005vh, Thies:2006ti}.
	It was also discussed, how the bosonic fluctuations alter the phenomenology of the \gls{gn} model, if the \infN{} approximation is relaxed to a finite number $\Nf$ of fermions \cite{Karsch:1986hm,Dashen:1974xz,Harrington:1974tf}.
	Only recently some of the authors and their collaborators returned to this intriguing and non-trivial question \cite{Stoll:2021ori,Pannullo:2019bfn,Pannullo:2019prx,Lenz:2020bxk,Lenz:2020cuv}.
	Currently, it seems as if bosonic fluctuations at finite $\Nf$ vaporize any form of homogeneous condensate (in the infinite volume limit), such that the \gls{hbp} vanishes completely, except potentially for ${T = 0}$ \cite{Stoll:2021ori}.
	For extended discussions, we refer to these works and recent related works in the chiral \gls{gn} model \cite{Lenz:2021kzo,Horie:2021wnn}.
	
	In this publication, however, we are primarily interested in technical aspects of the stability analysis and, therefore, stick to the \infN{} limit, where exact reference solutions are available.
	Still, we come back to implications of the findings of this work on calculations that involve fluctuations of the bosonic sigma mode. 
\subsection{The grand canonical potential}
\label{sec:gp}

	In the following paragraphs we recapitulate some well-known formulae and results for the (mean-field) effective potential and the associated gap equation.
	The latter will serve as a renormalization condition for the effective potential as well as the bosonic two-point function and related wave-function renormalization.
	Additionally, the renormalized effective potential is also needed for the (numerical) determination of the physical point (the homogeneous minimum $\sminhom$), where the bosonic two-point function and wave-function renormalization are evaluated in the stability analysis.
	Still, all results of this subsection can be found elsewhere, \textit{e.g.}, in the \Rcite{Stoll:2021ori,Schon:2000qy,Pannullo:2019}, and are only presented for the sake of completeness and consistency.\\

	The ($\frac{1}{\Nf}$-rescaled) effective potential in the mean-field approximation is defined as
		\begin{equation}
		    \gp [ \s ] \equiv \tfrac{1}{\beta V} \, \seff [ \s ] \, ,	\label{eq:effective_potential}
		\end{equation}
	where $V$ is the infinite spatial ``volume'' of the spacetime manifold (the infinite axial extent of the cylinder).
	Furthermore, the grand canonical potential $\grandp ( \mu, T )$ is defined as the mean-field effective potential \cref{eq:effective_potential} evaluated at its global minimum $\smin ( \mu, T, \xst )$, \textit{i.e.},
		\begin{equation}
			\grandp ( \mu, T ) \equiv \gp [ \smin ( \mu, T, \xst ) ] \, .
		\end{equation}
	In this work, the effective potential is merely considered for homogeneous background fields ${\sigma ( \xst ) = \shom = \mathrm{const.}}$, such that the mean-field potential $\gp [ \s ]$ reduces to a function $\gphom ( \shom, \mu, T )$.
	Hence, the homogeneous grand potential is given by 
		\begin{equation}
			\grandphom ( \mu, T ) = \gphom ( \sminhom ( \mu, T ), \mu, T ) \, ,
		\end{equation}
	where $\sminhom ( \mu, T )$ is the global homogeneous minimum\footnote{As discussed in Section \ref{subsec:phenomenology} there are points in the $\mu$-$T$ plane, where this minimum is not unique, when allowing exclusively for spatially homogeneous condensation.
	This is the case exactly on the first-order phase transition line between the \gls{hbp} and \gls{sp}.}
	of the effective potential at some $\mu$ and $T$.
	In this case, the eigenvalues of the Dirac operator
		\begin{align}
			\dophom ( \xst, \st{y} ) = \beta ( 2 \uppi ) \, \delta^{(2)} ( \xst - \st{y} ) \, \big( \slashed{\partial}_\st{y} - \gamma^2 \, \mu + \shom \big)
		\end{align}
	can be calculated analytically and one obtains from \cref{eq:S_eff}
	\begin{widetext}
		\begin{align}
			 \gphom ( \shom, \mu, T ) =& \tfrac{1}{2 \rscoupling} \, \shom^2 - \tfrac{1}{\uppi} \int_{0}^{\infty} \D p \, \tfrac{1}{\beta} \sum_{n = - \infty}^{\infty}  \ln \Big(\left(\nu_n - \ii \mu\right)^2 + E_p^2 \Big) =  \vphantom{\Bigg(\Bigg)}	\label{eq:gphom}
			\\
			 =&\tfrac{1}{2 \rscoupling} \, \shom^2 - \tfrac{1}{\uppi} \int_{0}^{\infty} \D p \, \Big\{ E_p 
			  + \tfrac{1}{\beta} \Big[ \ln \Big( 1 + \e^{- \beta ( \energy{p} + \mu )} \Big) + \ln \Big( 1 + \e^{- \beta (\energy{p} - \mu)} \Big) \Big] \Big\} \, ,	\nonumber
		\end{align}
	\end{widetext}
	\textit{cf.} \Rcite{Schon:2000qy,Harrington:1974tf}, with the energies ${{\energy{p} = \sqrt{p^2 + \shom^2}}}$ and fermionic Matsubara frequencies \cite{Matsubara:1955ws} ${{\nu_n = 2 \uppi \, T \, \big( n - \frac{1}{2} \big)}}$. Since the vacuum contribution of the integrand in the second line is divergent in the \gls{uv}, the expression is regularized, with a sharp momentum cutoff $\Lambda$\footnote{Other regularization schemes are possible but in the renormalized limit $\Lambda^2/\snull^2 \to \infty$ all observables are regularization scheme independent  in the present one-loop/mean-field computations as long as the observables are computed in a consistent manner, \textit{viz.} in a unified regularization scheme.
	}.
	We choose a fixed vacuum expectation value (Fermion mass) $\snull$ of $\sigma$ as the renormalization condition using the gap equation in the vacuum. Other choices for a renormalization condition like, \textit{e.g.}, fixed sigma curvature mass or critical temperature, can be related to our choice by simple rescalings since the \gls{gn} model has only one scale in the $\infN$ limit in vacuum. The generic gap equation is defined as the extremal condition 
		\begin{equation}
			\tfrac{\D}{\D \shom} \, \gphom ( \shom, \mu, T ) \big|_{\shom = \sminhom ( \mu, T )} \overset{!}{=} 0 \, ,	\label{eq:gap_eq}
		\end{equation}
    which evaluates to \cite{Wolff:1985av,Dashen:1974xz}
		\begin{equation}
		   \sminhom ( \mu, T ) \, \big[ \tfrac{1}{\rscoupling} - \ell_1 ( \sminhom ( \mu, T ), \mu, T )  \big] = 0
		   \label{eq:gap} 
		\end{equation}
	with
		\begin{align}
			& \ell_1 ( \shom, \mu, T ) \equiv	\vphantom{\Bigg(\Bigg)}	\label{eq:l1def}
			\\
			\equiv \, & \tfrac{2}{\uppi} \int_{0}^{\infty} \D p \, \tfrac{1}{\beta} \sum_{n = - \infty}^{\infty} \frac{1}{\left(\nu_n - \ii \mu\right)^2 + E_p^2} =	\vphantom{\Bigg(\Bigg)}	\nonumber
			\\ 
			= \, & \tfrac{1}{\uppi} \int_{0}^{\infty} \D p \, \tfrac{1}{\energy{p}} [ 1 - n ( \energy{p}, \mu ) - n ( \energy{p}, - \mu ) ] \, .	\vphantom{\Bigg(\Bigg)}	\nonumber
		\end{align}
	After summation over $n$ one introduces the Fermi-Dirac distribution function \cite{Fermi:1926,Fermi:1999ncp,Dirac:1926jz}
	\begin{equation}
		n(\energy{}, \mu) \equiv \frac{1}{\e^{\beta(\energy{} + \mu)} + 1}.
	\end{equation}
	Evaluating \Eqref{eq:gap} in the vacuum (${T=\mu=0}$) and excluding the trivial solution ${\shom = 0}$, one finds the explicit renormalization condition \cite{Wolff:1985av}
		\begin{align}
			\tfrac{1}{\rscoupling} = \tfrac{1}{\uppi} \int_{0}^{\infty} \D p \, \frac{1}{\sqrt{p^2+\snull^2}} \, ,	\label{eq:renorm_cond}
		\end{align}
	where 
		\begin{align}
			\snull \equiv \sminhom ( \mu, T ) \big|_{\mu = 0, T = 0}	\label{eq:snull}
		\end{align}
	is the vacuum expectation value of $\s$ in the mean-field approximation.
	Consistently with \cref{eq:gphom} the \gls{uv} divergence of the integrand in \cref{eq:renorm_cond} is regularized via a sharp momentum cutoff $\Lambda$, which evaluates to \cite{Gross:1974jv,Jacobs:1974ys,Wolff:1985av}
		\begin{equation}
			\tfrac{1}{\rscoupling} = \tfrac{1}{\uppi} \arcoth \left( \sqrt{ 1 + \tfrac{\snull^2}{\Lambda^2} } \, \right) \, .	\label{eq:running_coupling}
		\end{equation}
	This result for the four-Fermi coupling $g^2$ reflects the asymptotically free behavior of the original \gls{gn} model \cref{eq:gn-model}, \textit{cf.}\ \Rcite{Gross:1974jv,ZinnJustin:2002ru}, since $g^2$ vanishes in the renormalized limit $\Lambda^2/\snull^2 \to \infty$, and is a notion of dimensional transmutation.
	However, inserting \cref{eq:running_coupling} into \Eqref{eq:gphom}, the $\Lambda$-dependent contributions cancel each other for $\Lambda^2/\snull^2 \to \infty$.
	One obtains the universal, renormalized result
		\begin{align}
			\gphom ( \shom, \mu, T ) = \, & \tfrac{1}{2 \uppi} \bigg\{ \tfrac{1}{2} \, \shom^2 \Big[ \ln \Big( \tfrac{\shom^2}{\snull^2} \Big) - 1 \Big] +	\vphantom{\Bigg(\Bigg)}	\label{eq:gphom_renorm}
			\\
			& - \int_{0}^{\infty} \D p \, \tfrac{p^2}{\energy{p}} [ n ( \energy{p}, \mu ) + n ( \energy{p}, - \mu ) ] \bigg\} \, ,	\vphantom{\Bigg(\Bigg)}	\nonumber
		\end{align}
	where the medium contributions in the second term under the momentum integral stem from partial integration of the two posterior terms in the integrand in \Eqref{eq:gphom}\footnote{The surface term vanishes in the renormalized limit $\Lambda^2/\snull^2 \to \infty$.}.
	In order to find the homogeneous minimum $\sminhom ( \mu, T )$ of $\gphom ( \shom, \mu, T )$ for all $\mu$ and $T$ \Eqref{eq:gphom_renorm} is minimized (numerically), which yields the well-known homogeneous phase diagram with a first-order phase transition as shown in  \cref{fig:analytical_pd}.

\subsection{Stability analysis of the spatially homogeneous \texorpdfstring{$\Z_2$}{Z 2} symmetric phase}
\label{sec:stability}

	In this section we turn to the theoretical considerations behind the stability analysis to detect spatially inhomogeneous phases.
	Our derivation is based on \Rcite{Rechenberger:2018talk}, but similar discussions can be found in \Rcite{Carignano:2019ivp,Buballa:2020nsi,Nakano:2004cd,Buballa:2018hux}.

	In order to relax the assumption of spatially homogeneous condensation and to search for a spatially \gls{ip}, one has to find the global $\ex$-dependent minima $\Sigma ( \mu, T, \ex )$ of the functional $\gp [ \sigma ( \ex ) ]$ for all possible field configurations $\sigma ( \ex )$.
	Generically -- as was already discussed in \cref{sec:intro}  -- this is an extremely challenging task, both analytically and numerically.
	In case of the \gls{gn} model in $(1 + 1)$ dimensions this problem was nevertheless solved analytically in \Rcite{Schnetz:2004vr,Schnetz:2005ih,Schnetz:2005vh} resulting in the full phase diagram of the \gls{gn} model which we discussed in \cref{subsec:phenomenology}.
		
	For more involved models, no analytic procedures are known to find self-consistent and energetically preferred solutions for inhomogeneous condensate. 
	Beyond mean-field studies of explicitly inhomogeneous phases using functional methods like the \gls{frg} are not feasible due to the complex nature of the involved propagators which are non-diagonal in momentum space.
	One exception which leverages specific analytic properties of a known ansatz function, \textit{viz.} of the \gls{cdw}, is discussed in \Rcite{Steil:2022phd,Steil:2021RGMF}.
	Also a full direct numeric minimization of the effective potential becomes extremely challenging and expensive due to the high dimensionality of the optimizations problem for a lot of models.
	On top of that, model-specific problems may arise, such that there is no ``one-fits-all'' approach for direct detection of spatially inhomogeneous condensates.
	
	Due to these challenges, the idea of an indirect detection of inhomogeneous condensation arose, which is based on analyzing the stability of the spatially homogeneous ground state against inhomogeneous perturbations.
	This was already discussed and/or applied in various contexts \cite{Fu:2019hdw,Winstel:2021yok,Buballa:2020nsi,Winstel:2019zfn,Buballa:2018hux} and to some extent also in the context of the (chiral) \gls{gn} model \cite{Thies:2003kk,Thies:2019ejd}.
	The proposed approach allows to search for a sufficient condition for an \gls{ip}, \textit{i.e.}, if $\sminhom ( \mu, T )$ is unstable against spatially inhomogeneous perturbation, the true ground state has to be inhomogeneous.
	In general, $\sminhom ( \mu, T )$ could, however, be stable against inhomogeneous perturbations, but by a functional minimization of $U [ \sigma ]$ one may still find an inhomogeneous ground state.
	In this scenario the local, homogeneous and the global, inhomogeneous minimum are separated by a ``potential barrier'', which prevents a stability analysis and calls for global minimization approaches.
	This implies that the stability analysis is expected to work properly in the vicinity of second-order phase boundaries between homogeneous and inhomogeneous phases.
	Nonetheless, this model and approximation independent method is still a powerful tool for the search of exotic phases of matter.\\
	
	On a formal level the stability analysis is based on a functional (Taylor) expansion of the effective action ${\Gamma [ \s ] = \seff [ \s ]}$ about a spatially homogeneous background field $\shom$ in powers of an inhomogeneous perturbation $\dsx{\ex}$.
	A spatially homogeneous ground state is considered to be unstable, if the second-order coefficient of this expansion exhibits some unstable direction in field space, if it is evaluated on the spatially homogeneous minimum $\sminhom ( \mu, T )$.
	The second-order Taylor coefficient is the bosonic two-point function $\gtwo$, which is analyzed in momentum space.
	An unstable direction in field space corresponds to a negative value $\gtwo$ for external spatial momentum $q$, which can be associated to the wave-vector of a spatially oscillating energetically preferred ground state.
	This implies a lower ground state energy for the inhomogeneous phase when compared to the homogeneous one assuming the higher-order contributions of the expansion beyond the second-order are either positive or negligible.
	Whether or not this assumptions are met depends on the model, the expansion point and the magnitude of the inhomogeneous oscillations as we will discuss in detail in \cref{sec:results}.
	
	Hence, the stability analysis corresponds to searching for a sign change in $\gtwo$ at some external momentum $q$ in the $\mu$-$T$-plane.
	
	Since the aim of this work is a proof of concept for this method using the well-known solution of the \gls{gn} model, we will introduce the method and needed formulae in detail in the following paragraphs.
	This theoretical discussion already foreshadows limitations of the methods.
	
	We close this section with a discussion on the wave-function renormalization, which is the second-order Taylor coefficient/moment of the bosonic two-point function in momentum space about ${q=0}$.

\subsubsection{The bosonic two-point function}

	We start the derivation of the bosonic two-point function by expanding the effective action \cref{eq:S_eff} about the homogeneous background field by splitting\footnote{Note that we do not start at the energetically lowest homogeneous state.
		All derived formulae are valid for an arbitrary $\shom$, although statements about a possible, inhomogeneous ground state can only be made after setting ${\shom = \sminhom ( \mu, T )}$.}
		\begin{align}
			\s(\ex) = \shom + \dsx{\ex} \, ,
		\end{align}
	with the inhomogeneous perturbation $\dsx{\ex}$.
	As a consequence, the Dirac operator can also be split into the sum of inverse propagator of free fermions with constant mass $\shom$ and a $\ds$-dependent term, \textit{i.e.},
		\begin{equation}
			\dop ( \xst, \st{y} ) = \dophom ( \xst, \st{y} ) + \beta ( 2 \uppi ) \, \delta^2 ( \xst - \st{y} ) \, \I_2 \, \dsx{\ex} \, .
		\end{equation}
	For the fermion-loop contribution to the effective action in \cref{eq:S_eff} this implies,
		\begin{align}
			\ln\Det ( \dop ) = \, & \ln \Det \Big( \dophom \, \big( 1 + \dophom^{-1} \ds \big) \Big) =	\vphantom{\Bigg(\Bigg)}
			\\
			= \, & \Tr \ln ( \dophom ) + \Tr \ln \big( 1 + \dophom^{-1} \ds \big) =	\vphantom{\Bigg(\Bigg)}	\nonumber
			\\
			= \, & \Tr \ln ( \dophom ) - \sum_{n = 1}^{\infty} \tfrac{1}{n} \Tr \big( - \dophom^{-1} \ds \big)^n \, ,	\vphantom{\Bigg(\Bigg)}	\nonumber
		\end{align}
	where we slightly eased our notation for the sake of readability.
	Consequently, the full effective action \cref{eq:S_eff} can be rewritten as follows,
		\begin{align}
			\seff = \sum_{n = 0}^{\infty} \seff^{( n )} \, ,
		\end{align}
	where $\seff^{(n)}$ contains the $n$-th order contributions in $\ds$.
	Specifically, we find for the three lowest-order terms\footnote{Note that we recover the effective potential for a spatially homogeneous background field as the lowest order contribution $\seff^{(0)} \equiv \beta V \gphom$, see \cref{sec:gp}.}
		\begin{align}
			\seff^{( 0 )} = \, &  \tfrac{\beta V}{2 \rscoupling} \, \shom^2 - \Tr \ln ( \dophom )  \, ,	\vphantom{\Bigg(\Bigg)}
			\\
			\seff^{( 1 )} = \, &  \tfrac{\beta}{\rscoupling} \, \shom \int_{- \infty}^{\infty} \D x \, \ds - \Tr \big( \dophom^{- 1} \ds \big)  \, ,	\vphantom{\Bigg(\Bigg)}	\label{eq:Seff1}
			\\
			\seff^{( 2 )} = \, &  \tfrac{\beta}{2 \rscoupling} \int_{- \infty}^{\infty} \D x \, \ds^2 + \tfrac{1}{2} \Tr \big( \dophom^{- 1} \ds \dophom^{- 1} \ds \big)  \, .	\vphantom{\Bigg(\Bigg)}	\label{eq:Seff2}
		\end{align}
	The functional traces above are defined as
		\begin{align} 
			& \Tr \Big( \big( \ds \dophom^{-1} \big)^n \Big) \equiv	\vphantom{\Bigg(\Bigg)}
			\\
			\equiv \, & \int \prod_{j = 1}^{n} \D^2 x^{( j )} \, \tr \big( \dsx{\ex^{( 1 )}} \, \dophom^{-1} ( \xst^{( 1 )}, \xst^{( 2 )} ) \cdots	\vphantom{\Bigg(\Bigg)}	\nonumber
			\\
			& \qquad \, \cdots \dsx{\ex^{( n )}} \, \dophom^{- 1} ( \xst^{( n )}, \xst^{( 1 )} ) \big) \, ,	\vphantom{\Bigg(\Bigg)}	\nonumber
		\end{align}
	where $\textrm{tr}$ denotes a trace in spinor space. These expressions are evaluated in momentum space using the Fourier representation,
		\begin{align}
			& \dophom^{-1} ( \xst, \st{y} ) \equiv	\vphantom{\bigg(\bigg)}
			\\
			\equiv \, & \tfrac{1}{\beta V} \sumint \e^{\ii\st{p} \cdot ( \xst - \st{y} )} \, \dophom^{-1} ( \st{p} ) =	\vphantom{\bigg(\bigg)}	\nonumber
			\\
			= \, & \tfrac{1}{\beta} \int_{- \infty}^{\infty} \frac{\D p}{2 \uppi}\,  \sum_{n = - \infty}^{\infty} \e^{\ii [ \nu_n ( \ta{x} - \ta{y} ) + p ( \xOne{x} - \xOne{y} ) ]} \, \dophom^{-1} ( \nu_n, p ) \, ,	\vphantom{\bigg(\bigg)}	\nonumber
		\end{align}
	where
		\begin{align}
			\label{eq:fermion_prop}
			&	\dophom^{-1} ( \nu_n, p ) = \frac{- \ii \gamma^\nu \tilde p_\nu + \I_2 \, \shom}{ \st{\tilde p}^2 + \shom^2} \, ,
			&&	\st{\tilde p} \equiv
			\left(
				\begin{array}{c}
					\nu_n - \ii \mu
					\\
					p
				\end{array}
			\right) \, ,
		\end{align}
	is the Euclidean propagator in momentum space. The Fourier representation of the inhomogeneous fluctuations is
		\begin{align}
			\dsx{\ex} \equiv \int_{- \infty}^{\infty} \frac{\D q}{2 \uppi} \, \e^{\ii q \ex} \, \dstx{q} \, .
		\end{align}
	For the first-order contribution to the effective action \labelcref{eq:Seff1} one derives
		\begin{align}
			\seff^{( 1 )} = \dstx{0} \, \tfrac{\beta}{\rscoupling} \, \shom \, \big[ 1 -  \rscoupling \, \ell_1 ( \shom, \mu, T ) \big] \, , 
		\end{align}
	which vanishes, if it is evaluated at ${\shom = \sminhom ( \mu, T )}$\footnote{This finding is intuitive, since the first-order contributions are only proportional to the Fourier coefficient with zero momentum, which are expected to vanish due to the extremal condition on $\seff^{(0)}[\sminhom]$.
	}, due to the homogeneous gap equation \eqref{eq:gap}.
	Because the first-order contribution vanishes at the homogeneous minimum $\sminhom ( \mu, T )$, instabilities towards inhomogeneous fluctuations can be detected by analyzing the second-order coefficient
		\begin{align}
			\seff^{( 2 )} = \tfrac{\beta}{2} \int_{- \infty}^{\infty} \frac{\D q}{2 \uppi} \, \gtwovardef \, \dstx {q} \, \dstx {- q} \, 	\label{eq:Seff2q}
		\end{align}
	at ${\shom = \sminhom ( \mu, T )}$.
	Thereby
		\begin{align}
			& \gtwovardef \equiv	\vphantom{\bigg(\bigg)}	\label{eq:Gamma}
			\\
			\equiv \, & \tfrac{1}{g^2} +
			\begin{gathered}
				\includegraphics{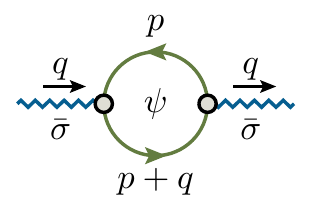}
			\end{gathered}\nonumber
			\\
			= \, &  \tfrac{1}{\rscoupling} - \ell_1 ( \shom, \mu, T ) - \ell_2 ( \shom, \mu, T, q ) \, 	\vphantom{\bigg(\bigg)}	\nonumber
		\end{align}
	is the bosonic two-point function.
	The fermion propagators are given by \cref{eq:fermion_prop} and the bosonic legs are amputated when going from \cref{eq:Seff2q} to \cref{eq:Gamma}.
	Here, we again used \cref{eq:l1def} and additionally defined
\begin{widetext}
		\begin{align}
			\ell_2 ( \shom, \mu, T, q ) \equiv \tfrac{2}{\uppi} \int_{0}^{\infty} \D p \, \tfrac{1}{\beta} \sum_{n = - \infty}^{\infty} \frac{p \, q - 2 \, \shom^2}{( \nu_n - \ii \mu )^2 + \energy{p + q}^2} \frac{1}{( \nu_n - \ii \mu )^2 + \energy{p}^2} \, . \label{eq:l2def}
		\end{align}
	Carrying out the Matsubara sum, one obtains
		\begin{align}
			\ell_2 ( \shom, \mu, T, q ) = \, & - \Big( \tfrac{q^2}{2} + \shom^2 \Big) \, \tfrac{2}{\uppi} \int_{0}^{\infty} \D p \, \tfrac{1}{\energy{p}} \, \Big( \tfrac{1}{\energy{p + q}^2 - \energy{p}^2} + \tfrac{1}{\energy{p - q}^2 - \energy{p}^2} \Big) \, [ 1 - n ( \energy{p}, \mu ) - n ( \energy{p}, - \mu ) ] \, .	\label{eq:ell2}
		\end{align}
	Evaluating \cref{eq:l2def} at ${q = 0}$ first, the summation over Matsubara frequencies yields\footnote{The original reason why it is challenging to derive \cref{eq:ell2_q0} directly from \cref{eq:ell2} is that the poles of the two propagators in \cref{eq:l2def} become degenerate for ${q = 0}$.
	} 
		\begin{align}
			\ell_2(\shom, \mu, T, 0) = - \shom^2 \, \tfrac{2}{\uppi} \int_{0}^{\infty} \D p \, \tfrac{1}{4 \energy{p}^3} \, \Big\{ 1 - ( 1 + \beta \energy{p} ) \, \big[ n ( \energy{p}, \mu ) + n ( \energy{p}, - \mu ) \big] + \beta \energy{p} \, \big[ n^2 ( \energy{p}, \mu ) + n^2 ( \energy{p}, - \mu ) \big] \Big\} \ .	\label{eq:ell2_q0}
		\end{align}
	Note that Eq.~\eqref{eq:Gamma} for $\gtwovardef$ is still ill-conditioned due to the \gls{uv} divergency in $\ell_1 ( \shom, \mu, T )$. This divergence has to be, in accordance with \cref{sec:gp}, treated with the same momentum cutoff regularization, \textit{viz.}, the sharp cutoff $\Lambda$. Again, we renormalize via the gap equation \eqref{eq:renorm_cond} in the vacuum, such that $\frac{1}{g^2} -\ell_1 ( \shom, \mu, T )$ remains finite. The resulting expression for the renormalized bosonic two-point function is\footnote{For similar results in other (higher-dimensional) models, \textit{cf.}\ Refs.~\cite{Carignano:2019ivp,Buballa:2020nsi}.
	}
		\begin{align}
			\gtwovardef = \tfrac{1}{\uppi} \Big\{ \tfrac{1}{2} \ln \Big( \tfrac{\shom^2}{\snull^2} \Big) + \int_{0}^{\infty} \D p \, \tfrac{1}{\energy{p}} \, [ n ( \energy{p}, \mu ) + n ( \energy{p}, - \mu ) ] - \ell_2 ( \shom, \mu, T, q )  \Big\} \, .	\label{eq:gtwo_renorm}
		\end{align}
\end{widetext}
	This result can be further simplified, if at least one of the four arguments $\shom$, $\mu$, $T$, and $q$ is zero.
	All possible cases and the respective simplifications of $\gtwo$ are listed in \cref{tab:gamma_limits} and the corresponding symbolic results can be found in \cref{app:the_bosonic_two-point_function}.
	A detailed derivation and discussion of the different cases is presented in \Rcite{Koenigstein:2022phd}.
	
	Negative values of $\gtwovar{\sminhom ( \mu, T )}{\mu}{T}{q}$ indicate an instability of the homogeneous minimum $\sminhom ( \mu, T )$ with respect to inhomogeneous perturbations of momentum $q$.
	Consequently, the two-point function can be used to search for inhomogeneous ground states for arbitrary $\mu$ and $T$.
	This can be done by analyzing $\gtwovar{\sminhom ( \mu, T )}{\mu}{T}{q}$ as a function of $q$ for each point $( \mu, T )$ in the phase diagram.
	In practice one searches for regions where $\gtwo(q)$ is negative.
	
	We close this discussion by noting, that the stability analysis via the bosonic two-point function, as it is presented in this section, goes beyond an improved Ginzburg-Landau expansion of the effective action \cref{eq:S_eff}, where a finite order in derivative terms (couplings) is taken into account \cite{Basar:2009fg,Nickel:2009ke,Abuki:2011pf}.
	The bosonic two-point function \cref{eq:gtwo_renorm} retains its full momentum structure, which makes our approach applicable for wave vectors $q$ of all magnitudes without the limitation to small $q$.
	Additionally, $\gtwo$ does not even need to be analytic for all $q$.
	This was already pointed out in Ref.~\cite{Carignano:2019ivp} and can be explicitly seen in \cref{fig:gamma2_compare_2} in \cref{sec:results}.
	It is also the reason, why the stability analysis is still predicting instabilities of the homogeneous condensate correctly for extremely small and even vanishing temperatures, see below.

\subsubsection{The bosonic wave-function renormalization}

	In the past it was speculated, if it might be sufficient to study the curvature of $\gtwovar{\sminhom ( \mu, T )}{\mu}{T}{q}$ at ${q = 0}$, \textit{i.e.}, the second-order coefficient of a Taylor expansion of the two-point function in momentum space about ${q=0}$, which is usually referred to as wave-function renormalization \cite{Dashen:1974xz,Buballa:2018hux,Carignano:2019ivp},
		\begin{equation}
			\zwave{\shom}{\mu}{T} \equiv \tfrac{1}{2} \tfrac{\D^2}{\D q^2} \, \gtwovardef \Big|_{q = 0} \, .	\label{eq:wave-function_renormalization_definition}
		\end{equation}
	It was speculated that a negative bosonic wave-function renormalization might be sufficient to destabilize spatially homogeneous ground states and to energetically favor gradients in the spatial profile of the ground state over a spatially uniform ground state field configuration.
	
	As pointed out, \textit{e.g.}, in \Rcite{Roscher:2015xha,Fu:2019hdw} and during informal discussions at conferences, a negative wave-function renormalization might only be some rather vague hint towards the possibility for spatial modulations of the ground state, but it is by no means a sufficient criterion, because higher-order momentum dependencies of the bosonic two-point function might again disfavor spatially inhomogeneous condensation over homogeneous condensation.
	Consequently, a study of the full momentum structure of the two-point function is necessary.
	
	Nevertheless, the wave-function renormalization is an extremely important quantity in \glspl{qft}, as it directly enters the dispersion relations \cite{Pisarski:2021qof,Rennecke:2021ovl}. It is worth to derive a stand-alone symbolic formula instead of merely extracting $\zwave{\sminhom ( \mu, T )}{\mu}{T}$ numerically from $\gtwovar{\sminhom ( \mu, T )}{\mu}{T}{q}$.
	From \cref{eq:wave-function_renormalization_definition} one finds \cite{Rechenberger:2018talk,Koenigstein:2022phd}
		\begin{align}
			\zwave{\shom}{\mu}{T} = \, & \tfrac{1}{\uppi} \int_{0}^{\infty} \D p \, \tfrac{1}{\beta} \sum_{n = - \infty}^{\infty} \frac{1}{[ ( \nu_n + \mathrm{i} \mu )^2 + \energy{p}^2 ]^2} \times	\vphantom{\Bigg(\Bigg)}	\nonumber
			\\
			& \quad  \times \bigg( 1 - \frac{\frac{4}{3} \, \shom^2}{( \nu_n + \mathrm{i} \mu )^2 + \energy{p}^2} \bigg) \, .	\vphantom{\Bigg(\Bigg)}	\label{eq:wave-function_renormalization}
		\end{align}
	The Matsubara frequency summations can again be evaluated using the residue theorem and contour integration by means of pen and paper calculations and one finally obtains \cref{eq:z_sigma_mu_t} from \cref{app:the_bosonic_wave-function_renormalization}.
	Also this result can be further simplified for the different scenarios, where $\shom$, $\mu$, and $T$ take either non-vanishing or vanishing values, by (partially) executing the remaining momentum integrations.
	In \cref{tab:z_limits} we list all possible combinations of $\shom$, $\mu$, and $T$ being (non-)zero.
	More details on the explicit derivation and some consistency checks are again presented in \Rcite{Koenigstein:2022phd}.
	
	The physically relevant wave-function renormalization is again obtained, if \cref{eq:wave-function_renormalization} is ultimately evaluated at the homogeneous ground state ${\shom = \smin ( \mu, T )}$.
	
	\setlength{\extrarowheight}{3pt}
	\begin{table}
		\caption{\label{tab:gamma_limits} The table lists the equation numbers of the explicit expressions of \cref{app:the_bosonic_two-point_function} for the momentum-dependent bosonic two-point function $\Gamma^{(2)} ( \sigma, \mu, T, q )$ in different limits. The formulae for $\Gamma^{(2)} ( \sigma, \mu, T, q )$ are simplified in terms of known functions as far as possible.
		}
		\begin{ruledtabular}
			\begin{tabular}{c c c c c}
				$T$							&	$\s$						&	$\mu$		&	$q \neq 0$						&	$q = 0$
				\\
				\hline
				\multirow{4}{*}{$\neq 0$}	&	\multirow{2}{*}{$\neq 0$}	&	$\neq 0$	&	\cref{eq:gamma2_sigma_mu_t_q}	&	\cref{eq:gamma2_sigma_mu_t_0}
				\\
				\cline{3-5}
											&								&	$= 0$		&	\cref{eq:gamma2_sigma_0_t_q}	& \cref{eq:gamma2_sigma_0_t_0}
				\\
				\cline{2-5}
											&	\multirow{2}{*}{$= 0$}		&	$\neq 0$	&	\cref{eq:gamma2_0_mu_t_q}		& \cref{eq:gamma2_0_mu_t_0}
				\\
				\cline{3-5}
											&								&	$= 0$		&	\cref{eq:gamma2_0_0_t_q}		& \cref{eq:gamma2_0_0_t_0}
				\\
				\hline
				\multirow{4}{*}{$= 0$}		&	\multirow{2}{*}{$\neq 0$}	&	$\neq 0$	&	\cref{eq:gamma2_sigma_mu_0_q}	& \cref{eq:gamma2_sigma_mu_0_0}
				\\
				\cline{3-5}
											&								&	$= 0$		&	\cref{eq:gamma2_sigma_0_0_q}	& \cref{eq:gamma2_sigma_0_0_0}
				\\
				\cline{2-5}
											&	\multirow{2}{*}{$= 0$}		&	$\neq 0$	&	\cref{eq:gamma2_0_mu_0_q}		& \cref{eq:gamma2_0_mu_0_0}
				\\
				\cline{3-5}
											&								&	$= 0$		&	\cref{eq:gamma2_0_0_0_q}		&	``$- \infty$''
			\end{tabular}
		\end{ruledtabular}
	\end{table}

	\setlength{\extrarowheight}{3pt}
	\begin{table}
		\caption{\label{tab:z_limits} The table lists the equation numbers of the explicit expressions of \cref{app:the_bosonic_wave-function_renormalization} for the bosonic wave-function renormalization, $\zwave{\shom}{\mu}{T}$ in different limits. The formulae for $\zwave{\shom}{\mu}{T}$ are simplified in terms of known functions as far as possible.
		}
		\begin{ruledtabular}
			\begin{tabular}{c c c c}
				$T$							&	$\s$						&	$\mu$		&
				\\
				\hline
				\multirow{4}{*}{$\neq 0$}	&	\multirow{2}{*}{$\neq 0$}	&	$\neq 0$	&	\cref{eq:z_sigma_mu_t}
				\\
				\cline{3-4}
											&								&	$= 0$		&	\cref{eq:z_sigma_0_t}
				\\
				\cline{2-4}
											&	\multirow{2}{*}{$= 0$}		&	$\neq 0$	&	\cref{eq:z_0_mu_t}
				\\
				\cline{3-4}
											&								&	$= 0$		&	\cref{eq:z_0_0_t}
				\\
				\hline
				\multirow{4}{*}{$= 0$}		&	\multirow{2}{*}{$\neq 0$}	&	$\neq 0$	&	\cref{eq:z_sigma_mu_0}
				\\
				\cline{3-4}
											&								&	$= 0$		&	\cref{eq:z_sigma_0_0}
				\\
				\cline{2-4}
											&	\multirow{2}{*}{$= 0$}		&	$\neq 0$	&	\cref{eq:z_0_mu_0}
				\\
				\cline{3-4}
											&								&	$= 0$		&	``$+ \infty$''
			\end{tabular}
		\end{ruledtabular}
	\end{table}

\section{Results}
\label{sec:results}

Finally, we turn to the actual results and the promised proof of concept.
We start in \cref{subsec:momentum_structure} by presenting the $q$-dependence of the bosonic two-point function at various points in the $\mu$-$T$-plane.
The discussion of this momentum structure provides deeper insights in the (physical pairing) mechanisms and the operating principle behind the stability analysis.
Furthermore, we come back to these results, when we comment on recent calculations in the \gls{gn} model beyond the mean-field approximation \cite{Stoll:2021ori,Pannullo:2019bfn,Pannullo:2019prx,Lenz:2020bxk,Lenz:2020cuv}.
Based on the analysis in  \cref{subsec:momentum_structure} the actual stability analysis of the homogeneous phase in the $\mu$-$T$-plane can be performed and presented in \cref{subsec:phase_diagram_stability_analysis}.
Here, we demonstrate that this method is actually able to detect the well-known second-order phase-transition line between the \gls{ip} and the \gls{sp}, but also comment on its shortcomings.
Afterwards in \cref{subsec:wavevector}, the momentum profile of the bosonic two-point function, \textit{i.e.}, the dominant wave vector, is compared with the analytic solutions from \Rcite{Schnetz:2004vr}.
Finally, we close the discussion of our results by presenting results for bosonic wave-function renormalization in the $\mu$-$T$-plane in \cref{subsec:wave-function renormalization}.
We again comment on the insufficiency of the wave-function renormalization as a single measure for the detection of spatially inhomogeneous condensates. Furthermore, we discuss possibile implications on the quality of the mean-field approximation based on our quantitative calculations.
 
\subsection{The momentum structure of the bosonic two-point function}
\label{subsec:momentum_structure}

\begin{figure}
			\includegraphics[width=\columnwidth]{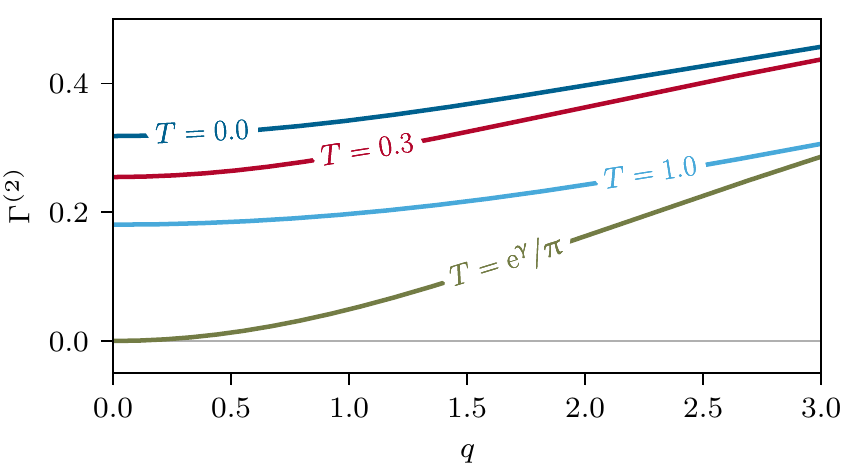}
			\caption{\label{fig:gamma2_compare_0}The bosonic two-point function $\gtwovar{\shom}{\mu}{T}{q}$ as a function of the external momentum $q$ at vanishing chemical potential ${\mu = 0}$ and fixed temperatures ${T/\snull \in \{ 0.0, 0.3, \e^\upgamma / \uppi , 1.0 \}}$ evaluated at the respective homogeneous minimum ${\shom = \sminhom ( \mu, T )}$.
			}
		\end{figure}

\begin{figure}
			\includegraphics[width=\columnwidth]{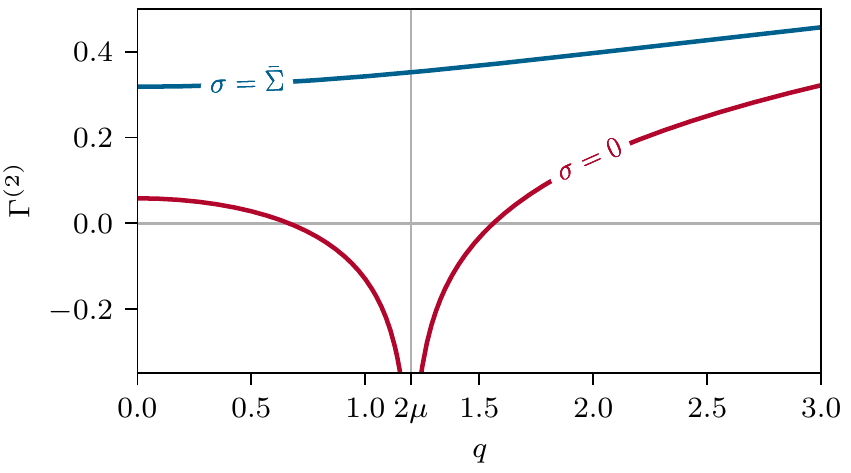}
			\caption{\label{fig:gamma2_compare_1}The bosonic two-point function $\gtwovar{\shom}{\mu}{T}{q}$ as a function of the external momentum $q$ at constant chemical potential ${\mu/\snull = 0.6}$ and vanishing temperature ${T = 0}$ evaluated at the homogeneous global minimum ${\shom = \sminhom ( \mu, 0 ) \neq 0}$ and the homogeneous local minimum ${\shom = 0.0}$.
				The unphysical red curve (stemming from an evaluation away from the homogeneous ground state, \textit{viz.} ${\shom = 0.0}$) has a pole at ${q = 2\, \mu}$.
			}
		\end{figure}

\begin{figure}
			\includegraphics[width=\columnwidth]{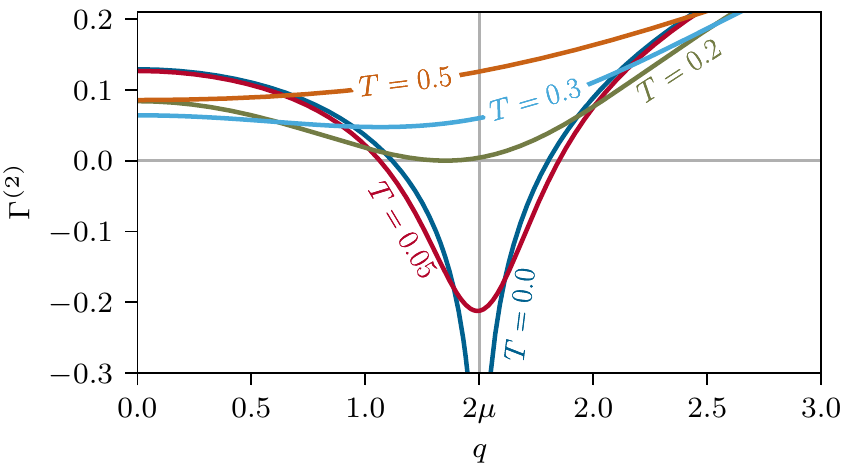}
			\caption{\label{fig:gamma2_compare_2}The bosonic two-point function $\gtwovar{\shom}{\mu}{T}{q}$ as a function of the external momentum $q$ at constant chemical potential ${\mu/\snull = 0.75}$ and fixed temperatures ${T/\snull \in \{ 0.0, 0.05, 0.2, 0.3, 0.5 \}}$ evaluated at the respective homogeneous minimum ${\shom = \sminhom ( \mu, T )=0}$.
				The curve for ${T = 0.0}$ has a pole at ${q = 2 \, \mu}$, \textit{cf.} Eq.~\eqref{eq:gamma2_0_mu_0_q}.
			}
		\end{figure}

\begin{figure}
			\includegraphics[width=\columnwidth]{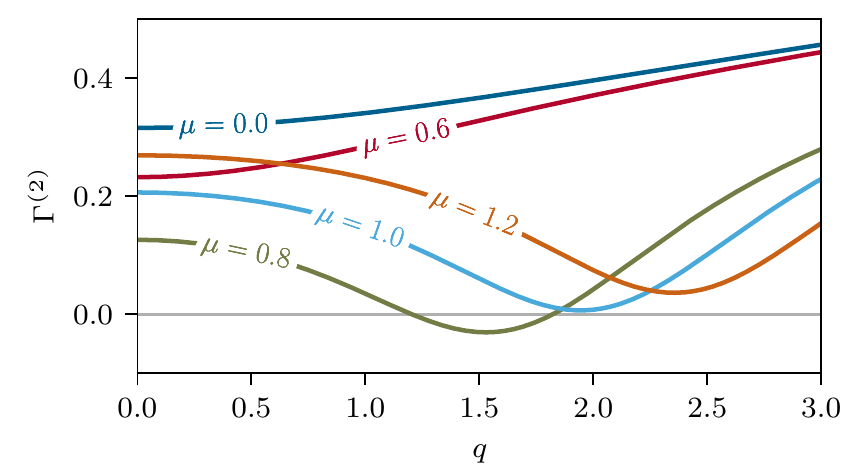}
			\caption{\label{fig:gamma2_compare_3}The bosonic two-point function $\gtwovar{\shom}{\mu}{T}{q}$ as a function of the external momentum $q$ at constant temperature ${T/\snull = 0.15}$ and fixed chemical potentials ${\mu/\snull \in \{ 0.0, 0.6, 0.8, 1.0, 1.2 \}}$ evaluated at the respective homogeneous minimum ${\shom = \sminhom ( \mu, T )}$.
			}
		\end{figure}

	The entire idea of the stability analysis is based on the momentum structure/dispersion of the bosonic two-point function.
	Therefore, this subsection contains a detailed discussion of the various possible shapes of $\gtwo$, which occur in the Gross-Neveu model at different points $( \mu, T )$ in the phase diagram.
	Our discussion is based on \cref{fig:gamma2_compare_0,fig:gamma2_compare_1,fig:gamma2_compare_2,fig:gamma2_compare_3,fig:gamma2_sigma_q}, which were directly produced by (numeric) evaluation of \cref{eq:gtwo_renorm} or its simplified versions, see \cref{tab:gamma_limits} and \cref{app:the_bosonic_two-point_function}.
	If needed, the corresponding spatially homogeneous ground state $\sminhom ( \mu, T )$ was determined (numerically) by minimization of \cref{eq:gphom_renorm}.
	
	We begin our discussion at vanishing chemical potential ${\mu = 0}$.
	The corresponding plots for $\gtwovar{\sminhom ( 0, T )}{0}{T}{q}$ are presented in \cref{fig:gamma2_compare_0}.
	One finds that the bosonic two-point function is always positive and convex for all external momenta at ${\mu = 0}$.
	This is the case for zero and non-zero temperature, in the $\Z_2$ symmetry broken and $\Z_2$ symmetric phase.
	Consequently, the spatially homogeneous minimum is stable against inhomogeneous perturbations.
	Furthermore, this might also imply that a low order derivative expansion of the bosonic effective action, \textit{e.g.}, in the context of \gls{frg} calculations \cite{Stoll:2021ori}, should be a decent approximation and capture the relevant momentum dependencies.
	Additionally, this confirms that it is unlikely to generate crystalline like ground states at zero density\footnote{Remember that the stability analysis is not sufficient to exclude inhomogeneous condensation.}. 
	Only at the phase transition at ${T/\snull = \e^\upgamma/\uppi}$ the curve for $\gtwo$ has a single root at ${q = 0}$, which is expected, since the bosonic curvature mass vanishes at this phase transition \cite{Dolan:1973qd,Weinberg:1974hy,Harrington:1974tf,Thies:2003kk}.
	
	Next, \cref{fig:gamma2_compare_1} is discussed, where we plot $\gtwovar{\shom}{\mu}{T}{q}$ at constant chemical potential ${\mu/\snull = 0.6}$ and vanishing temperature ${T = 0}$ for two evaluation points $\shom$ in the constant background field space.
	As can be seen from \cref{fig:analytical_pd} this $\mu$-$T$-point lies in the \gls{hbp} implying that the true homogeneous ground state ${\sminhom ( \mu, 0 ) \neq 0}$.
	The two curves in \cref{fig:gamma2_compare_1} show that it is crucial to evaluate $\gtwovardef$ at the correct homogeneous minimum ${\shom = \sminhom ( \mu, 0 ) \neq 0}$ as the evaluation at ${\shom = 0}$ leads to negative values of $\gtwo$ giving a false signal of instability.
	This seems somewhat obvious, especially for the rather simple \gls{gn} model in mean-field approximation.
	However, for example in more involved \gls{frg} model calculations as in  \Rcite{Eser:2018jqo,Eser:2019pvd,Cichutek:2020bli,Divotgey:2019xea,Tripolt:2017zgc,Pawlowski:2014zaa,Grossi:2021ksl,Otto:2019zjy,Otto:2020hoz,Dupuis:2020fhh,Eser:2021ivo} and especially for advanced truncations, it is sometimes not obvious to determine the correct evaluation point in field-space for correlation functions -- at least during the renormalization group flow.
	
	After covering the simple scenarios, we turn to \cref{fig:gamma2_compare_2}, where we plot the behavior of $\gtwo$ for different temperatures but constant chemical potential ${\mu/\snull = 0.75}$.
	As discussed in \cref{subsec:phenomenology} and \cref{fig:analytical_pd}, these $\mu$-$T$-points are all located in the \gls{sp} and in the \gls{ip}, such that the correct evaluation point in background field space is always the trivial homogeneous minimum ${\shom = \sminhom ( \mu, T ) = 0}$.
	As expected we find a manifestly positive and convex $\gtwo$ at very high temperatures, where thermal fluctuations are likely to vaporize any kind of crystal like structures and condensates, because the temperature $T$ and not the chemical potential $\mu$ is the dominating external energy scale.
	On the other hand, for moderate temperatures one finds that $\gtwo$ develops a non-trivial minimum at some non-zero $q$, which indicates that the energy scale that is set by $\mu$ gains in importance.
	However, this non-trivial minimum does not destabilize the spatially homogeneous ground state if $\gtwo$ stays manifestly positive.
	Only below a certain threshold for the temperature (here $T/\snull \approx 0.2$), where the minimum of $\gtwo$ turns negative, an instability is observed implying a breaking of the $\Z_2$ symmetry and translational invariance by some lower lying ground state $\smin ( \mu, T, \ex )$.
	Exactly at the temperature threshold the new $\ex$-dependent ground state $\smin ( \mu, T, \ex )$ is anticipated to exhibit a single wave vector $\qmin$, namely the single touching root of $\gtwo$.
	The latter is equal to the minimum of the two-point function.
	This is discussed in detail in \cref{subsec:wavevector}.
	In the following $\qmin$ generically denotes the location of the minimum of $\gtwo$ in $q$ direction, \textit{i.e.}, 
	\begin{align}
		\label{eq:qmin}
		\qmin \equiv \mathrm{argmin}_q \, \gtwovar{\sminhom( \mu, T )}{\mu}{T}{q}\, .
	\end{align}
	Further decreasing the temperature, we observe in \cref{fig:gamma2_compare_2} that the $q$-range where $\gtwo$ is negative grows.
	Additionally, the minimum of $\gtwo$ gets more and more negative and ultimately turns into a pole at ${q = 2 \, \mu}$ for ${T = 0}$.	
	The roots of $\gtwo$ which are poles of the propagator $1/\gtwo$ signal a resonance in the respective anti-fermion-fermion two-point function $\langle\bar{\psi}\,\psi \rangle$. 
	This resonance is associated with an anti-fermion-fermion bound state in which an anti-fermion and a fermion of opposite chirality are paired with a non-zero total momentum forming an inhomogeneous chiral condensate.
	More details and qualitative as well as quantitative discussions of this pairing mechanism can be found in Refs.~\cite{Kojo:2009ha,Kojo:2011cn,Buballa:2014tba}.
	The preferred momenta for the anti-fermion-fermion pairs are from the momentum range of negative $\gtwo(q)$ with the dominant frequency $Q$ typically associated with the minimum of $\gtwo$, see Eq.~\eqref{eq:qmin}.
	The dominant frequency of $Q\sim 2\,\mu$ at low and especially zero temperature is typical for such inhomogeneous condensates as the anti-fermion-fermion pairs are formed in vicinity of the Fermi surface \cite{Kojo:2009ha,Kojo:2011cn,Buballa:2014tba}.
	Apart from the identification of the dominant frequency $Q$ the course of $\gtwo(q)$ for $\gtwo<0$ between the roots (including the pole at ${q=2\,\mu}$ for ${T=0}$) is not very instructive because the employed stability analysis using a homogeneous expansion point is incapable of capturing the full physics of the inhomogeneous chiral condensate in this momentum regime.
	A notable exception occurs when we have a single touching root, in \cref{fig:gamma2_compare_2} the case for $T\approx0.2$, signaling the onset of instability of the homogeneous phase in favor of an inhomogeneous phase with an explicit single momentum mode $Q$ instead of a spectrum.
	We will discuss this further in the following \cref{subsec:phase_diagram_stability_analysis}.

	Lastly, \cref{fig:gamma2_compare_3} is considered, where again $\gtwo$ is presented at different points in the phase diagram.
	In contrast to the previous discussion, we do not vary the temperature at constant chemical potential, but fix ${T/\snull = 0.15}$ and study curves at various chemical potentials.
	As can be seen from \cref{fig:analytical_pd} the slice through the phase diagram at ${T/\snull = 0.15}$ is chosen, because of its rich phenomenology at different $\mu$.
	Starting from zero density at ${\mu = 0}$, a convex and manifestly positive function course of $\gtwo$ is observed.
	Increasing $\mu$ the bosonic curvature mass (the value of $\gtwo$ at ${q = 0}$) is lowered, but $\gtwo$ stays convex.
	As soon as one leaves the \gls{hbp} and crosses the first-order phase boundary to the $\Z_2$ symmetric phase (for spatially homogeneous condensates), \textit{cf.}  \cref{fig:analytical_pd}, $\gtwo$ immediately develops a non-trivial negative minimum (for ${\mu/\snull = 0.8}$ at $q/\snull \approx 1.6$ in \cref{fig:gamma2_compare_3}), which indicates that spatially inhomogeneous condensation is energetically favorable and $\mu$ completely dominates the dynamics as an external energy scale, \textit{i.e.}, one enters the \gls{ip}.
	However, further increasing $\mu$ at non-zero $T$ ultimately shifts the $\gtwo$-profile to larger values, such that at $\mu/\snull \approx 1.0$ the minimal value of $\gtwo$ turns positive again, see \cref{fig:gamma2_compare_3}.
	This means that by further increasing $\mu$ we again cross a phase transition line and enter ultimately the $\Z_2$ symmetric and translation invariant phase.

	At this point we remark, that the $q$-profiles for $\gtwo$ in \cref{fig:gamma2_compare_0,fig:gamma2_compare_1,fig:gamma2_compare_2,fig:gamma2_compare_3} are very similar to courses of $\gtwo$ that were sketched in Fig.~5 of Ref.~\cite{Roscher:2015xha} or the ones calculated and displayed in different contexts in Fig.~5 of Ref.~\cite{Tripolt:2017zgc}, Fig.~2 of Ref.~\cite{Buballa:2018hux}, and Fig.~8 of Ref.~\cite{Nakano:2004cd}.\\

\begin{figure}
			\includegraphics[width=1\columnwidth]{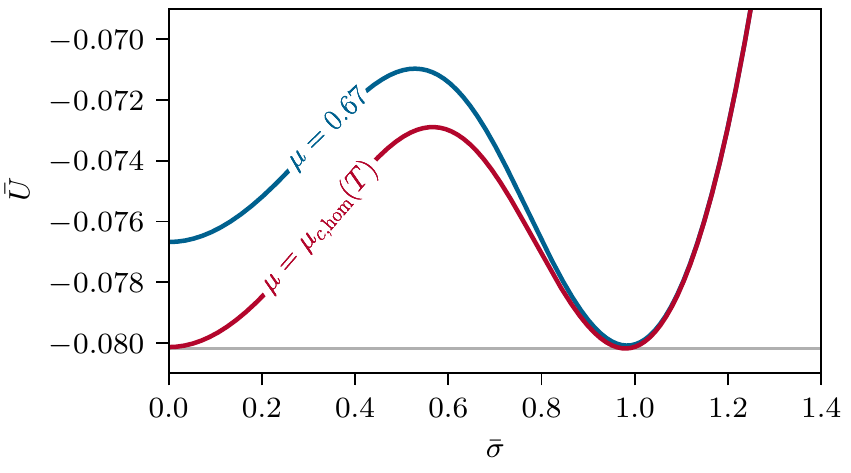}
			\caption{\label{fig:ueff} The homogeneous effective potential $\gphom ( \shom, \mu, T )$  as a function of the homogeneous background field $\shom$ at constant temperature ${T/\snull = 0.1}$ and fixed chemical potentials $\mu/\snull \in \{ 0.67, \mu_{c,\text{hom}}(T)\}$, where $\mu_{c,\text{hom}}(T)\approx0.686$ is the critical chemical potential of the homogeneous phase transition at this temperature.
			}
		\end{figure}

\begin{figure}
			\includegraphics[width=1\columnwidth]{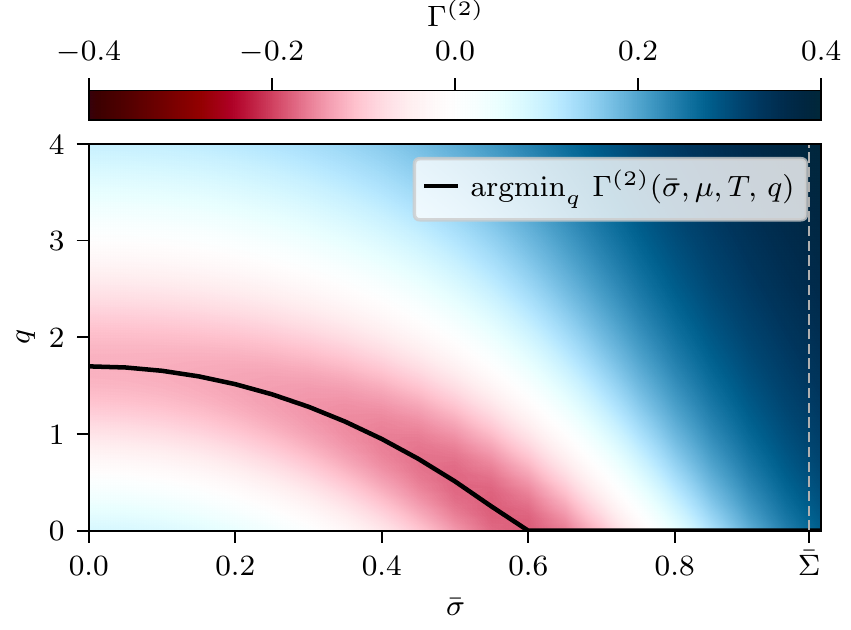}
			\caption{\label{fig:gamma2_sigma_q} The bosonic two-point function $\gtwovar{\shom}{\mu}{T}{q}$ in the $\shom$-$q$-plane for the point ${( \mu, T )/\snull = ( 0.67, 0.1 )}$ in the phase diagram.
				The solid black line marks the non-trivial minima.
			}
		\end{figure}

	We conclude this subsection with a discussion of a shortcoming of the stability analysis.
	To do so, a point in the phase diagram  is studied that is located extremely close to the first-order phase transition line in \cref{fig:analytical_pd}, but still only just corresponds to the \gls{hbp}, if only spatially homogeneous condensation is considered.
	At ${( \mu, T )/\snull = ( 0.67, 0.1 )}$ the correct homogeneous minimum of the effective potential $\gphom$ is located at ${\bar{\sigma} = \sminhom/\snull \approx 1.0}$, while the point ${\shom = 0}$ corresponds to a local minimum, which is of similar depth, see \cref{fig:ueff}.
	However, it is known from the exact solution \cite{Schnetz:2004vr,Schnetz:2005ih,Schnetz:2005vh}, see \cref{fig:analytical_pd}, that this point in the $\mu$-$T$-plane actually corresponds to the \gls{ip}, if one allows for spatial modulations of the ground state.
	For selected sample points we experienced during this subsection that the stability analysis seems to work well, if the expansion point is the trivial homogeneous minimum of the effective potential in the $\Z_2$ symmetric phase, thus ${\sminhom ( \mu, T ) = 0}$.
	Naturally the question arises whether or not the stability analysis maintains its predictive power even with non-trivial spatially homogeneous expansion points $\sminhom ( \mu, T ) \neq 0$ are considered.
	Hence, we present $\gtwovardef$ at ${( \mu, T )/\snull = ( 0.67, 0.1 )}$ as a function of $\shom$ and $q$ in \cref{fig:gamma2_sigma_q}.
	Evaluating $\gtwo$ at large values of $\shom$, \textit{e.g.}, at the correct homogeneous minimum and expansion point $\sminhom/\snull \approx 1$, the bosonic two-point function is manifestly positive and does not signal any instability.

	The reason is that the non-trivial homogeneous minimum and the spatially oscillating minimum are separated by a kind of ``potential barrier'', as the effective potential increases when studying small perturbations about the homogeneous minima.
	Formally, the correct expansion point is no longer unique.
	There are two degenerate homogeneous minima and therefore two possible expansion points with the trivial minimum and a potential barrier in between, see \cref{fig:ueff} upper curve.
	We observe that the two homogeneous minima are no longer saddle points with an unstable direction in momentum space, when studying inhomogeneous perturbations.
	In fact the analytic solution for the ground state in the \gls{ip} in terms of Jacobi elliptic functions turns into rather pronounced kinks close to the correct second-order phase transition to the \gls{hbp}.
	This means that the condensate almost oscillates between the two homogeneous minima $\pm \sminhom ( \mu, T )$ and cannot be described as a small perturbation/oscillation around just one of the two non-trivial minima.
	Finding an instability with large oscillations about ${\shom = 0}$ would require even larger $\dsx{\ex}$ when expanding around one of the two homogeneous minima.
	However, having large perturbations $\dsx{\ex}$ about $\pm \sminhom(\mu,T)$ would require to always change the expansion point during an oscillation.
	Furthermore, large $\dsx{\ex}$ would call for the inclusion of basically all higher order coefficients in the expansion, see \Rcite{Braun:2014fga}.
	The reason is the increase in the effective potential when perturbing around the homogeneous minima with small $\dsx{\ex}$ before the effective potential decreases in the vicinity of the inhomogeneous ground state when studying large $\dsx{\ex}$. Due to this behavior higher and higher order coefficients are required in the expansion in order to reproduce this behavior when moving in the $\mu$-$T$ plane towards the phase boundary between the \gls{ip} and the \gls{hbp}, as described in \Rcite{Braun:2014fga,Roscher:2013cma}.

	In summary, we observe that for this model the stability analysis fails to detect the inhomogeneous phase as long as the correct expansion point $\sminhom ( \mu, T ) \neq 0$.
	This was already partially discussed in \Rcite{Thies:2003kk} and observed in \Rcite{deForcrand:2006zz}, where a similar analysis of the \gls{gn} was done on a finite lattice.
	In the latter reference, it was stated that this ``potential barrier'' was a result of the finite volume, but our present results in an infinite volume suggest that this is a generic problem of the stability analysis independent of the considered volume.
 
\subsection{The phase diagram from the stability analysis}
\label{subsec:phase_diagram_stability_analysis}

	Based on our previous discussion, we turn to the central result of this work.
	Within the following paragraphs it is demonstrated and briefly discussed that the stability analysis correctly detects the well-known phase transition line between the \gls{sp} and the \gls{ip}, but fails in the region between the \gls{hbp}$\pbArrow$\gls{ip} phase boundary and the homogeneous first-order phase transition, \textit{cf.} \cref{fig:analytical_pd} and the related discussion in \Rcite{Thies:2003kk}.

\begin{figure}
		\includegraphics[width=\columnwidth]{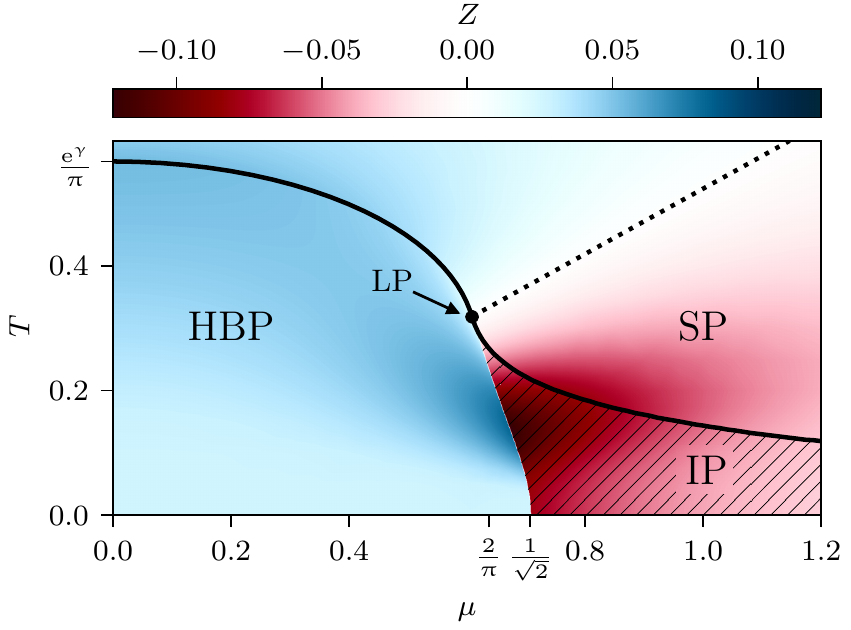}
		\caption{\label{fig:pd_stability}The bosonic wave-function renormalization $\zwave{\sminhom ( \mu, T )}{\mu}{T}$ (heat map), line of vanishing wave-function renormalization ${Z ( \sminhom ( \mu, T ), \mu, T ) = 0}$ (thick black dashed line), and the line of vanishing bosonic two-point function ${\gtwovar{\sminhom ( \mu, T )}{\mu}{T}{\qmin} = 0}$ (thick, black solid line) in the $\mu$-$T$-plane. In region marked by the diagonal hatching using thin black solid lines (bottom-right corner) we find $\gtwovar{\sminhom ( \mu, T )}{\mu}{T}{\qmin} < 0$, \textit{i.e.}, the homogeneous minimum is unstable with respect to an inhomogeneous perturbation.
		}
	\end{figure}

	As we argued before, we can trust this method in the regions of the phase diagram where the minimum and correct expansion point in field space is at ${\shom = \sminhom ( \mu, T ) = 0}$ and especially where the inhomogeneous condensate oscillates with a small amplitude about the expansion point.
	This is the case in the \gls{gn} model at the phase boundary between the \gls{ip} and the \gls{sp}.
	Thus, it is expected that the exact phase boundary and the line of instability obtained via the stability analysis match.
	This is supported by our (numerical) results that are plotted in \cref{fig:pd_stability}.
	The solid black line is the line where $\gtwo$ has a single root at ${q = \qmin}$ in the external momentum, \textit{i.e.}, ${\gtwovar{\sminhom ( \mu, T )}{\mu}{T}{\qmin} = 0}$ only for one wave vector ${q = \qmin}$, see also Ref.~\cite{Nakano:2004cd}.
	The line extends from the \gls{lp} to larger $\mu$ and is numerically identical\footnote{The phase boundaries coincide within the numerical resolution in the $\mu$-$T$-plane ($\sim0.005\,\snull$) to which both were computed. An increase in numerical resolution in the $\mu$-$T$-plane for the stability analysis and the full/exact solution will not change this agreement. Higher order terms in the stability analysis are not required to reproduce this boundary since the amplitude of the spatial inhomogeneous chiral condensate $\smin ( \mu, T, \ex )$ vanishes at this boundary.} to the exact phase boundary, which is shown in \cref{fig:analytical_pd}.
	
	Interestingly (but actually not really surprisingly) also the second-order phase boundary between the \gls{sp} and \gls{hbp} is correctly detected using $\gtwo$.
	The reason is that the bosonic curvature mass vanishes along this phase transition line \cite{Dolan:1973qd,Weinberg:1974hy,Harrington:1974tf,Thies:2003kk}.
	The curvature mass, however, is defined as ${\gtwovar{\sminhom ( \mu, T )}{\mu}{T}{q}}$ evaluated at vanishing external momentum ${q = 0}$.
	The minimum of $\gtwo$ in $q$-direction is located at ${q = 0}$ above the \gls{lp}, \textit{viz.} for $T\geq T_L$, as discussed in the previous subsection, which explains the recovery of the \gls{sp}$\pbArrow$\gls{hbp} phase boundary from the employed two-point function.
	
	Nonetheless, at the phase boundary of the \gls{hbp} and \gls{ip} the amplitude of the inhomogeneous condensate is large and the inhomogeneous condensate almost oscillates between the values of the homogeneous minima, \textit{i.e.}, between $\pm \sminhom(\mu, T)$, \textit{cf.}\ \cref{fig:inhomogeneous_condensate}.
	As soon as one crosses the first-order phase transition and needs to switch to one of these minima as the formal correct expansion point, the initial assumption of the stability analysis of small perturbations about the expansion point is violated and one finds a deviation from the exact result.
	However, the related analysis in \Rcite{Braun:2014fga} is able to qualitatively describe this phase boundary.
	This is possibly due to the fact that the employed \textit{fermion doubler trick} takes also higher orders in the expansion of the effective potential into account.
	Although, these higher-order coefficients do not match the bosonic $n$-point functions $\Gamma^{(n)}$ necessarily, it hints to the fact that an evaluation of these higher-order $n$-point functions might improve the present analysis in the problematic region next to the \gls{hbp}$\pbArrow$\gls{ip} phase boundary, \textit{cf.}\ \Rcite{Roscher:2013cma}.
	
	The additional color map in \cref{fig:pd_stability} shows the value of the wave-function renormalization $\zwave{\shom}{\mu}{T}$, \cref{eq:wave-function_renormalization}, evaluated at the true homogeneous minimum ${\shom = \sminhom ( \mu, T )}$.
	It is calculated via the appropriate formulae from \cref{app:the_bosonic_wave-function_renormalization} that are listed in \cref{tab:z_limits}.
	We also cross-checked that these results coincide with results, which are obtained by a numeric evaluation of the $q$-derivatives of $\gtwo$ in \cref{eq:wave-function_renormalization_definition}.
	In \cref{eq:wave-function_renormalization_definition} the wave-function renormalization is defined as the curvature in $q$-direction of $\gtwo$ at ${q = 0}$.
	In the \gls{sp} the wave-function renormalization is given by Eq.~\eqref{eq:z_0_mu_t} such that the ${(Z=0)}$-line is given by ${\mu/T=\mu_L/T_L\simeq1.910}$\footnote{This specific result for the ${(Z=0)}$-line has a connection \cite{Steil:2022phd} to results of a Ginzburg-Landau analysis in the \gls{gn} model, see, \textit{e.g.}, Refs.~\cite{Ahmed:2018tcs,Stoll:2021ori,Steil:2022phd}. }.
	It is immediately clear that a negative $\zwave{\sminhom ( \mu, T )}{\mu}{T}$ can only be an indication that an inhomogeneous perturbation might lower the action, because negative curvature of $\gtwo$ at ${q = 0}$ does not guarantee that the function has a root.
	This scenario is found in the region between the $(Z=0)$-line and the \gls{sp}$\pbArrow$\gls{ip} phase boundary right of the \gls{lp}, where the wave-function renormalization is negative, but the spatially homogeneous ground state is stable.
	A similar region was recently found in fully-fledged \gls{frg} calculations for \gls{qcd} \cite{Fu:2019hdw} and is also discussed in \Rcite{Rennecke:2021ovl,Pisarski:2020gkx,Pisarski:2021qof, Pisarski:2020dnx} and referred to as moat and Lifshitz regimes\footnote{
		Moat or Lifshitz regimes are regions of negative wave function renormalization, which signals a dispersion relation with a minimum at a non-zero momentum \cite{Rennecke:2021ovl,Pisarski:2021qof, Pisarski:2020gkx,Pisarski:2020dnx}. The expression moat regime \cite{Pisarski:2021qof,Rennecke:2021ovl} goes back to the dispersion relations encountered in these phases, \textit{cf.}\ \cref{fig:gamma2_compare_2}, resembling the  deep, broad ditch -- the moat -- in front of a castle wall.
	}.
	Regions of inhomogeneous phases can be included in such moat regimes, like in the present study, but they do not have to be present since $Z<0$ is just a necessary but not sufficient condition for instability of the homogeneous phase in favor of inhomogeneous condensation.
	Other exotic phases of matter, like a quantum spin liquid \cite{Pisarski:2020dnx}, might be possible and energetically preferred over a typical homogeneous static ground state in the moat regime -- if the particle content and space-time dimensionality of the model is more involved.
	
	In summary, an inhomogeneous field configuration with momentum $q$ that lowers the effective action can only be guaranteed to exist through our analysis, when $\gtwovar{\sminhom ( \mu, T )}{\mu}{T}{q} < 0$ and ${\sminhom ( \mu, T ) = 0}$, which corresponds to the hatched region (bottom, right) in \cref{fig:pd_stability}.
		
\subsection{The wave vector of the inhomogeneous perturbation and the wave vector of the true inhomogeneous condensate}
\label{subsec:wavevector}

	Even though the stability analysis is expected to work only for very small perturbations about a vanishing homogeneous condensate, we found that it even correctly predicts inhomogeneous condensation at points to the right of the homogeneous first-order phase transition at extremely small temperatures which are far away from the second-order \gls{sp}$\pbArrow$\gls{ip} phase transition line.
	At these points one still uses the appropriate expansion point ${\sminhom ( \mu, T ) = 0}$, but the perturbations are no longer small and the true condensate has a spectrum of wave vectors instead of a single frequency/wave vector, \textit{cf.}\ \cref{fig:inhomogeneous_condensate}.
	
	One might thus wonder, if the single wave vector $\qmin$ at the phase transition line actually matches the wave-vector of the true solution, \textit{i.e.}, the dominating wave vector of the Jacobi elliptic functions.
	
	Therefore, this section is used to compare the dominating wave vector of the correct inhomogeneous condensate minimizing the effective
    action 
    \begin{equation}
    	\qs \equiv \mathrm{argmax}_q \, \tilde{\smin} ( \mu, T, q )
    \end{equation}
     with the wave vector that minimizes the two-point function $\qmin$ as defined in \cref{eq:qmin}.
     While $\qmin$ is the direction of the largest curvature of the action at the saddle point, it does not necessarily coincide with $\qs$.
	
\begin{figure}
			\includegraphics[width=1\columnwidth]{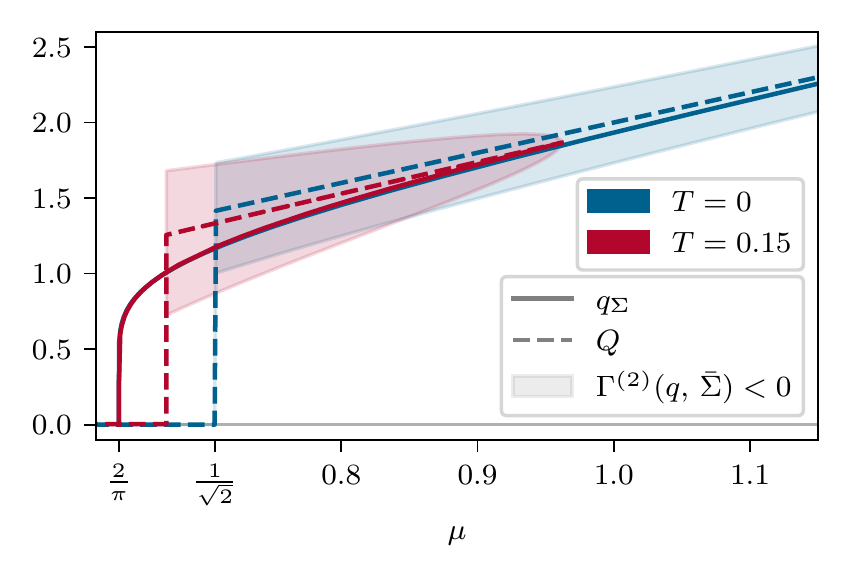}
			\caption{\label{fig:stability_wavevector_vs_condensate_wavevector_mu_scan}The minimum of the bosonic two-point function $\qmin ( \mu )$ and the dominating wave vector of the true inhomogeneous condensate $\qs ( \mu )$ as a function of the chemical potential at constant temperatures $T/\snull \in \{ 0.0, 0.15 \}$.
				The colored regions mark the range of momenta $q$, where $\gtwovar{\sminhom ( \mu, T )}{\mu}{T}{q} < 0$.
			}
		\end{figure}

	In \cref{fig:stability_wavevector_vs_condensate_wavevector_mu_scan} these two quantities  are plotted for two different temperatures.
	At ${T = 0}$, $\qmin$ approaches $\qs$ for increasing chemical potential\footnote{Plots similar to \cref{fig:stability_wavevector_vs_condensate_wavevector_mu_scan} of the wave vector of some inhomogeneous condensate plotted over baryon density (chemical potential), can be found in, \textit{e.g.}, Fig.~2 of Ref.~\cite{Dautry:1979bk}, Fig.~2 of Ref.~\cite{Kutschera:1989yz}, Figs.~6 \& 7 of Ref.~\cite{Kutschera:1990xk}.} and at ${T/\snull = 0.15}$ the two momenta match at the phase boundary.
	This is expected as the amplitude of the inhomogeneous condensate $\smin ( \mu, T, \ex )$ at this point is infinitesimal and therefore the stability analysis becomes exact.
	At small chemical potential -- as already discussed before -- the stability analysis does not detect an inhomogeneous phase unless ${\sminhom ( \mu, T ) = 0}$, right of the homogeneous first-order phase transition.
	At intermediate chemical potential, $\qmin$ and $\qs$ do not agree.
	However, $\qs$ is within the interval where $\gtwo < 0$ is predicted by the stability analysis, which means that the latter at least captures the dominating wave vectors.
	
\begin{figure*}
			\includegraphics[width=\textwidth]{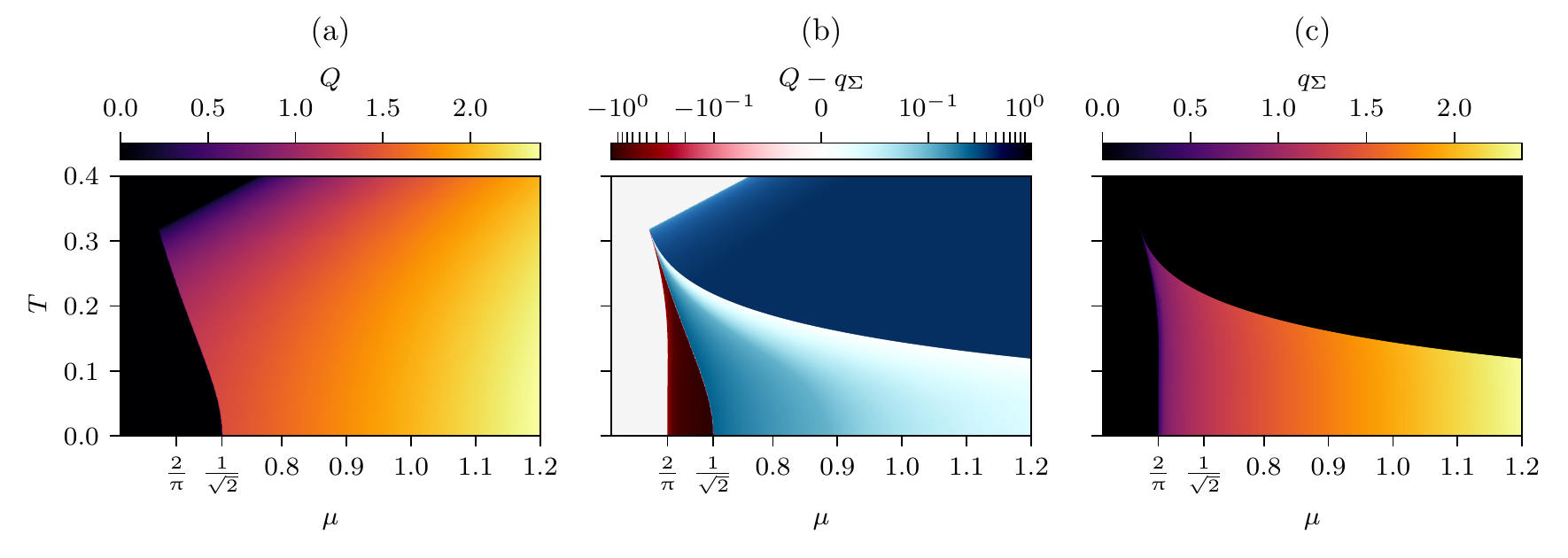}
			\caption{\label{fig:stability_wavevector_vs_condensate_wavevector_mu_T} (a) $\qmin$, (b) $\qmin-\qs$ and (c) $\qs$ in the $(\mu,T)$-plane. Note that the middle plot (b) has a colormap that is linear around $0$ and logarithmic for values $|(\qmin-\qs)/\snull|>0.1$.}
		\end{figure*}

	In \cref{fig:stability_wavevector_vs_condensate_wavevector_mu_T} we again compare $\qmin$ and $\qs$.
	This time we plot $\qmin$, $\qmin - \qs$, and $\qs$ in the $\mu$-$T$-plane in using different color maps.
	The previously discussed trend extends to the whole temperature range.
	The difference $\qmin - \qs$ approaches zero close to the \gls{ip}$\pbArrow$\gls{sp} boundary and its magnitude is the largest close to the \gls{hbp}$\pbArrow$\gls{ip} boundary, where $\qmin$ is zero (because the stability analysis is ill-conditioned) and $\qs$ is maximal.
	On the other hand, $\qmin$ is also non-zero in the region of $Z < 0$ above the phase transition line, but does not correspond to an inhomogeneous perturbation that lowers the action, since $\gtwo$ is manifestly positive.
	We want to emphasize that this does not mark a failure of the employed method, but is rather just an effect of the negative wave-function renormalization $Z$.
	A discussion similar to our elaboration on \cref{fig:stability_wavevector_vs_condensate_wavevector_mu_T} can be found in a different context in \Rcite{Nakano:2004cd}. 
\subsection{The bosonic wave-function renormalization}
\label{subsec:wave-function renormalization}

	Before closing our discussion we shortly return to our results for the bosonic wave-function renormalization and their implications.
	
	In \cref{fig:pd_stability} the bosonic wave-function renormalization $\zwave{\sminhom ( \mu, T )}{\mu}{T}$ was already presented in the entire $\mu$-$T$-plane.
	We stress again that negative values of $Z$ are not a sufficient criterion for instabilities of the homogeneous phase, which is also discussed in \Rcite{Fu:2019hdw}.
	However, in regions of the phase diagram where the stability analysis is expected to work, \textit{i.e.}, regions with ${\sminhom(\mu,T)=0}$, a negative wave-function renormalization is a necessary condition\footnote{For the sake of completeness, it is strictly speaking not a necessary condition.
	There might be more exotic dispersion relations that allow for positive $Z$, but regions of negative $\gtwo$ at larger external momentum $q$.} and central to the search of inhomogeneous phases as discussed in length in \Rcite{Rennecke:2021ovl,Pisarski:2021qof, Pisarski:2020gkx,Pisarski:2020dnx}.

	Apart from this, one can learn a lot from the values of the wave-function renormalization alone.	
	In a first rather rough approximation, we can use $Z$ as a measure for the importance of bosonic quantum fluctuations, because it accompanies the trivial quadratic momentum dependence of the bosonic field in the action -- the kinetic term -- which drives fluctuations.
	Inspecting the classical \gls{uv} action of the \gls{gn} model, we find that it lacks by construction a term like $\frac{Z}{2} \, ( \partial_\mu \phi )^2$ and also all other bosonic higher-order derivative terms, which could partially be associated to the higher-order Taylor coefficients/moments of $\gtwo$ in momentum space.
	Hence, in the classical action of the \gls{gn} model all these coefficients are initially zero, because there are no bosonic fluctuations in the \gls{uv} -- there are only non-interacting fermions -- and $\phi$ is only introduced as an auxiliary field.
	However, by integrating out all fermionic quantum fluctuations and interactions one finds that the system gets strongly coupled and anti-fermion-fermion pairs are bosonized and eventually condense, if external energy scales ($\mu$ \& $T$) are not too large \cite{Wolff:1985av,Harrington:1974tf,Dashen:1974xz}.
	Ultimately, also all of the bosonic derivative couplings are generated by integrating out the fermion fluctuations, as can be seen from our results for $Z$ and $\gtwo$.
	From an \gls{frg} perspective this is a rather natural finding, see \Rcite{Dupuis:2020fhh} and references therein.
	Though, the generation of all these bosonic kinetic couplings actually implies that the system tends to drive bosonic quantum fluctuations by itself, which is only hindered by the artificially suppression of the \infN{} limit.
	Therefore, one might conclude that our results for the bosonic wave-function renormalization (and the bosonic two-point function) in the \infN{} limit may -- at least to some extent -- predict the insufficiency of the mean-field approximation at finite $N$. 
	In consequence one might state, that at least in those areas of the phase diagram, where the bosonic wave-function renormalization significantly deviates from zero and rapidly changes its value (and sign) with $\mu$ and $T$, bosonic quantum fluctuations will play an important role, if the \infN{} approximation is relaxed and calculations are performed at finite $N$.
	Interestingly, such values are indeed observed, especially close to the first-order phase transition and right below the \gls{lp}.
	This is actually expected, since in these regions correlation length usually diverge and fluctuations of all orders become relevant.

	To better visualize the behavior of $Z$, we additionally plot slices through the color map of \cref{fig:pd_stability}:
		\begin{figure}
			\includegraphics{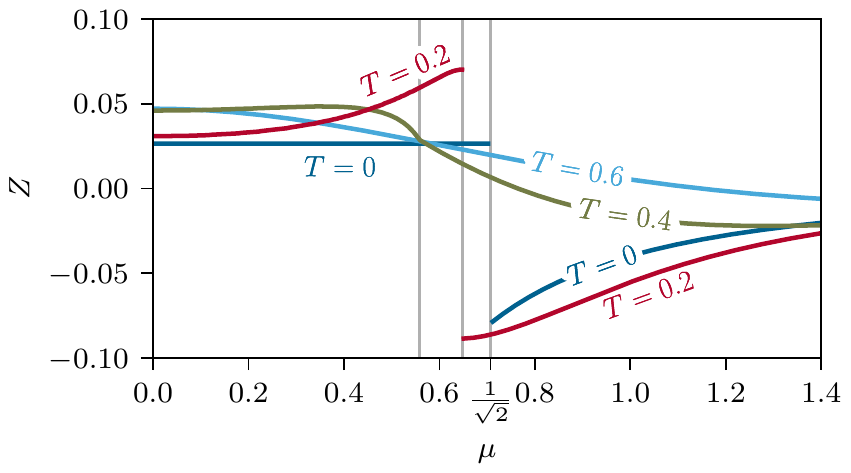}
			\caption{\label{fig:Z_slice_0}The bosonic wave-function renormalization $\zwave{\shom}{\mu}{T}$ as a function of the chemical potential at fixed temperatures ${T \in \{0.0, 0.2, 0.4, 0.6 \}}$ evaluated at the homogeneous minimum ${\shom = \sminhom ( \mu, T )}$.
			}
		\end{figure}
	\cref{fig:Z_slice_0} shows $Z$ as a function of $\mu$ at different fixed $T$, while \cref{fig:Z_slice_1} shows $Z$ as a function of temperature and different fixed $\mu$.
		\begin{figure}
			\includegraphics{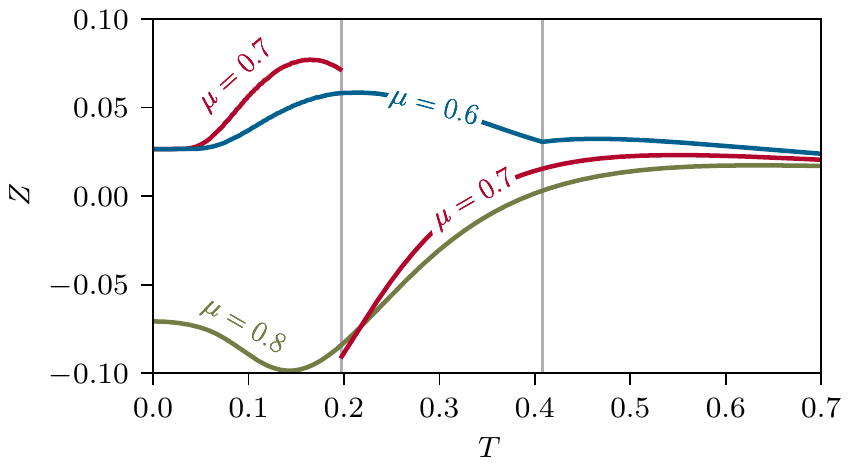}
			\caption{\label{fig:Z_slice_1} The bosonic wave-function renormalization $\zwave{\shom}{\mu}{T}$ as a function of temperature at fixed chemical potentials ${\mu \in \{ 0.55, 0.65, 0.75 \}}$ evaluated at the homogeneous minimum ${\shom = \sminhom ( \mu, T )}$.}
		\end{figure}
	The slices are chosen in a way to cover all interesting regions of the phase diagram.
	
	From both plots we observe that the wave-function renormalization increases (decreases) close to the phase transition line\footnote{Except for ${T=0}$, where $\zwave{\sminhom(\mu,0)}{\mu}{0}$ is independent of $\mu$ for $\sminhom(\mu,0)^2>\mu^2 \Leftrightarrow \mu < \tfrac{1}{\sqrt{2}}$, see \cref{eq:z_sigma_mu_0}. This is a notion of the silver blaze property.} and then jumps from positive to negative values at the phase transition.
	The region of drastically rising $Z$ is exactly the region adjacent to the first-order phase transition, where the stability analysis fails.
	Thus, it seems as if the wave-function renormalization already signals that fluctuations and gradient driven bosonic field configurations are of great importance in this region.
	Moreover, in this region the phase boundary described by the analysis in \Rcite{Braun:2014fga} deviates the most from the exact result.

\section{Conclusion and Outlook}
\label{sec:conclusion_and_outlook}

	\subsection{Conclusion}
	We have applied the stability analysis to the $(1 + 1)$-dimensional \gls{gn} model in the mean-field approximation and benchmarked its ability to detect inhomogeneous condensation against the exact solution.
	\\
	
	It was shown in \cref{subsec:phase_diagram_stability_analysis} that the stability analysis is able to accurately predict the second-order phase transition between the \gls{ip} and \gls{sp} as the amplitude of the inhomogeneous condensate $\smin$ at this phase boundary becomes infinitesimal and its functional form is described by a harmonic wave.
	Matching the initial expectation, the stability analysis fails to correctly detect the phase boundary between the \gls{hbp} and \gls{ip}, because large perturbations would be required, which, however, would drive the stability analysis ad absurdum.
	
	Therefore, the region of the \gls{ip}, where $\sminhom (\mu, T)\neq 0$, is completely undetected by the stability analysis.
	
	Moreover, we compared the wave-vector that minimizes the bosonic two-point function ${\qmin = \mathrm{argmin}_q \, \gtwovar{\sminhom( \mu, T )}{\mu}{T}{q}}$ with the dominating wave-vector of the inhomogeneous condensate ${\qs = \mathrm{argmax}_q \, \tilde{\smin} ( \mu, T, q )}$ in \cref{subsec:wavevector}.
	Inside the \gls{ip} close to the phase boundary between the \gls{sp} and \gls{ip} these two quantities agree very well due to the relatively small amplitudes of the sine-shaped condensate.
	Further away from this phase boundary, the amplitude of the inhomogeneous condensate is large, thus violating an assumption of the stability analysis.
	This is reflected in a small but finite tension of $\qmin$ and $\qs$.
	There are two regions where $\qmin$ and $\qs$ exhibit completely different behavior.
	Inside the \gls{ip} for $\mu<\mu_{c,\text{hom}}(T)$ the stability analysis fails - as previously described - by predicting ${\qmin=0}$ while $\qs>0$.
	The second region is located in the \gls{sp} where the wave-function renormalization is negative resulting in a finite $\qmin$, while ${\qs=0}$.
	However, in this region the stability analysis does not fail as an inhomogeneous phase is not detected, since $\gtwo > 0$ for all wave-vectors.
	
	Also, the bosonic wave-function renormalization $Z$ was investigated.
	The existence of a region where the wave-function renormalization is negative and the homogeneous minimum is stable under inhomogeneous perturbations, \textit{i.e.}, $Z<0$ and $\gtwovar{\sminhom (\mu, T)}{\mu}{T}{q} >0\ \forall q$, explicitly shows that a negative $Z$ is only a necessary condition for an inhomogeneous phase\footnote{This statement is limited to the regions where the stability analysis is expected to work, \textit{i.e.}, where ${\sminhom(\mu,T)=0}$.}.\\
	
	In summary, these findings show that the stability analysis can indeed be an appropriate tool in the search for second-order phase boundaries of inhomogeneous phases.
	By using the $1+1$-dimensional \gls{gn} model as a test ground the shortcomings of this methods were quantified and it was demonstrated that it can also give a reasonable estimate of quantities within the inhomogeneous phase like the dominating wave-vector of the condensate.
	These findings present a supplement and extension of the related analysis in \Rcite{Braun:2014fga}, where the importance of the bosonic two-point function was already emphasized, and earlier works by Thies \textit{et al.}, especially \Rcite{Thies:2003kk}, which is partially based on similar methods.
	
	\subsection{Outlook}
	A recent \gls{frg} calculation of the \gls{qcd} phase diagram found a region with a negative wave-function renormalization \cite{Fu:2019hdw}, which was also identified as a necessary condition for inhomogeneous phases in \Rcite{Rennecke:2021ovl,Pisarski:2021qof, Pisarski:2020gkx,Pisarski:2020dnx}.
	We showed, however, in our simple test case that it is vital to consider the full momentum structure of the bosonic two-point function $\gtwo$ and not only $Z<0$ in order to accurately determine if an inhomogeneous condensate is favored over a homogeneous one.
	With our results we want to motivate that future studies in the search for inhomogeneous phases should take the full momentum dependent two-point function into account.\\	
	
	In \cref{subsec:wave-function renormalization} we argued that the occurrence of large values of the wave-function renormalization $Z$ indicate that bosonic quantum fluctuations that are artificially suppressed in the $\infN{}$ limit play an important role in the dynamics of the system at finite $\Nf$.
	Of course, we are aware that at finite $\Nf$ the coefficients of the bosonic kinetic terms will also react to the bosonic fluctuations, which they drive.
	This can be modeled, \textit{e.g.}, via a comprehensive \gls{frg} calculation or by directly simulating the \gls{gn} model at finite $\Nf$.
	First calculations at finite $\Nf$ were performed by the authors and collaborators and indicate that bosonic fluctuations indeed play an important role in the correct description of the phase structure of the \gls{gn} model and might destabilize condensates of all types for all $\mu$ and $T > 0$ \cite{Stoll:2021ori,Lenz:2020bxk}.
	Notwithstanding these first studies of the \gls{gn} model at finite $\Nf$, further research and development is needed to gain a robust understanding of the phase structure, possible truncation artifacts in \gls{frg} computations, and finite size effects in lattice Monte-Carlo simulations.\\
	
	Furthermore, we also want to comment on another application within the context of spatially inhomogeneous condensation, where the wave-function renormalization (but also higher derivative couplings or even the full two-point function itself) might be of interest and use -- conceptually and technically independent of a specific model or theory and the method of computation.
	Having calculated an effective \gls{ir} action, like \cref{eq:gphom_renorm}, as well as some derivative couplings like $Z$ or even $\gtwo(q)$ itself from an arbitrary method and in an arbitrary truncation/approximation, one may use the quantum equation of motion to derive partial differential equations for the quantum (mean) fields that describe the state of least energy, \textit{i.e.}, the ground state.
	Assuming static solutions of these equations, but allowing for variations in spatial directions, one may solve these partial/ordinary\footnote{Whether these differential equations in Euclidean space-time present as partial or ordinary differential equations in space depends on the spatial dimension of the problem. A $1+1$ dimensional study using for example the \gls{gn} model at non-zero $\mu$ and $T$ is again a very promising starting point since static solutions in one spatial dimension are governed by simple ordinary differential equations.} differential equations to obtain an approximation to the ground-state of the system.
	Hereby, one uses the potential and the coefficients of the derivative terms, which were previously calculated by the respective method of choice.
	The quality of this approximation to the true ground state of the system strongly depends on the number and importance of the derivative couplings and the quality of the input effective potential.
	First promising results were already produced by the authors, but details and results on this approach will be presented elsewhere.

\begin{acknowledgments}
	A.K., L.P., S.R., M.J.S., and M.W.\ acknowledge the support of the \textit{Deutsche Forschungsgemeinschaft} (DFG, German Research Foundation) through the collaborative research center trans-regio  CRC-TR 211 ``Strong-interaction matter under extreme conditions''-- project number 315477589 -- TRR 211.
	
	A.K., L.P., M.J.S.,\ and M.W.\ acknowledge the support of the \textit{Helmholtz Graduate School for Hadron and Ion Research}.
	
	A.K.,\ M.J.S.,\ and M.W. acknowledge the support of the \textit{Giersch Foundation}.
	
	M.W. acknowledges support by the GSI Forschungs- und Entwicklungsvereinbarungen (GSIF \& E).
	
	A.K.\ acknowledges the support of the \textit{Friedrich-Naumann-Foundation for Freedom}.
	
	We thank J.~Braun, M.~Buballa, D.~H.~Rischke, and M.~Wagner for valuable comments, enlightening discussions, and initial ideas on this and related projects. Furthermore, we are grateful for their encouragement to realize this project as well as their supervision of PhD theses, which are associated to this project.
	
	We further thank L.~Kurth, R.~D.~Pisarski, J.~Stoll, and N.~Zorbach for valuable discussions.\\
		
	Numerical results in this work were produced with a \texttt{Python} code using various libraries \cite{reback2020pandas,mckinney-proc-scipy-2010,2020SciPy-NMeth,2020NumPy-Array} and a \texttt{C++} code using the \texttt{cubature} package \cite{cubature:2020} for numerical integration and a Nelder-Mead simplex method for repeated, local, numerical minimization \cite{Nelder:1965}.
	Symbolic computations and additional cross-checks of all numeric results were performed using \texttt{Mathematica} \cite{Mathematica:12.2}.
	The ``Feynman'' diagram in Eq.~\eqref{eq:Gamma} was generated using \texttt{Axodraw Version 2} \cite{Collins:2016aya}.
	All figures in this work were produced with the \texttt{matplotlib} package \cite{Hunter:2007}.
\end{acknowledgments} 

\appendix

\section{The bosonic two-point function}
\label{app:the_bosonic_two-point_function}

	In this appendix, we list all explicit expressions for the renormalized bosonic two-point function \labelcref{eq:Gamma} depending on the constant background field $\shom$, the chemical potential $\mu$, the temperature $T$, and the external momentum $q$.
	Details on the derivation of all expressions are presented in Ref.~\cite{Koenigstein:2022phd}.
	All results can be derived using sharp \gls{uv} and \gls{ir} cutoff regularizations, while the \gls{uv} divergences are resolved using the asymptotic freedom of the four-Fermi coupling, \cref{eq:running_coupling}.
	The challenging part of some of the derivations are the \gls{ir} divergences that occur for the special case ${\shom = 0}$.
	
	For $\shom \neq 0, \mu \neq 0, T \neq 0, q \neq 0$ we find
\begin{widetext}
		\begin{align}
			\gtwovar{\shom}{\mu}{T}{q} = \, & \tfrac{1}{\uppi} \bigg\{ \tfrac{1}{2} \ln \Big( \tfrac{\shom^2}{\snull^2} \Big) + \sqrt{ \tfrac{( 2 \shom )^2}{q^2} + 1 } \arcoth \Big( \sqrt{ \tfrac{( 2 \shom )^2}{q^2} + 1 } \Big) + \int_{0}^{\infty} \mathrm{d} p \, \tfrac{1}{\energy{p}} \, \big[ n ( \energy{p}, \mu ) + n ( \energy{p}, - \mu ) \big] +	\vphantom{\Bigg(\Bigg)}	\nonumber
			\\
			& \quad - 2 \, \left( \tfrac{q^2}{(2 \shom )^2} + 1  \right) \, \shom^2 \int_{0}^{\infty} \mathrm{d} p \, \tfrac{1}{\energy{p}} \Big( \tfrac{1}{2 p q+ q^2}  - \tfrac{1}{2 p q - q^2} \Big) \, \big[ n ( \energy{p}, \mu ) + n ( \energy{p}, - \mu ) \big] \bigg\} \, .	\vphantom{\Bigg(\Bigg)}	\label{eq:gamma2_sigma_mu_t_q}
		\end{align}
	Note that the remaining integral in the second line has a pole at ${p = \frac{| q |}{2}}$, which needs careful numeric treatment, but can be evaluated in terms of the Cauchy principal value. The same pole also appears in some of the following results.
	
	For ${q = 0}$ but $\shom \neq 0$, $\mu \neq 0$, and $T \neq 0$, we obtain,
		\begin{align}
			\gtwovar{\shom}{\mu}{T}{0} = \, & \tfrac{1}{\uppi} \bigg[ \tfrac{1}{2} \ln \Big( \tfrac{\shom^2}{\snull^2} \Big) + 1 + \int_{0}^{\infty} \mathrm{d} p \, \tfrac{1}{\energy{p}} \, \big[ n ( \energy{p}, \mu ) + n ( \energy{p}, - \mu ) \big] +	\vphantom{\Bigg(\Bigg)}	\label{eq:gamma2_sigma_mu_t_0}
			\\
			& \quad - \shom^2 \int_{0}^{\infty} \mathrm{d} p \, \tfrac{1}{\energy{p}^3} \Big\{ n ( \energy{p}, \mu ) + n ( \energy{p}, - \mu ) - \tfrac{\energy{p}}{T} \, \big[ n^2 ( \energy{p}, \mu ) + n^2 ( \energy{p}, - \mu ) - n ( \energy{p}, \mu ) - n ( \energy{p}, - \mu ) \big] \Big\} \bigg] \, .	\vphantom{\Bigg(\Bigg)}	\nonumber
		\end{align}
	On the other hand, for ${\mu = 0}$, \cref{eq:gamma2_sigma_mu_t_q} is given by
		\begin{align}
			\gtwovar{\shom}{0}{T}{q} = \, & \tfrac{1}{\uppi} \bigg[ \tfrac{1}{2} \ln \Big( \tfrac{\shom^2}{\snull^2} \Big) + \sqrt{ \tfrac{( 2 \shom )^2}{q^2} + 1 } \arcoth \Big( \sqrt{ \tfrac{( 2 \shom )^2}{q^2} + 1 } \Big) + 2 \int_{0}^{\infty} \mathrm{d} p \, \tfrac{1}{\energy{p}} \, n ( \energy{p}, 0 ) +	\vphantom{\Bigg(\Bigg)}	\label{eq:gamma2_sigma_0_t_q}
			\\
			& \quad - 4 \, \left( \tfrac{q^2}{(2 \shom )^2} + 1  \right) \, \shom^2 \int_{0}^{\infty} \mathrm{d} p \, \tfrac{1}{\energy{p}} \Big( \tfrac{1}{2 p q+ q^2}  - \tfrac{1}{2 p q - q^2} \Big) \, n ( \energy{p}, 0 ) \bigg] \, .	\vphantom{\Bigg(\Bigg)}	\nonumber
		\end{align}
	For ${\mu = 0}$ and ${q = 0}$ \cref{eq:gamma2_sigma_mu_t_0} simplifies to
		\begin{align}
			\gtwovar{\shom}{0}{T}{0} = \, & \tfrac{1}{\uppi} \bigg[ \tfrac{1}{2} \ln \Big( \tfrac{\shom^2}{\snull^2} \Big) + 1 + 2 \int_{0}^{\infty} \mathrm{d} p \, \tfrac{1}{\energy{p}} \, n ( \energy{p}, 0 ) -	\vphantom{\Bigg(\Bigg)}	\label{eq:gamma2_sigma_0_t_0}
			\\
			& \quad - 2 \, \shom^2 \int_{0}^{\infty} \mathrm{d} p \, \tfrac{1}{\energy{p}^3} \Big\{ n ( \energy{p}, 0 ) - \tfrac{\energy{p}}{T} \, \big[ n^2 ( \energy{p}, 0 ) - n ( \energy{p}, 0 ) \big] \Big\} \bigg] \, .	\vphantom{\Bigg(\Bigg)}	\nonumber
		\end{align}
	On the other hand, for ${\shom = 0}$ the bosonic two-point function reads,
		\begin{align}
			\gtwovar{0}{\mu}{T}{q} = \, & \tfrac{1}{\uppi} \bigg\{ \tfrac{1}{2} \ln \Big( \tfrac{( 2 T )^2}{\snull^2} \Big) - \upgamma - \mathrm{DLi}_0 \big( \tfrac{\mu}{T} \big) +	\vphantom{\Bigg(\Bigg)}	\label{eq:gamma2_0_mu_t_q}
			\\
			& \quad + \tfrac{q^2}{2} \int_{0}^{\infty} \mathrm{d} p \, \tfrac{1}{p} \Big( \tfrac{1}{2 p q + q^2} - \tfrac{1}{2 p q - q^2} \Big) \, \big[ 1 - n ( p, \mu ) - n ( p, - \mu ) \big] \bigg\} \, ,	\vphantom{\Bigg(\Bigg)}	\nonumber
		\end{align}
	where
		\begin{align}
			\mathrm{DLi}_{2n} ( y ) &\equiv \big[ \tfrac{\partial}{\partial s} \mathrm{Li}_s ( - \e^{y} ) + \tfrac{\partial}{\partial s} \mathrm{Li}_s ( - \e^{- y} ) \big]_{s = 2n} \, = \vphantom{\bigg(}	\label{eq:def_dli}\\
		&= -\delta_{0,n}(\,\log(2\uppi)+\upgamma\,) + (-1)^{1-n} (2\uppi)^{2n}\Re\psi^{(-2n)}\big(\tfrac{1}{2}+\tfrac{\mathrm{i}}{2\uppi}y\big) \, , \vphantom{\bigg(}	\label{eq:dli_polyGamma}
		\end{align}
	and $\mathrm{Li}_s ( z )$ is the polylogarithm\footnote{We experienced that the numerical evaluation of $\mathrm{DLi}_{2n} ( y )$ for $2n\leq0$ is more stable and precise, if it is further expressed and numerically evaluated in terms of the polygamma functions $\psi^{(-2n)}(z)$ with complex argument $z$ using Eq.~\eqref{eq:dli_polyGamma} \cite{Steil:2022phd}. }, $\upgamma$ is the Euler-Mascheroni constant, and Eq.~\eqref{eq:dli_polyGamma} holds only for $2n\leq0$.
	
	For ${q = 0}$ the previous expression turns into
		\begin{align}
			\gtwovar{0}{\mu}{T}{0} = \, & \tfrac{1}{\uppi} \, \Big[ \tfrac{1}{2} \ln \Big( \tfrac{( 2 T )^2}{\snull^2} \Big) - \upgamma - \mathrm{DLi}_0 \big( \tfrac{\mu}{T} \big) \Big] \, .	\label{eq:gamma2_0_mu_t_0}
		\end{align}
	On the other hand, for ${\mu = 0}$ \cref{eq:gamma2_0_mu_t_q} transforms to
		\begin{align}
			\gtwovar{0}{0}{T}{q} = \, & \tfrac{1}{\uppi} \bigg\{ \tfrac{1}{2} \ln \Big( \tfrac{( \uppi T )^2}{\snull^2} \Big) - \upgamma + \tfrac{q^2}{2} \int_{0}^{\infty} \mathrm{d} p \, \tfrac{1}{p} \Big( \tfrac{1}{2 p q + q^2} - \tfrac{1}{2 p q - q^2} \Big) \, \Big[ 1 - 2 \, n ( p, 0 ) \Big] \bigg\} \, .	\vphantom{\Bigg(\Bigg)}	\label{eq:gamma2_0_0_t_q}
		\end{align}
	Furthermore, \cref{eq:gamma2_0_mu_t_0} in the limit $\frac{\mu}{T} \to 0$ reads,
		\begin{align}
			\gtwovar{0}{0}{T}{0} = \, & \tfrac{1}{\uppi} \Big[ \tfrac{1}{2} \ln \Big( \tfrac{( \uppi T )^2}{( h \s_0 )^2} \Big) - \upgamma \Big] \, .	\label{eq:gamma2_0_0_t_0}
		\end{align}
	However, starting with \cref{eq:gamma2_sigma_mu_t_q} and considering ${T = 0}$ first, we obtain,
		\begin{align}
			& \gtwovar{\shom}{\mu}{0}{q} =	\vphantom{\Bigg(\Bigg)}	\label{eq:gamma2_sigma_mu_0_q}
			\\
			= \, & \tfrac{1}{\uppi} \bigg[ \tfrac{1}{2} \ln \Big( \tfrac{\shom^2}{\snull^2} \Big) + \sqrt{ 1 + \tfrac{( 2 \shom )^2}{q^2} } \arcoth \Big( \sqrt{ 1 + \tfrac{( 2 \shom )^2}{q^2} } \Big) \bigg] +
			\left\{
				\begin{matrix}
					\tfrac{1}{\uppi} \artanh \Big( \sqrt{1 - \tfrac{\shom^2}{\mu^2}} \Big)	&	\vphantom{\bigg(\bigg)}
					\\
					0 \, ,	\vphantom{\bigg(\bigg)}
				\end{matrix}
			\right\} -	\nonumber
			\\
			& \quad - 
			\left\{
				\begin{matrix}
					 \tfrac{1}{2 \uppi} \sqrt{1 + \tfrac{( 2 \shom )^2}{q^2}} \left[ \artanh \left( \frac{\frac{2 \shom^2}{\mu q} + \sqrt{1 - \frac{( \shom )^2}{\mu^2}}}{\sqrt{1 + \frac{( 2 \shom )^2}{q^2}}} \right) + \artanh \left( \frac{- \frac{2 ( \shom )^2}{\mu q} + \sqrt{1 - \frac{( \shom )^2}{\mu^2}}}{\sqrt{1 + \frac{( 2 \shom )^2}{q^2}}} \right) \right] \, ,	&	\text{if} \quad \shom^2 < \mu^2 \, ,
					\\
					0 \, , &	\text{if} \quad \shom^2 \geq \mu^2 \, .	\vphantom{\bigg(\bigg)}
				\end{matrix}
			\right.	\nonumber
		\end{align}
	Either setting ${q = 0}$ in the previous expression or studying \cref{eq:gamma2_sigma_mu_t_0} for ${T = 0}$ we gain,
		\begin{align}
			\gtwovar{\shom}{\mu}{0}{0} = \, & \tfrac{1}{\uppi} \Big[ \tfrac{1}{2} \ln \Big( \tfrac{\shom^2}{\snull^2} \Big) + 1 \Big] +
			\begin{cases}
				\tfrac{1}{\uppi} \artanh \Big( \sqrt{1 - \tfrac{\shom^2}{\mu^2}} \Big) - \tfrac{1}{\uppi} \Big( 1 - \frac{\shom^2}{\mu^2} \Big)^{- \frac{1}{2}} \, ,	&	\text{if} \quad \shom^2 < \mu^2 \, ,	\vphantom{\bigg(\bigg)}
				\\
				0 \, ,	&	\text{if} \quad \shom^2 \geq \mu^2 \, .	\vphantom{\bigg(\bigg)}
			\end{cases}	\label{eq:gamma2_sigma_mu_0_0}
		\end{align}
\end{widetext}
	For ${\mu = 0}$ \cref{eq:gamma2_sigma_mu_0_q} reads
		\begin{align}
			& \gtwovar{\shom}{0}{0}{q} =	\vphantom{\Bigg(\Bigg)}	\label{eq:gamma2_sigma_0_0_q}
			\\
			= \, & \tfrac{1}{\uppi} \bigg[ \tfrac{1}{2} \ln \Big( \tfrac{\shom^2}{\snull^2} \Big) + \sqrt{1 + \tfrac{( 2 \shom )^2}{q^2}} \arcoth \Big( \sqrt{1 + \tfrac{( 2 \shom )^2}{q^2}} \Big) \bigg] \, .	\vphantom{\Bigg(\Bigg)}	\nonumber
		\end{align}
	In the limit $\frac{\mu}{\shom} \to 0$ of \cref{eq:gamma2_sigma_mu_0_0} or for $\frac{q}{\shom} \to 0$ of \cref{eq:gamma2_sigma_0_0_q} we find,
		\begin{align}
			\gtwovar{\shom}{0}{0}{0} = \, & \tfrac{1}{\uppi} \Big[ \tfrac{1}{2} \ln \Big( \tfrac{\shom^2}{\snull^2} \Big) + 1 \Big] \, .	\label{eq:gamma2_sigma_0_0_0}
		\end{align}
	From \cref{eq:gamma2_sigma_mu_0_q} for $\shom \to 0$ it can be derived,
		\begin{align}
			\gtwovar{0}{\mu}{0}{q} = \, & \tfrac{1}{2 \uppi} \ln \Big( \tfrac{| 4 \mu^2 - q^2 |}{\snull^2} \Big) \, .	\label{eq:gamma2_0_mu_0_q}
		\end{align}
	For ${q = 0}$ this simplifies to
		\begin{align}
			\gtwovar{0}{\mu}{0}{0} = \, & \tfrac{1}{2 \uppi} \ln \Big( \tfrac{4 \mu^2}{\snull^2} \Big) \, ,	\label{eq:gamma2_0_mu_0_0}
		\end{align}
	while for ${\mu = 0}$ we find,
		\begin{align}
			\gtwovar{0}{0}{0}{q} = \, & \tfrac{1}{2 \uppi} \ln \Big( \tfrac{q^2}{\snull^2} \Big) \, .	\label{eq:gamma2_0_0_0_q}
		\end{align} 
\section{The bosonic wave-function renormalization}
\label{app:the_bosonic_wave-function_renormalization}

	In this appendix the renormalized expressions for the bosonic wave-function renormalization $\zwave{\shom}{\mu}{T}$ are presented, which are listed in \cref{tab:z_limits}, depending on the position in background field space $\shom$, the chemical potential $\mu$, and the temperature $T$.
	A detailed discussion is again presented in \Rcite{Koenigstein:2022phd}.
	
	The wave-function renormalization for $\shom \neq 0$, $\mu \neq 0$, $T \neq 0$ reads
\begin{widetext}
		\begin{align}
			& \zwave{\shom}{\mu}{T} =	\vphantom{\Bigg(\Bigg)}	\label{eq:z_sigma_mu_t}
			\\
			= \, & \tfrac{1}{4 \uppi} \bigg[ \tfrac{1}{3} \tfrac{1}{\shom^2} - \int_{0}^{\infty} \mathrm{d} p \, \tfrac{1}{\energy{p}^3} \Big\{ n ( \energy{p}, \mu ) + n ( \energy{p}, - \mu ) - \tfrac{\energy{p}}{T} \, \big[ n^2 ( \energy{p}, \mu ) + n^2 ( \energy{p}, - \mu ) - n ( \energy{p}, \mu ) - n ( \energy{p}, - \mu ) \big] \Big\} +	\vphantom{\Bigg(\Bigg)}	\nonumber
			\\
			& \quad + \shom^2 \int_{0}^{\infty} \mathrm{d} p \, \tfrac{1}{\energy{p}^5} \Big\{ n ( \energy{p}, \mu ) + n ( \energy{p}, - \mu ) - \tfrac{\energy{p}}{T} \, \big[ n^2 ( \energy{p}, \mu ) + n^2 ( \energy{p}, - \mu ) - n ( \energy{p}, \mu ) - n ( \energy{p}, - \mu ) \big] +	\vphantom{\Bigg(\Bigg)}	\nonumber
			\\
			& \qquad + \tfrac{1}{3} \tfrac{\energy{p}^2}{T^2} \, \big[ 2 \, n^3 ( \energy{p}, \mu ) + 2 \, n^3 ( \energy{p}, - \mu ) - 3 \, n^2 ( \energy{p}, \mu ) - 3 \, n^2 ( \energy{p}, - \mu ) + n ( \energy{p}, \mu ) + n ( \energy{p}, - \mu ) \big] \Big\} \bigg]	 \, . \vphantom{\Bigg(\Bigg)}	\nonumber
		\end{align}
	For ${\mu = 0}$ this expression simplifies and we find
		\begin{align}
			\zwave{\shom}{0}{T} = \, & \tfrac{1}{4 \uppi} \bigg[ \tfrac{1}{3} \tfrac{1}{\shom^2} - 2 \int_{0}^{\infty} \mathrm{d} p \, \tfrac{1}{\energy{p}^3} \Big\{ n ( \energy{p}, 0 ) - \tfrac{\energy{p}}{T} \, \big[ n^2 ( \energy{p}, 0 ) - n ( \energy{p}, 0 ) \big] \Big\} +	\vphantom{\Bigg(\Bigg)}	\label{eq:z_sigma_0_t}
			\\
			& \quad + 2 \, \shom^2 \int_{0}^{\infty} \mathrm{d} p \, \tfrac{1}{\energy{p}^5} \Big\{  n ( \energy{p}, 0 ) - \tfrac{\energy{p}}{T} \, \big[ n^2 ( \energy{p}, 0 ) - n ( \energy{p}, 0 ) \big] + \tfrac{1}{3} \tfrac{\energy{p}^2}{T^2} \, \big[ 2 \, n^3 ( \energy{p}, 0 ) - 3 \, n^2 ( \energy{p}, 0 ) + n ( \energy{p}, 0 ) \big] \Big\} \bigg] \, . \vphantom{\Bigg(\Bigg)}	\nonumber
		\end{align}
\end{widetext}
	On the other hand, for ${\shom = 0}$, but $\mu \neq 0$ and $T \neq 0$, the bosonic wave-function renormalization reads
		\begin{align}
			\zwave{0}{\mu}{T} = - \tfrac{1}{8 \uppi} \tfrac{1}{T^2} \, \mathrm{DLi}_{-2} \big( \tfrac{\mu}{T} \big) \, ,	\label{eq:z_0_mu_t}
		\end{align}
	where definition \labelcref{eq:def_dli} was used. For ${\mu = 0}$, this simplifies to
		\begin{align}
			\zwave{0}{0}{T} = \tfrac{7}{16 \uppi^3} \tfrac{1}{T^2} \, \zeta ( 3 ) \, ,	\label{eq:z_0_0_t}
		\end{align}
	which is consistent with Eq.~(3.17) of Ref.~\cite{Dashen:1974xz}.
	Here $\zeta ( s )$ is the Riemann zeta function.
	However, studying the limit $T \to 0$ first, we find
		\begin{align}
			& \zwave{\shom}{\mu}{0} =	\vphantom{\bigg(\bigg)}	\label{eq:z_sigma_mu_0}
			\\
			= \, & \tfrac{1}{12 \uppi} \tfrac{1}{\shom^2} -
			\begin{cases}
				\tfrac{1}{12 \uppi} \tfrac{1}{\shom^2} \Big( 1 - \tfrac{\shom^2}{\mu^2} \Big)^{- \frac{3}{2}} \, ,	&	\text{if} \quad \shom^2 < \mu^2 \, ,	\vphantom{\bigg(\bigg)}
				\\
				0	&	\text{if} \quad \shom^2 \geq \mu^2 \, .	\vphantom{\bigg(\bigg)}
			\end{cases}	\nonumber
		\end{align}
	Furthermore, for ${\shom = 0}$ this simplifies to
		\begin{align}
			\zwave{0}{\mu}{0} = - \tfrac{1}{8 \uppi} \tfrac{1}{\mu^2} \, ,	\label{eq:z_0_mu_0}
		\end{align}
	while for ${\mu = 0}$ one obtains
		\begin{align}
			\zwave{\shom}{0}{0} = \tfrac{1}{12 \uppi} \tfrac{1}{\shom^2} \, ,	\label{eq:z_sigma_0_0}
		\end{align}
	which is consistent with Eq.~(3.19) of Ref.~\cite{Dashen:1974xz}. 

\bibliography{bibliography} 

\begin{thebibliography}{159}%
\makeatletter
\providecommand \@ifxundefined [1]{%
 \@ifx{#1\undefined}
}%
\providecommand \@ifnum [1]{%
 \ifnum #1\expandafter \@firstoftwo
 \else \expandafter \@secondoftwo
 \fi
}%
\providecommand \@ifx [1]{%
 \ifx #1\expandafter \@firstoftwo
 \else \expandafter \@secondoftwo
 \fi
}%
\providecommand \natexlab [1]{#1}%
\providecommand \enquote  [1]{``#1''}%
\providecommand \bibnamefont  [1]{#1}%
\providecommand \bibfnamefont [1]{#1}%
\providecommand \citenamefont [1]{#1}%
\providecommand \href@noop [0]{\@secondoftwo}%
\providecommand \href [0]{\begingroup \@sanitize@url \@href}%
\providecommand \@href[1]{\@@startlink{#1}\@@href}%
\providecommand \@@href[1]{\endgroup#1\@@endlink}%
\providecommand \@sanitize@url [0]{\catcode `\\12\catcode `\$12\catcode
  `\&12\catcode `\#12\catcode `\^12\catcode `\_12\catcode `\%12\relax}%
\providecommand \@@startlink[1]{}%
\providecommand \@@endlink[0]{}%
\providecommand \url  [0]{\begingroup\@sanitize@url \@url }%
\providecommand \@url [1]{\endgroup\@href {#1}{\urlprefix }}%
\providecommand \urlprefix  [0]{URL }%
\providecommand \Eprint [0]{\href }%
\providecommand \doibase [0]{https://doi.org/}%
\providecommand \selectlanguage [0]{\@gobble}%
\providecommand \bibinfo  [0]{\@secondoftwo}%
\providecommand \bibfield  [0]{\@secondoftwo}%
\providecommand \translation [1]{[#1]}%
\providecommand \BibitemOpen [0]{}%
\providecommand \bibitemStop [0]{}%
\providecommand \bibitemNoStop [0]{.\EOS\space}%
\providecommand \EOS [0]{\spacefactor3000\relax}%
\providecommand \BibitemShut  [1]{\csname bibitem#1\endcsname}%
\let\auto@bib@innerbib\@empty
\bibitem [{\citenamefont {Rechenberger}(2018)}]{Rechenberger:2018talk}%
  \BibitemOpen
  \bibfield  {author} {\bibinfo {author} {\bibfnamefont {S.}~\bibnamefont
  {Rechenberger}},\ }\href@noop {} {\bibinfo {title} {{Inhomogeneous phases at
  high density: In Search for Instabilities}}},\ \bibinfo {howpublished} {{Talk
  at the 1st retreat of the CRC-TR 211 presented by D.~H.~Rischke, March 12-16,
  ZiF Bielefeld and corresponding private notes}} (\bibinfo {year}
  {2018})\BibitemShut {NoStop}%
\bibitem [{\citenamefont {Braun}\ \emph {et~al.}(2015)\citenamefont {Braun},
  \citenamefont {Finkbeiner}, \citenamefont {Karbstein},\ and\ \citenamefont
  {Roscher}}]{Braun:2014fga}%
  \BibitemOpen
  \bibfield  {author} {\bibinfo {author} {\bibfnamefont {J.}~\bibnamefont
  {Braun}}, \bibinfo {author} {\bibfnamefont {S.}~\bibnamefont {Finkbeiner}},
  \bibinfo {author} {\bibfnamefont {F.}~\bibnamefont {Karbstein}},\ and\
  \bibinfo {author} {\bibfnamefont {D.}~\bibnamefont {Roscher}},\ }\bibfield
  {title} {\bibinfo {title} {{Search for inhomogeneous phases in fermionic
  models}},\ }\href {https://doi.org/10.1103/PhysRevD.91.116006} {\bibfield
  {journal} {\bibinfo  {journal} {Phys. Rev. D}\ }\textbf {\bibinfo {volume}
  {91}},\ \bibinfo {pages} {116006} (\bibinfo {year} {2015})},\ \Eprint
  {https://arxiv.org/abs/1410.8181} {arXiv:1410.8181 [hep-ph]} \BibitemShut
  {NoStop}%
\bibitem [{\citenamefont {Thies}\ and\ \citenamefont
  {Urlichs}(2003)}]{Thies:2003kk}%
  \BibitemOpen
  \bibfield  {author} {\bibinfo {author} {\bibfnamefont {M.}~\bibnamefont
  {Thies}}\ and\ \bibinfo {author} {\bibfnamefont {K.}~\bibnamefont
  {Urlichs}},\ }\bibfield  {title} {\bibinfo {title} {{Revised phase diagram of
  the Gross-Neveu model}},\ }\href {https://doi.org/10.1103/PhysRevD.67.125015}
  {\bibfield  {journal} {\bibinfo  {journal} {Phys. Rev. D}\ }\textbf {\bibinfo
  {volume} {67}},\ \bibinfo {pages} {125015} (\bibinfo {year} {2003})},\
  \Eprint {https://arxiv.org/abs/hep-th/0302092} {arXiv:hep-th/0302092}
  \BibitemShut {NoStop}%
\bibitem [{\citenamefont {Nakano}\ and\ \citenamefont
  {Tatsumi}(2005)}]{Nakano:2004cd}%
  \BibitemOpen
  \bibfield  {author} {\bibinfo {author} {\bibfnamefont {E.}~\bibnamefont
  {Nakano}}\ and\ \bibinfo {author} {\bibfnamefont {T.}~\bibnamefont
  {Tatsumi}},\ }\bibfield  {title} {\bibinfo {title} {{Chiral symmetry and
  density wave in quark matter}},\ }\href
  {https://doi.org/10.1103/PhysRevD.71.114006} {\bibfield  {journal} {\bibinfo
  {journal} {Phys. Rev. D}\ }\textbf {\bibinfo {volume} {71}},\ \bibinfo
  {pages} {114006} (\bibinfo {year} {2005})},\ \Eprint
  {https://arxiv.org/abs/hep-ph/0411350} {arXiv:hep-ph/0411350} \BibitemShut
  {NoStop}%
\bibitem [{\citenamefont {Boehmer}\ \emph {et~al.}(2007)\citenamefont
  {Boehmer}, \citenamefont {Thies},\ and\ \citenamefont
  {Urlichs}}]{Boehmer:2007ea}%
  \BibitemOpen
  \bibfield  {author} {\bibinfo {author} {\bibfnamefont {C.}~\bibnamefont
  {Boehmer}}, \bibinfo {author} {\bibfnamefont {M.}~\bibnamefont {Thies}},\
  and\ \bibinfo {author} {\bibfnamefont {K.}~\bibnamefont {Urlichs}},\
  }\bibfield  {title} {\bibinfo {title} {{Tricritical behavior of the massive
  chiral Gross-Neveu model}},\ }\href
  {https://doi.org/10.1103/PhysRevD.75.105017} {\bibfield  {journal} {\bibinfo
  {journal} {Phys. Rev. D}\ }\textbf {\bibinfo {volume} {75}},\ \bibinfo
  {pages} {105017} (\bibinfo {year} {2007})},\ \Eprint
  {https://arxiv.org/abs/hep-th/0702201} {arXiv:hep-th/0702201} \BibitemShut
  {NoStop}%
\bibitem [{\citenamefont {Boehmer}\ \emph {et~al.}(2008)\citenamefont
  {Boehmer}, \citenamefont {Fritsch}, \citenamefont {Kraus},\ and\
  \citenamefont {Thies}}]{Boehmer:2008uq}%
  \BibitemOpen
  \bibfield  {author} {\bibinfo {author} {\bibfnamefont {C.}~\bibnamefont
  {Boehmer}}, \bibinfo {author} {\bibfnamefont {U.}~\bibnamefont {Fritsch}},
  \bibinfo {author} {\bibfnamefont {S.}~\bibnamefont {Kraus}},\ and\ \bibinfo
  {author} {\bibfnamefont {M.}~\bibnamefont {Thies}},\ }\bibfield  {title}
  {\bibinfo {title} {{Phase structure of the massive chiral Gross-Neveu model
  from Hartree-Fock}},\ }\href {https://doi.org/10.1103/PhysRevD.78.065043}
  {\bibfield  {journal} {\bibinfo  {journal} {Phys. Rev. D}\ }\textbf {\bibinfo
  {volume} {78}},\ \bibinfo {pages} {065043"} (\bibinfo {year} {2008})},\
  \Eprint {https://arxiv.org/abs/0807.2571} {arXiv:0807.2571 [hep-th]}
  \BibitemShut {NoStop}%
\bibitem [{\citenamefont {Basar}\ \emph {et~al.}(2009)\citenamefont {Basar},
  \citenamefont {Dunne},\ and\ \citenamefont {Thies}}]{Basar:2009fg}%
  \BibitemOpen
  \bibfield  {author} {\bibinfo {author} {\bibfnamefont {G.}~\bibnamefont
  {Basar}}, \bibinfo {author} {\bibfnamefont {G.~V.}\ \bibnamefont {Dunne}},\
  and\ \bibinfo {author} {\bibfnamefont {M.}~\bibnamefont {Thies}},\ }\bibfield
   {title} {\bibinfo {title} {{Inhomogeneous condensates in the thermodynamics
  of the chiral ${\mathrm{NJL}}_{2}$ model}},\ }\href
  {https://doi.org/10.1103/PhysRevD.79.105012} {\bibfield  {journal} {\bibinfo
  {journal} {Phys. Rev. D}\ }\textbf {\bibinfo {volume} {79}},\ \bibinfo
  {pages} {105012} (\bibinfo {year} {2009})},\ \Eprint
  {https://arxiv.org/abs/0903.1868} {arXiv:0903.1868 [hep-th]} \BibitemShut
  {NoStop}%
\bibitem [{\citenamefont {Nickel}(2009{\natexlab{a}})}]{Nickel:2009ke}%
  \BibitemOpen
  \bibfield  {author} {\bibinfo {author} {\bibfnamefont {D.}~\bibnamefont
  {Nickel}},\ }\bibfield  {title} {\bibinfo {title} {{How many phases meet at
  the chiral critical point?}},\ }\href
  {https://doi.org/10.1103/PhysRevLett.103.072301} {\bibfield  {journal}
  {\bibinfo  {journal} {Phys. Rev. Lett.}\ }\textbf {\bibinfo {volume} {103}},\
  \bibinfo {pages} {072301} (\bibinfo {year} {2009}{\natexlab{a}})},\ \Eprint
  {https://arxiv.org/abs/0902.1778} {arXiv:0902.1778 [hep-ph]} \BibitemShut
  {NoStop}%
\bibitem [{\citenamefont {Abuki}\ \emph {et~al.}(2012)\citenamefont {Abuki},
  \citenamefont {Ishibashi},\ and\ \citenamefont {Suzuki}}]{Abuki:2011pf}%
  \BibitemOpen
  \bibfield  {author} {\bibinfo {author} {\bibfnamefont {H.}~\bibnamefont
  {Abuki}}, \bibinfo {author} {\bibfnamefont {D.}~\bibnamefont {Ishibashi}},\
  and\ \bibinfo {author} {\bibfnamefont {K.}~\bibnamefont {Suzuki}},\
  }\bibfield  {title} {\bibinfo {title} {{Crystalline chiral condensates off
  the tricritical point in a generalized Ginzburg-Landau approach}},\ }\href
  {https://doi.org/10.1103/PhysRevD.85.074002} {\bibfield  {journal} {\bibinfo
  {journal} {Phys. Rev. D}\ }\textbf {\bibinfo {volume} {85}},\ \bibinfo
  {pages} {074002} (\bibinfo {year} {2012})},\ \Eprint
  {https://arxiv.org/abs/1109.1615} {arXiv:1109.1615 [hep-ph]} \BibitemShut
  {NoStop}%
\bibitem [{\citenamefont {de~Forcrand}\ and\ \citenamefont
  {Wenger}(2006)}]{deForcrand:2006zz}%
  \BibitemOpen
  \bibfield  {author} {\bibinfo {author} {\bibfnamefont {P.}~\bibnamefont
  {de~Forcrand}}\ and\ \bibinfo {author} {\bibfnamefont {U.}~\bibnamefont
  {Wenger}},\ }\bibfield  {title} {\bibinfo {title} {{New baryon matter in the
  lattice Gross-Neveu model}},\ }\href {https://doi.org/10.22323/1.032.0152}
  {\bibfield  {journal} {\bibinfo  {journal} {PoS}\ }\textbf {\bibinfo {volume}
  {LAT2006}},\ \bibinfo {pages} {152} (\bibinfo {year} {2006})},\ \Eprint
  {https://arxiv.org/abs/hep-lat/0610117} {arXiv:hep-lat/0610117} \BibitemShut
  {NoStop}%
\bibitem [{\citenamefont {Wagner}(2007)}]{Wagner:2007he}%
  \BibitemOpen
  \bibfield  {author} {\bibinfo {author} {\bibfnamefont {M.}~\bibnamefont
  {Wagner}},\ }\bibfield  {title} {\bibinfo {title} {{Fermions in the
  pseudoparticle approach}},\ }\href
  {https://doi.org/10.1103/PhysRevD.76.076002} {\bibfield  {journal} {\bibinfo
  {journal} {Phys. Rev. D}\ }\textbf {\bibinfo {volume} {76}},\ \bibinfo
  {pages} {076002} (\bibinfo {year} {2007})},\ \Eprint
  {https://arxiv.org/abs/0704.3023} {arXiv:0704.3023 [hep-lat]} \BibitemShut
  {NoStop}%
\bibitem [{\citenamefont {Tripolt}\ \emph {et~al.}(2018)\citenamefont
  {Tripolt}, \citenamefont {Schaefer}, \citenamefont {von Smekal},\ and\
  \citenamefont {Wambach}}]{Tripolt:2017zgc}%
  \BibitemOpen
  \bibfield  {author} {\bibinfo {author} {\bibfnamefont {R.-A.}\ \bibnamefont
  {Tripolt}}, \bibinfo {author} {\bibfnamefont {B.-J.}\ \bibnamefont
  {Schaefer}}, \bibinfo {author} {\bibfnamefont {L.}~\bibnamefont {von
  Smekal}},\ and\ \bibinfo {author} {\bibfnamefont {J.}~\bibnamefont
  {Wambach}},\ }\bibfield  {title} {\bibinfo {title} {{Low-temperature behavior
  of the quark-meson model}},\ }\href
  {https://doi.org/10.1103/PhysRevD.97.034022} {\bibfield  {journal} {\bibinfo
  {journal} {Phys. Rev. D}\ }\textbf {\bibinfo {volume} {97}},\ \bibinfo
  {pages} {034022} (\bibinfo {year} {2018})},\ \Eprint
  {https://arxiv.org/abs/1709.05991} {arXiv:1709.05991 [hep-ph]} \BibitemShut
  {NoStop}%
\bibitem [{\citenamefont {Winstel}\ \emph {et~al.}(2020)\citenamefont
  {Winstel}, \citenamefont {Stoll},\ and\ \citenamefont
  {Wagner}}]{Winstel:2019zfn}%
  \BibitemOpen
  \bibfield  {author} {\bibinfo {author} {\bibfnamefont {M.}~\bibnamefont
  {Winstel}}, \bibinfo {author} {\bibfnamefont {J.}~\bibnamefont {Stoll}},\
  and\ \bibinfo {author} {\bibfnamefont {M.}~\bibnamefont {Wagner}},\
  }\bibfield  {title} {\bibinfo {title} {{Lattice investigation of an
  inhomogeneous phase of the 2 + 1-dimensional Gross-Neveu model in the limit
  of infinitely many flavors}},\ }\href
  {https://doi.org/10.1088/1742-6596/1667/1/012044} {\bibfield  {journal}
  {\bibinfo  {journal} {J. Phys. Conf. Ser.}\ }\textbf {\bibinfo {volume}
  {1667}},\ \bibinfo {pages} {012044} (\bibinfo {year} {2020})},\ \Eprint
  {https://arxiv.org/abs/1909.00064} {arXiv:1909.00064 [hep-lat]} \BibitemShut
  {NoStop}%
\bibitem [{\citenamefont {Carignano}\ and\ \citenamefont
  {Buballa}(2020)}]{Carignano:2019ivp}%
  \BibitemOpen
  \bibfield  {author} {\bibinfo {author} {\bibfnamefont {S.}~\bibnamefont
  {Carignano}}\ and\ \bibinfo {author} {\bibfnamefont {M.}~\bibnamefont
  {Buballa}},\ }\bibfield  {title} {\bibinfo {title} {{Inhomogeneous chiral
  condensates in three-flavor quark matter}},\ }\href
  {https://doi.org/10.1103/PhysRevD.101.014026} {\bibfield  {journal} {\bibinfo
   {journal} {Phys. Rev. D}\ }\textbf {\bibinfo {volume} {101}},\ \bibinfo
  {pages} {014026} (\bibinfo {year} {2020})},\ \Eprint
  {https://arxiv.org/abs/1910.03604} {arXiv:1910.03604 [hep-ph]} \BibitemShut
  {NoStop}%
\bibitem [{\citenamefont {Buballa}\ and\ \citenamefont
  {Carignano}(2019{\natexlab{a}})}]{Buballa:2018hux}%
  \BibitemOpen
  \bibfield  {author} {\bibinfo {author} {\bibfnamefont {M.}~\bibnamefont
  {Buballa}}\ and\ \bibinfo {author} {\bibfnamefont {S.}~\bibnamefont
  {Carignano}},\ }\bibfield  {title} {\bibinfo {title} {{Inhomogeneous chiral
  phases away from the chiral limit}},\ }\href
  {https://doi.org/10.1016/j.physletb.2019.02.045} {\bibfield  {journal}
  {\bibinfo  {journal} {Phys. Lett. B}\ }\textbf {\bibinfo {volume} {791}},\
  \bibinfo {pages} {361} (\bibinfo {year} {2019}{\natexlab{a}})},\ \Eprint
  {https://arxiv.org/abs/1809.10066} {arXiv:1809.10066 [hep-ph]} \BibitemShut
  {NoStop}%
\bibitem [{\citenamefont {Thies}(2020{\natexlab{a}})}]{Thies:2019ejd}%
  \BibitemOpen
  \bibfield  {author} {\bibinfo {author} {\bibfnamefont {M.}~\bibnamefont
  {Thies}},\ }\bibfield  {title} {\bibinfo {title} {{Phase structure of the $(1
  + 1)$-dimensional Nambu-Jona-Lasinio model with isospin}},\ }\href
  {https://doi.org/10.1103/PhysRevD.101.014010} {\bibfield  {journal} {\bibinfo
   {journal} {Phys. Rev. D}\ }\textbf {\bibinfo {volume} {101}},\ \bibinfo
  {pages} {014010} (\bibinfo {year} {2020}{\natexlab{a}})},\ \Eprint
  {https://arxiv.org/abs/1911.11439} {arXiv:1911.11439 [hep-th]} \BibitemShut
  {NoStop}%
\bibitem [{\citenamefont {Buballa}\ \emph {et~al.}(2021)\citenamefont
  {Buballa}, \citenamefont {Kurth}, \citenamefont {Wagner},\ and\ \citenamefont
  {Winstel}}]{Buballa:2020nsi}%
  \BibitemOpen
  \bibfield  {author} {\bibinfo {author} {\bibfnamefont {M.}~\bibnamefont
  {Buballa}}, \bibinfo {author} {\bibfnamefont {L.}~\bibnamefont {Kurth}},
  \bibinfo {author} {\bibfnamefont {M.}~\bibnamefont {Wagner}},\ and\ \bibinfo
  {author} {\bibfnamefont {M.}~\bibnamefont {Winstel}},\ }\bibfield  {title}
  {\bibinfo {title} {{Regulator dependence of inhomogeneous phases in the ( 2+1
  )-dimensional Gross-Neveu model}},\ }\href
  {https://doi.org/10.1103/PhysRevD.103.034503} {\bibfield  {journal} {\bibinfo
   {journal} {Phys. Rev. D}\ }\textbf {\bibinfo {volume} {103}},\ \bibinfo
  {pages} {034503} (\bibinfo {year} {2021})},\ \Eprint
  {https://arxiv.org/abs/2012.09588} {arXiv:2012.09588 [hep-lat]} \BibitemShut
  {NoStop}%
\bibitem [{\citenamefont {Buballa}\ \emph {et~al.}(2020)\citenamefont
  {Buballa}, \citenamefont {Carignano},\ and\ \citenamefont
  {Kurth}}]{Buballa:2020xaa}%
  \BibitemOpen
  \bibfield  {author} {\bibinfo {author} {\bibfnamefont {M.}~\bibnamefont
  {Buballa}}, \bibinfo {author} {\bibfnamefont {S.}~\bibnamefont {Carignano}},\
  and\ \bibinfo {author} {\bibfnamefont {L.}~\bibnamefont {Kurth}},\ }\bibfield
   {title} {\bibinfo {title} {{Inhomogeneous phases in the quark-meson model
  with explicit chiral-symmetry breaking}},\ }\href
  {https://doi.org/10.1140/epjst/e2020-000101-x"} {\bibfield  {journal}
  {\bibinfo  {journal} {Eur. Phys. J. ST}\ }\textbf {\bibinfo {volume} {229}},\
  \bibinfo {pages} {3371} (\bibinfo {year} {2020})},\ \Eprint
  {https://arxiv.org/abs/2006.02133} {arXiv:2006.02133 [hep-ph]} \BibitemShut
  {NoStop}%
\bibitem [{\citenamefont {Winstel}\ \emph {et~al.}(2021)\citenamefont
  {Winstel}, \citenamefont {Pannullo},\ and\ \citenamefont
  {Wagner}}]{Winstel:2021yok}%
  \BibitemOpen
  \bibfield  {author} {\bibinfo {author} {\bibfnamefont {M.}~\bibnamefont
  {Winstel}}, \bibinfo {author} {\bibfnamefont {L.}~\bibnamefont {Pannullo}},\
  and\ \bibinfo {author} {\bibfnamefont {M.}~\bibnamefont {Wagner}},\
  }\bibfield  {title} {\bibinfo {title} {{Phase diagram of the 2+1-dimensional
  Gross-Neveu model with chiral imbalance}},\ }in\ \href@noop {} {\emph
  {\bibinfo {booktitle} {{38th International Symposium on Lattice Field
  Theory}}}}\ (\bibinfo {year} {2021})\ \Eprint
  {https://arxiv.org/abs/2109.04277} {arXiv:2109.04277 [hep-lat]} \BibitemShut
  {NoStop}%
\bibitem [{\citenamefont {Wilczek}(2012)}]{Wilczek:2012jt}%
  \BibitemOpen
  \bibfield  {author} {\bibinfo {author} {\bibfnamefont {F.}~\bibnamefont
  {Wilczek}},\ }\bibfield  {title} {\bibinfo {title} {{Quantum Time
  Crystals}},\ }\href {https://doi.org/10.1103/PhysRevLett.109.160401}
  {\bibfield  {journal} {\bibinfo  {journal} {Phys. Rev. Lett.}\ }\textbf
  {\bibinfo {volume} {109}},\ \bibinfo {pages} {160401} (\bibinfo {year}
  {2012})},\ \Eprint {https://arxiv.org/abs/1202.2539} {arXiv:1202.2539
  [quant-ph]} \BibitemShut {NoStop}%
\bibitem [{\citenamefont {Shapere}\ and\ \citenamefont
  {Wilczek}(2012)}]{Shapere:2012nq}%
  \BibitemOpen
  \bibfield  {author} {\bibinfo {author} {\bibfnamefont {A.}~\bibnamefont
  {Shapere}}\ and\ \bibinfo {author} {\bibfnamefont {F.}~\bibnamefont
  {Wilczek}},\ }\bibfield  {title} {\bibinfo {title} {{Classical Time
  Crystals}},\ }\href {https://doi.org/10.1103/PhysRevLett.109.160402}
  {\bibfield  {journal} {\bibinfo  {journal} {Phys. Rev. Lett.}\ }\textbf
  {\bibinfo {volume} {109}},\ \bibinfo {pages} {160402} (\bibinfo {year}
  {2012})},\ \Eprint {https://arxiv.org/abs/1202.2537} {arXiv:1202.2537
  [cond-mat.other]} \BibitemShut {NoStop}%
\bibitem [{\citenamefont {Gruner}(1994)}]{Gruner:1994zz}%
  \BibitemOpen
  \bibfield  {author} {\bibinfo {author} {\bibfnamefont {G.}~\bibnamefont
  {Gruner}},\ }\bibfield  {title} {\bibinfo {title} {{The dynamics of
  spin-density waves}},\ }\href {https://doi.org/10.1103/RevModPhys.66.1}
  {\bibfield  {journal} {\bibinfo  {journal} {Rev. Mod. Phys.}\ }\textbf
  {\bibinfo {volume} {66}},\ \bibinfo {pages} {1} (\bibinfo {year}
  {1994})}\BibitemShut {NoStop}%
\bibitem [{\citenamefont {Bulgac}\ and\ \citenamefont
  {Forbes}(2008)}]{Bulgac:2008tm}%
  \BibitemOpen
  \bibfield  {author} {\bibinfo {author} {\bibfnamefont {A.}~\bibnamefont
  {Bulgac}}\ and\ \bibinfo {author} {\bibfnamefont {M.~M.}\ \bibnamefont
  {Forbes}},\ }\bibfield  {title} {\bibinfo {title} {{A Unitary Fermi
  Supersolid: The Larkin-Ovchinnikov Phase}},\ }\href
  {https://doi.org/10.1103/PhysRevLett.101.215301} {\bibfield  {journal}
  {\bibinfo  {journal} {Phys. Rev. Lett.}\ }\textbf {\bibinfo {volume} {101}},\
  \bibinfo {pages} {215301} (\bibinfo {year} {2008})},\ \Eprint
  {https://arxiv.org/abs/0804.3364} {arXiv:0804.3364 [cond-mat.supr-con]}
  \BibitemShut {NoStop}%
\bibitem [{\citenamefont {Radzihovsky}(2012)}]{Radzihovsky:2011zz}%
  \BibitemOpen
  \bibfield  {author} {\bibinfo {author} {\bibfnamefont {L.}~\bibnamefont
  {Radzihovsky}},\ }\bibfield  {title} {\bibinfo {title} {{Quantum
  liquid-crystal order in resonant atomic gases}},\ }\href
  {https://doi.org/10.1016/j.physc.2012.04.014} {\bibfield  {journal} {\bibinfo
   {journal} {Physica C}\ }\textbf {\bibinfo {volume} {481}},\ \bibinfo {pages}
  {189} (\bibinfo {year} {2012})},\ \Eprint {https://arxiv.org/abs/1112.0773}
  {arXiv:1112.0773 [cond-mat.quant-gas]} \BibitemShut {NoStop}%
\bibitem [{\citenamefont {Gubbels}\ and\ \citenamefont
  {Stoof}(2013)}]{Gubbels:2012kcu}%
  \BibitemOpen
  \bibfield  {author} {\bibinfo {author} {\bibfnamefont {K.~B.}\ \bibnamefont
  {Gubbels}}\ and\ \bibinfo {author} {\bibfnamefont {H.~T.~C.}\ \bibnamefont
  {Stoof}},\ }\bibfield  {title} {\bibinfo {title} {{Imbalanced Fermi gases at
  unitarity}},\ }\href {https://doi.org/10.1016/j.physrep.2012.11.004}
  {\bibfield  {journal} {\bibinfo  {journal} {Phys. Rept.}\ }\textbf {\bibinfo
  {volume} {525}},\ \bibinfo {pages} {255} (\bibinfo {year} {2013})},\ \Eprint
  {https://arxiv.org/abs/1205.0568} {arXiv:1205.0568 [cond-mat.quant-gas]}
  \BibitemShut {NoStop}%
\bibitem [{\citenamefont {Roscher}\ \emph {et~al.}(2014)\citenamefont
  {Roscher}, \citenamefont {Braun},\ and\ \citenamefont
  {Drut}}]{Roscher:2013cma}%
  \BibitemOpen
  \bibfield  {author} {\bibinfo {author} {\bibfnamefont {D.}~\bibnamefont
  {Roscher}}, \bibinfo {author} {\bibfnamefont {J.}~\bibnamefont {Braun}},\
  and\ \bibinfo {author} {\bibfnamefont {J.~E.}\ \bibnamefont {Drut}},\
  }\bibfield  {title} {\bibinfo {title} {{Inhomogeneous phases in
  one-dimensional mass- and spin-imbalanced Fermi gases}},\ }\href
  {https://doi.org/10.1103/PhysRevA.89.063609} {\bibfield  {journal} {\bibinfo
  {journal} {Phys. Rev. A}\ }\textbf {\bibinfo {volume} {89}},\ \bibinfo
  {pages} {063609} (\bibinfo {year} {2014})},\ \Eprint
  {https://arxiv.org/abs/1311.0179} {arXiv:1311.0179 [cond-mat.quant-gas]}
  \BibitemShut {NoStop}%
\bibitem [{\citenamefont {Baarsma}\ and\ \citenamefont
  {Stoof}(2013)}]{Baarsma:2013in}%
  \BibitemOpen
  \bibfield  {author} {\bibinfo {author} {\bibfnamefont {J.~E.}\ \bibnamefont
  {Baarsma}}\ and\ \bibinfo {author} {\bibfnamefont {H.~T.~C.}\ \bibnamefont
  {Stoof}},\ }\href@noop {} {\bibinfo {title} {Inhomogeneous superfluid phases
  in the unitary li6-k40 mixture}} (\bibinfo {year} {2013}),\ \Eprint
  {https://arxiv.org/abs/1212.5450} {arXiv:1212.5450 [cond-mat.quant-gas]}
  \BibitemShut {NoStop}%
\bibitem [{\citenamefont {Fulde}\ and\ \citenamefont
  {Ferrell}(1964)}]{Fulde:1964zz}%
  \BibitemOpen
  \bibfield  {author} {\bibinfo {author} {\bibfnamefont {P.}~\bibnamefont
  {Fulde}}\ and\ \bibinfo {author} {\bibfnamefont {R.~A.}\ \bibnamefont
  {Ferrell}},\ }\bibfield  {title} {\bibinfo {title} {{Superconductivity in a
  Strong Spin-Exchange Field}},\ }\href
  {https://doi.org/10.1103/PhysRev.135.A550} {\bibfield  {journal} {\bibinfo
  {journal} {Phys. Rev.}\ }\textbf {\bibinfo {volume} {135}},\ \bibinfo {pages}
  {A550} (\bibinfo {year} {1964})}\BibitemShut {NoStop}%
\bibitem [{\citenamefont {Larkin}\ and\ \citenamefont
  {Ovchinnikov}(1964)}]{Larkin:1964wok}%
  \BibitemOpen
  \bibfield  {author} {\bibinfo {author} {\bibfnamefont {A.~I.}\ \bibnamefont
  {Larkin}}\ and\ \bibinfo {author} {\bibfnamefont {Y.~N.}\ \bibnamefont
  {Ovchinnikov}},\ }\bibfield  {title} {\bibinfo {title} {{Nonuniform state of
  superconductors}},\ }\href@noop {} {\bibfield  {journal} {\bibinfo  {journal}
  {Zh. Eksp. Teor. Fiz.}\ }\textbf {\bibinfo {volume} {47}},\ \bibinfo {pages}
  {1136} (\bibinfo {year} {1964})}\BibitemShut {NoStop}%
\bibitem [{\citenamefont {Dautry}\ and\ \citenamefont
  {Nyman}(1979)}]{Dautry:1979bk}%
  \BibitemOpen
  \bibfield  {author} {\bibinfo {author} {\bibfnamefont {F.}~\bibnamefont
  {Dautry}}\ and\ \bibinfo {author} {\bibfnamefont {E.~M.}\ \bibnamefont
  {Nyman}},\ }\bibfield  {title} {\bibinfo {title} {{Pion condensation and the
  $\sigma$-model in liquid neutron matter}},\ }\href
  {https://doi.org/10.1016/0375-9474(79)90518-9} {\bibfield  {journal}
  {\bibinfo  {journal} {Nucl. Phys. A}\ }\textbf {\bibinfo {volume} {319}},\
  \bibinfo {pages} {323} (\bibinfo {year} {1979})}\BibitemShut {NoStop}%
\bibitem [{\citenamefont {Kutschera}\ \emph
  {et~al.}(1990{\natexlab{a}})\citenamefont {Kutschera}, \citenamefont
  {Broniowski},\ and\ \citenamefont {Kotlorz}}]{Kutschera:1989yz}%
  \BibitemOpen
  \bibfield  {author} {\bibinfo {author} {\bibfnamefont {M.}~\bibnamefont
  {Kutschera}}, \bibinfo {author} {\bibfnamefont {W.}~\bibnamefont
  {Broniowski}},\ and\ \bibinfo {author} {\bibfnamefont {A.}~\bibnamefont
  {Kotlorz}},\ }\bibfield  {title} {\bibinfo {title} {{Quark matter with
  neutral pion condensate}},\ }\href
  {https://doi.org/10.1016/0370-2693(90)91421-7} {\bibfield  {journal}
  {\bibinfo  {journal} {Phys. Lett. B}\ }\textbf {\bibinfo {volume} {237}},\
  \bibinfo {pages} {159} (\bibinfo {year} {1990}{\natexlab{a}})}\BibitemShut
  {NoStop}%
\bibitem [{\citenamefont {Broniowski}\ \emph {et~al.}(1991)\citenamefont
  {Broniowski}, \citenamefont {Kotlorz},\ and\ \citenamefont
  {Kutschera}}]{Broniowski:1990dy}%
  \BibitemOpen
  \bibfield  {author} {\bibinfo {author} {\bibfnamefont {W.}~\bibnamefont
  {Broniowski}}, \bibinfo {author} {\bibfnamefont {A.}~\bibnamefont
  {Kotlorz}},\ and\ \bibinfo {author} {\bibfnamefont {M.}~\bibnamefont
  {Kutschera}},\ }\bibfield  {title} {\bibinfo {title} {{Quarks with a pion
  condensate. A new phase of matter}},\ }\href@noop {} {\bibfield  {journal}
  {\bibinfo  {journal} {Acta Phys. Polon. B}\ }\textbf {\bibinfo {volume}
  {22}},\ \bibinfo {pages} {145} (\bibinfo {year} {1991})}\BibitemShut
  {NoStop}%
\bibitem [{\citenamefont {Kutschera}\ \emph
  {et~al.}(1990{\natexlab{b}})\citenamefont {Kutschera}, \citenamefont
  {Broniowski},\ and\ \citenamefont {Kotlorz}}]{Kutschera:1990xk}%
  \BibitemOpen
  \bibfield  {author} {\bibinfo {author} {\bibfnamefont {M.}~\bibnamefont
  {Kutschera}}, \bibinfo {author} {\bibfnamefont {W.}~\bibnamefont
  {Broniowski}},\ and\ \bibinfo {author} {\bibfnamefont {A.}~\bibnamefont
  {Kotlorz}},\ }\bibfield  {title} {\bibinfo {title} {{Quark matter with pion
  condensate in an effective chiral model}},\ }\href
  {https://doi.org/10.1016/0375-9474(90)90128-9} {\bibfield  {journal}
  {\bibinfo  {journal} {Nucl. Phys. A}\ }\textbf {\bibinfo {volume} {516}},\
  \bibinfo {pages} {566} (\bibinfo {year} {1990}{\natexlab{b}})}\BibitemShut
  {NoStop}%
\bibitem [{\citenamefont {Deryagin}\ \emph {et~al.}(1992)\citenamefont
  {Deryagin}, \citenamefont {Grigoriev},\ and\ \citenamefont
  {Rubakov}}]{Deryagin:1992rw}%
  \BibitemOpen
  \bibfield  {author} {\bibinfo {author} {\bibfnamefont {D.~V.}\ \bibnamefont
  {Deryagin}}, \bibinfo {author} {\bibfnamefont {D.~Y.}\ \bibnamefont
  {Grigoriev}},\ and\ \bibinfo {author} {\bibfnamefont {V.~A.}\ \bibnamefont
  {Rubakov}},\ }\bibfield  {title} {\bibinfo {title} {{Standing wave ground
  state in high density, zero temperature QCD at large $N_\mathrm{c}$}},\
  }\href {https://doi.org/10.1142/S0217751X92000302} {\bibfield  {journal}
  {\bibinfo  {journal} {Int. J. Mod. Phys. A}\ }\textbf {\bibinfo {volume}
  {7}},\ \bibinfo {pages} {659} (\bibinfo {year} {1992})}\BibitemShut {NoStop}%
\bibitem [{\citenamefont {Kojo}\ \emph {et~al.}(2010)\citenamefont {Kojo},
  \citenamefont {Hidaka}, \citenamefont {McLerran},\ and\ \citenamefont
  {Pisarski}}]{Kojo:2009ha}%
  \BibitemOpen
  \bibfield  {author} {\bibinfo {author} {\bibfnamefont {T.}~\bibnamefont
  {Kojo}}, \bibinfo {author} {\bibfnamefont {Y.}~\bibnamefont {Hidaka}},
  \bibinfo {author} {\bibfnamefont {L.}~\bibnamefont {McLerran}},\ and\
  \bibinfo {author} {\bibfnamefont {R.~D.}\ \bibnamefont {Pisarski}},\
  }\bibfield  {title} {\bibinfo {title} {{Quarkyonic Chiral Spirals}},\ }\href
  {https://doi.org/10.1016/j.nuclphysa.2010.05.053} {\bibfield  {journal}
  {\bibinfo  {journal} {Nucl. Phys. A}\ }\textbf {\bibinfo {volume} {843}},\
  \bibinfo {pages} {37} (\bibinfo {year} {2010})},\ \Eprint
  {https://arxiv.org/abs/0912.3800} {arXiv:0912.3800 [hep-ph]} \BibitemShut
  {NoStop}%
\bibitem [{\citenamefont {Kojo}\ \emph {et~al.}(2012)\citenamefont {Kojo},
  \citenamefont {Hidaka}, \citenamefont {Fukushima}, \citenamefont {McLerran},\
  and\ \citenamefont {Pisarski}}]{Kojo:2011cn}%
  \BibitemOpen
  \bibfield  {author} {\bibinfo {author} {\bibfnamefont {T.}~\bibnamefont
  {Kojo}}, \bibinfo {author} {\bibfnamefont {Y.}~\bibnamefont {Hidaka}},
  \bibinfo {author} {\bibfnamefont {K.}~\bibnamefont {Fukushima}}, \bibinfo
  {author} {\bibfnamefont {L.~D.}\ \bibnamefont {McLerran}},\ and\ \bibinfo
  {author} {\bibfnamefont {R.~D.}\ \bibnamefont {Pisarski}},\ }\bibfield
  {title} {\bibinfo {title} {{Interweaving Chiral Spirals}},\ }\href
  {https://doi.org/10.1016/j.nuclphysa.2011.11.007} {\bibfield  {journal}
  {\bibinfo  {journal} {Nucl. Phys. A}\ }\textbf {\bibinfo {volume} {875}},\
  \bibinfo {pages} {94} (\bibinfo {year} {2012})},\ \Eprint
  {https://arxiv.org/abs/1107.2124} {arXiv:1107.2124 [hep-ph]} \BibitemShut
  {NoStop}%
\bibitem [{\citenamefont {Buballa}\ and\ \citenamefont
  {Carignano}(2015)}]{Buballa:2014tba}%
  \BibitemOpen
  \bibfield  {author} {\bibinfo {author} {\bibfnamefont {M.}~\bibnamefont
  {Buballa}}\ and\ \bibinfo {author} {\bibfnamefont {S.}~\bibnamefont
  {Carignano}},\ }\bibfield  {title} {\bibinfo {title} {{Inhomogeneous chiral
  condensates}},\ }\href {https://doi.org/10.1016/j.ppnp.2014.11.001}
  {\bibfield  {journal} {\bibinfo  {journal} {Prog. Part. Nucl. Phys.}\
  }\textbf {\bibinfo {volume} {81}},\ \bibinfo {pages} {39} (\bibinfo {year}
  {2015})},\ \Eprint {https://arxiv.org/abs/1406.1367} {arXiv:1406.1367
  [hep-ph]} \BibitemShut {NoStop}%
\bibitem [{\citenamefont {Carignano}\ and\ \citenamefont
  {Buballa}(2012)}]{Carignano:2012sx}%
  \BibitemOpen
  \bibfield  {author} {\bibinfo {author} {\bibfnamefont {S.}~\bibnamefont
  {Carignano}}\ and\ \bibinfo {author} {\bibfnamefont {M.}~\bibnamefont
  {Buballa}},\ }\bibfield  {title} {\bibinfo {title} {{Two-dimensional chiral
  crystals in the NJL model}},\ }\href
  {https://doi.org/10.1103/PhysRevD.86.074018} {\bibfield  {journal} {\bibinfo
  {journal} {Phys. Rev. D}\ }\textbf {\bibinfo {volume} {86}},\ \bibinfo
  {pages} {074018} (\bibinfo {year} {2012})},\ \Eprint
  {https://arxiv.org/abs/1203.5343} {arXiv:1203.5343 [hep-ph]} \BibitemShut
  {NoStop}%
\bibitem [{\citenamefont {Carignano}\ \emph {et~al.}(2014)\citenamefont
  {Carignano}, \citenamefont {Buballa},\ and\ \citenamefont
  {Schaefer}}]{Carignano:2014jla}%
  \BibitemOpen
  \bibfield  {author} {\bibinfo {author} {\bibfnamefont {S.}~\bibnamefont
  {Carignano}}, \bibinfo {author} {\bibfnamefont {M.}~\bibnamefont {Buballa}},\
  and\ \bibinfo {author} {\bibfnamefont {B.-J.}\ \bibnamefont {Schaefer}},\
  }\bibfield  {title} {\bibinfo {title} {{Inhomogeneous phases in the
  quark-meson model with vacuum fluctuations}},\ }\href
  {https://doi.org/10.1103/PhysRevD.90.014033} {\bibfield  {journal} {\bibinfo
  {journal} {Phys. Rev. D}\ }\textbf {\bibinfo {volume} {90}},\ \bibinfo
  {pages} {014033} (\bibinfo {year} {2014})},\ \Eprint
  {https://arxiv.org/abs/1404.0057} {arXiv:1404.0057 [hep-ph]} \BibitemShut
  {NoStop}%
\bibitem [{\citenamefont {Carignano}\ \emph {et~al.}(2018)\citenamefont
  {Carignano}, \citenamefont {Schramm},\ and\ \citenamefont
  {Buballa}}]{Carignano:2018hvn}%
  \BibitemOpen
  \bibfield  {author} {\bibinfo {author} {\bibfnamefont {S.}~\bibnamefont
  {Carignano}}, \bibinfo {author} {\bibfnamefont {M.}~\bibnamefont {Schramm}},\
  and\ \bibinfo {author} {\bibfnamefont {M.}~\bibnamefont {Buballa}},\
  }\bibfield  {title} {\bibinfo {title} {{Influence of vector interactions on
  the favored shape of inhomogeneous chiral condensates}},\ }\href
  {https://doi.org/10.1103/PhysRevD.98.014033} {\bibfield  {journal} {\bibinfo
  {journal} {Phys. Rev. D}\ }\textbf {\bibinfo {volume} {98}},\ \bibinfo
  {pages} {014033} (\bibinfo {year} {2018})},\ \Eprint
  {https://arxiv.org/abs/1805.06203} {arXiv:1805.06203 [hep-ph]} \BibitemShut
  {NoStop}%
\bibitem [{\citenamefont {Buballa}\ and\ \citenamefont
  {Carignano}(2019{\natexlab{b}})}]{Buballa:2019clw}%
  \BibitemOpen
  \bibfield  {author} {\bibinfo {author} {\bibfnamefont {M.}~\bibnamefont
  {Buballa}}\ and\ \bibinfo {author} {\bibfnamefont {S.}~\bibnamefont
  {Carignano}},\ }\bibfield  {title} {\bibinfo {title} {{Influence of quark
  masses and strangeness degrees of freedom on inhomogeneous chiral phases}},\
  }\href {https://doi.org/10.22323/1.336.0202} {\bibfield  {journal} {\bibinfo
  {journal} {PoS}\ }\textbf {\bibinfo {volume} {Confinement2018}},\ \bibinfo
  {pages} {202} (\bibinfo {year} {2019}{\natexlab{b}})},\ \Eprint
  {https://arxiv.org/abs/1906.03241} {arXiv:1906.03241 [hep-ph]} \BibitemShut
  {NoStop}%
\bibitem [{\citenamefont {Nambu}\ and\ \citenamefont
  {Jona-Lasinio}(1961{\natexlab{a}})}]{Nambu:1961tp}%
  \BibitemOpen
  \bibfield  {author} {\bibinfo {author} {\bibfnamefont {Y.}~\bibnamefont
  {Nambu}}\ and\ \bibinfo {author} {\bibfnamefont {G.}~\bibnamefont
  {Jona-Lasinio}},\ }\bibfield  {title} {\bibinfo {title} {{Dynamical model of
  elementary particles based on an analogy with superconductivity. I}},\ }\href
  {https://doi.org/10.1103/PhysRev.122.345} {\bibfield  {journal} {\bibinfo
  {journal} {Phys. Rev.}\ }\textbf {\bibinfo {volume} {122}},\ \bibinfo {pages}
  {345} (\bibinfo {year} {1961}{\natexlab{a}})}\BibitemShut {NoStop}%
\bibitem [{\citenamefont {Nambu}\ and\ \citenamefont
  {Jona-Lasinio}(1961{\natexlab{b}})}]{Nambu:1961fr}%
  \BibitemOpen
  \bibfield  {author} {\bibinfo {author} {\bibfnamefont {Y.}~\bibnamefont
  {Nambu}}\ and\ \bibinfo {author} {\bibfnamefont {G.}~\bibnamefont
  {Jona-Lasinio}},\ }\bibfield  {title} {\bibinfo {title} {{Dynamical model of
  elementary particles based on an analogy with superconductivity. II}},\
  }\href {https://doi.org/10.1103/PhysRev.124.246} {\bibfield  {journal}
  {\bibinfo  {journal} {Phys. Rev.}\ }\textbf {\bibinfo {volume} {124}},\
  \bibinfo {pages} {246} (\bibinfo {year} {1961}{\natexlab{b}})}\BibitemShut
  {NoStop}%
\bibitem [{\citenamefont {Klevansky}(1992)}]{Klevansky:1992qe}%
  \BibitemOpen
  \bibfield  {author} {\bibinfo {author} {\bibfnamefont {S.~P.}\ \bibnamefont
  {Klevansky}},\ }\bibfield  {title} {\bibinfo {title} {{The Nambu-Jona-Lasinio
  model of quantum chromodynamics}},\ }\href
  {https://doi.org/10.1103/RevModPhys.64.649} {\bibfield  {journal} {\bibinfo
  {journal} {Rev. Mod. Phys.}\ }\textbf {\bibinfo {volume} {64}},\ \bibinfo
  {pages} {649} (\bibinfo {year} {1992})}\BibitemShut {NoStop}%
\bibitem [{\citenamefont {Gell-Mann}\ and\ \citenamefont
  {Levy}(1960)}]{GellMann:1960np}%
  \BibitemOpen
  \bibfield  {author} {\bibinfo {author} {\bibfnamefont {M.}~\bibnamefont
  {Gell-Mann}}\ and\ \bibinfo {author} {\bibfnamefont {M.~M.}\ \bibnamefont
  {Levy}},\ }\bibfield  {title} {\bibinfo {title} {{The axial vector current in
  beta decay}},\ }\href {https://doi.org/10.1007/BF02859738} {\bibfield
  {journal} {\bibinfo  {journal} {Nuovo Cim.}\ }\textbf {\bibinfo {volume}
  {16}},\ \bibinfo {pages} {705} (\bibinfo {year} {1960})}\BibitemShut
  {NoStop}%
\bibitem [{\citenamefont {Scavenius}\ \emph {et~al.}(2001)\citenamefont
  {Scavenius}, \citenamefont {Mocsy}, \citenamefont {Mishustin},\ and\
  \citenamefont {Rischke}}]{Scavenius:2000qd}%
  \BibitemOpen
  \bibfield  {author} {\bibinfo {author} {\bibfnamefont {O.}~\bibnamefont
  {Scavenius}}, \bibinfo {author} {\bibfnamefont {A.}~\bibnamefont {Mocsy}},
  \bibinfo {author} {\bibfnamefont {I.~N.}\ \bibnamefont {Mishustin}},\ and\
  \bibinfo {author} {\bibfnamefont {D.~H.}\ \bibnamefont {Rischke}},\
  }\bibfield  {title} {\bibinfo {title} {{Chiral phase transition within
  effective models with constituent quarks}},\ }\href
  {https://doi.org/10.1103/PhysRevC.64.045202} {\bibfield  {journal} {\bibinfo
  {journal} {Phys. Rev. C}\ }\textbf {\bibinfo {volume} {64}},\ \bibinfo
  {pages} {045202} (\bibinfo {year} {2001})},\ \Eprint
  {https://arxiv.org/abs/nucl-th/0007030} {arXiv:nucl-th/0007030} \BibitemShut
  {NoStop}%
\bibitem [{\citenamefont {Schaefer}\ and\ \citenamefont
  {Wambach}(2007)}]{Schaefer:2006ds}%
  \BibitemOpen
  \bibfield  {author} {\bibinfo {author} {\bibfnamefont {B.-J.}\ \bibnamefont
  {Schaefer}}\ and\ \bibinfo {author} {\bibfnamefont {J.}~\bibnamefont
  {Wambach}},\ }\bibfield  {title} {\bibinfo {title} {{Susceptibilities near
  the QCD (tri)critical point}},\ }\href
  {https://doi.org/10.1103/PhysRevD.75.085015} {\bibfield  {journal} {\bibinfo
  {journal} {Phys. Rev. D}\ }\textbf {\bibinfo {volume} {75}},\ \bibinfo
  {pages} {085015} (\bibinfo {year} {2007})},\ \Eprint
  {https://arxiv.org/abs/hep-ph/0603256} {arXiv:hep-ph/0603256} \BibitemShut
  {NoStop}%
\bibitem [{\citenamefont {Asakawa}\ and\ \citenamefont
  {Yazaki}(1989)}]{Asakawa:1989bq}%
  \BibitemOpen
  \bibfield  {author} {\bibinfo {author} {\bibfnamefont {M.}~\bibnamefont
  {Asakawa}}\ and\ \bibinfo {author} {\bibfnamefont {K.}~\bibnamefont
  {Yazaki}},\ }\bibfield  {title} {\bibinfo {title} {{Chiral Restoration at
  Finite Density and Temperature}},\ }\href
  {https://doi.org/10.1016/0375-9474(89)90002-X} {\bibfield  {journal}
  {\bibinfo  {journal} {Nucl. Phys. A}\ }\textbf {\bibinfo {volume} {504}},\
  \bibinfo {pages} {668} (\bibinfo {year} {1989})}\BibitemShut {NoStop}%
\bibitem [{\citenamefont {Buballa}(2005)}]{Buballa:2003qv}%
  \BibitemOpen
  \bibfield  {author} {\bibinfo {author} {\bibfnamefont {M.}~\bibnamefont
  {Buballa}},\ }\bibfield  {title} {\bibinfo {title} {{NJL model analysis of
  quark matter at large density}},\ }\href
  {https://doi.org/10.1016/j.physrep.2004.11.004} {\bibfield  {journal}
  {\bibinfo  {journal} {Phys. Rept.}\ }\textbf {\bibinfo {volume} {407}},\
  \bibinfo {pages} {205} (\bibinfo {year} {2005})},\ \Eprint
  {https://arxiv.org/abs/hep-ph/0402234} {arXiv:hep-ph/0402234} \BibitemShut
  {NoStop}%
\bibitem [{\citenamefont {Nickel}(2009{\natexlab{b}})}]{Nickel:2009wj}%
  \BibitemOpen
  \bibfield  {author} {\bibinfo {author} {\bibfnamefont {D.}~\bibnamefont
  {Nickel}},\ }\bibfield  {title} {\bibinfo {title} {{Inhomogeneous phases in
  the Nambu-Jona-Lasino and quark-meson model}},\ }\href
  {https://doi.org/10.1103/PhysRevD.80.074025} {\bibfield  {journal} {\bibinfo
  {journal} {Phys. Rev. D}\ }\textbf {\bibinfo {volume} {80}},\ \bibinfo
  {pages} {074025} (\bibinfo {year} {2009}{\natexlab{b}})},\ \Eprint
  {https://arxiv.org/abs/v0906.5295} {arXiv:v0906.5295 [hep-ph]} \BibitemShut
  {NoStop}%
\bibitem [{\citenamefont {Heinz}\ \emph {et~al.}(2015)\citenamefont {Heinz},
  \citenamefont {Giacosa},\ and\ \citenamefont {Rischke}}]{Heinz:2013hza}%
  \BibitemOpen
  \bibfield  {author} {\bibinfo {author} {\bibfnamefont {A.}~\bibnamefont
  {Heinz}}, \bibinfo {author} {\bibfnamefont {F.}~\bibnamefont {Giacosa}},\
  and\ \bibinfo {author} {\bibfnamefont {D.~H.}\ \bibnamefont {Rischke}},\
  }\bibfield  {title} {\bibinfo {title} {{Chiral density wave in nuclear
  matter}},\ }\href {https://doi.org/10.1016/j.nuclphysa.2014.09.027}
  {\bibfield  {journal} {\bibinfo  {journal} {Nucl. Phys. A}\ }\textbf
  {\bibinfo {volume} {933}},\ \bibinfo {pages} {34} (\bibinfo {year} {2015})},\
  \Eprint {https://arxiv.org/abs/1312.3244} {arXiv:1312.3244 [nucl-th]}
  \BibitemShut {NoStop}%
\bibitem [{\citenamefont {Heinz}\ \emph {et~al.}(2016)\citenamefont {Heinz},
  \citenamefont {Giacosa}, \citenamefont {Wagner},\ and\ \citenamefont
  {Rischke}}]{Heinz:2015lua}%
  \BibitemOpen
  \bibfield  {author} {\bibinfo {author} {\bibfnamefont {A.}~\bibnamefont
  {Heinz}}, \bibinfo {author} {\bibfnamefont {F.}~\bibnamefont {Giacosa}},
  \bibinfo {author} {\bibfnamefont {M.}~\bibnamefont {Wagner}},\ and\ \bibinfo
  {author} {\bibfnamefont {D.~H.}\ \bibnamefont {Rischke}},\ }\bibfield
  {title} {\bibinfo {title} {{Inhomogeneous condensation in effective models
  for QCD using the finite-mode approach}},\ }\href
  {https://doi.org/10.1103/PhysRevD.93.014007} {\bibfield  {journal} {\bibinfo
  {journal} {Phys. Rev. D}\ }\textbf {\bibinfo {volume} {93}},\ \bibinfo
  {pages} {014007} (\bibinfo {year} {2016})},\ \Eprint
  {https://arxiv.org/abs/1508.06057} {arXiv:1508.06057 [hep-ph]} \BibitemShut
  {NoStop}%
\bibitem [{\citenamefont {Braun}\ \emph {et~al.}(2016)\citenamefont {Braun},
  \citenamefont {Karbstein}, \citenamefont {Rechenberger},\ and\ \citenamefont
  {Roscher}}]{Braun:2015fva}%
  \BibitemOpen
  \bibfield  {author} {\bibinfo {author} {\bibfnamefont {J.}~\bibnamefont
  {Braun}}, \bibinfo {author} {\bibfnamefont {F.}~\bibnamefont {Karbstein}},
  \bibinfo {author} {\bibfnamefont {S.}~\bibnamefont {Rechenberger}},\ and\
  \bibinfo {author} {\bibfnamefont {D.}~\bibnamefont {Roscher}},\ }\bibfield
  {title} {\bibinfo {title} {{Crystalline ground states in Polyakov-loop
  extended Nambu\textendash{}Jona-Lasinio models}},\ }\href
  {https://doi.org/10.1103/PhysRevD.93.014032} {\bibfield  {journal} {\bibinfo
  {journal} {Phys. Rev. D}\ }\textbf {\bibinfo {volume} {93}},\ \bibinfo
  {pages} {014032} (\bibinfo {year} {2016})},\ \Eprint
  {https://arxiv.org/abs/1510.04012} {arXiv:1510.04012 [hep-ph]} \BibitemShut
  {NoStop}%
\bibitem [{\citenamefont {Lakaschus}\ \emph {et~al.}(2021)\citenamefont
  {Lakaschus}, \citenamefont {Buballa},\ and\ \citenamefont
  {Rischke}}]{Lakaschus:2020caq}%
  \BibitemOpen
  \bibfield  {author} {\bibinfo {author} {\bibfnamefont {P.}~\bibnamefont
  {Lakaschus}}, \bibinfo {author} {\bibfnamefont {M.}~\bibnamefont {Buballa}},\
  and\ \bibinfo {author} {\bibfnamefont {D.~H.}\ \bibnamefont {Rischke}},\
  }\bibfield  {title} {\bibinfo {title} {{Competition of inhomogeneous chiral
  phases and two-flavor color superconductivity in the NJL model}},\ }\href
  {https://doi.org/10.1103/PhysRevD.103.034030} {\bibfield  {journal} {\bibinfo
   {journal} {Phys. Rev. D}\ }\textbf {\bibinfo {volume} {103}},\ \bibinfo
  {pages} {034030} (\bibinfo {year} {2021})},\ \Eprint
  {https://arxiv.org/abs/2012.07520} {arXiv:2012.07520 [hep-ph]} \BibitemShut
  {NoStop}%
\bibitem [{\citenamefont {Fu}\ \emph {et~al.}(2020)\citenamefont {Fu},
  \citenamefont {Pawlowski},\ and\ \citenamefont {Rennecke}}]{Fu:2019hdw}%
  \BibitemOpen
  \bibfield  {author} {\bibinfo {author} {\bibfnamefont {W.-j.}\ \bibnamefont
  {Fu}}, \bibinfo {author} {\bibfnamefont {J.~M.}\ \bibnamefont {Pawlowski}},\
  and\ \bibinfo {author} {\bibfnamefont {F.}~\bibnamefont {Rennecke}},\
  }\bibfield  {title} {\bibinfo {title} {{QCD phase structure at finite
  temperature and density}},\ }\href
  {https://doi.org/10.1103/PhysRevD.101.054032} {\bibfield  {journal} {\bibinfo
   {journal} {Phys. Rev. D}\ }\textbf {\bibinfo {volume} {101}},\ \bibinfo
  {pages} {054032} (\bibinfo {year} {2020})},\ \Eprint
  {https://arxiv.org/abs/1909.02991} {arXiv:1909.02991 [hep-ph]} \BibitemShut
  {NoStop}%
\bibitem [{\citenamefont {Pisarski}\ \emph {et~al.}(2021)\citenamefont
  {Pisarski}, \citenamefont {Rennecke}, \citenamefont {Tsvelik},\ and\
  \citenamefont {Valgushev}}]{Pisarski:2020gkx}%
  \BibitemOpen
  \bibfield  {author} {\bibinfo {author} {\bibfnamefont {R.~D.}\ \bibnamefont
  {Pisarski}}, \bibinfo {author} {\bibfnamefont {F.}~\bibnamefont {Rennecke}},
  \bibinfo {author} {\bibfnamefont {A.~M.}\ \bibnamefont {Tsvelik}},\ and\
  \bibinfo {author} {\bibfnamefont {S.}~\bibnamefont {Valgushev}},\ }\bibfield
  {title} {\bibinfo {title} {{The Lifshitz Regime and its Experimental
  Signals}},\ }\href {https://doi.org/10.1016/j.nuclphysa.2020.121910}
  {\bibfield  {journal} {\bibinfo  {journal} {Nucl. Phys. A}\ }\textbf
  {\bibinfo {volume} {1005}},\ \bibinfo {pages} {121910} (\bibinfo {year}
  {2021})},\ \Eprint {https://arxiv.org/abs/2005.00045} {arXiv:2005.00045
  [nucl-th]} \BibitemShut {NoStop}%
\bibitem [{\citenamefont {Pisarski}\ \emph {et~al.}(2020)\citenamefont
  {Pisarski}, \citenamefont {Tsvelik},\ and\ \citenamefont
  {Valgushev}}]{Pisarski:2020dnx}%
  \BibitemOpen
  \bibfield  {author} {\bibinfo {author} {\bibfnamefont {R.~D.}\ \bibnamefont
  {Pisarski}}, \bibinfo {author} {\bibfnamefont {A.~M.}\ \bibnamefont
  {Tsvelik}},\ and\ \bibinfo {author} {\bibfnamefont {S.}~\bibnamefont
  {Valgushev}},\ }\bibfield  {title} {\bibinfo {title} {{How transverse thermal
  fluctuations disorder a condensate of chiral spirals into a quantum spin
  liquid}},\ }\href {https://doi.org/10.1103/PhysRevD.102.016015} {\bibfield
  {journal} {\bibinfo  {journal} {Phys. Rev. D}\ }\textbf {\bibinfo {volume}
  {102}},\ \bibinfo {pages} {016015} (\bibinfo {year} {2020})},\ \Eprint
  {https://arxiv.org/abs/2005.10259} {arXiv:2005.10259 [hep-ph]} \BibitemShut
  {NoStop}%
\bibitem [{\citenamefont {Pisarski}\ and\ \citenamefont
  {Rennecke}(2021)}]{Pisarski:2021qof}%
  \BibitemOpen
  \bibfield  {author} {\bibinfo {author} {\bibfnamefont {R.~D.}\ \bibnamefont
  {Pisarski}}\ and\ \bibinfo {author} {\bibfnamefont {F.}~\bibnamefont
  {Rennecke}},\ }\bibfield  {title} {\bibinfo {title} {{Signatures of Moat
  Regimes in Heavy-Ion Collisions}},\ }\href
  {https://doi.org/10.1103/PhysRevLett.127.152302} {\bibfield  {journal}
  {\bibinfo  {journal} {Phys. Rev. Lett.}\ }\textbf {\bibinfo {volume} {127}},\
  \bibinfo {pages} {152302} (\bibinfo {year} {2021})},\ \Eprint
  {https://arxiv.org/abs/2103.06890} {arXiv:2103.06890 [hep-ph]} \BibitemShut
  {NoStop}%
\bibitem [{\citenamefont {Rennecke}\ and\ \citenamefont
  {Pisarski}(2021)}]{Rennecke:2021ovl}%
  \BibitemOpen
  \bibfield  {author} {\bibinfo {author} {\bibfnamefont {F.}~\bibnamefont
  {Rennecke}}\ and\ \bibinfo {author} {\bibfnamefont {R.~D.}\ \bibnamefont
  {Pisarski}},\ }\bibfield  {title} {\bibinfo {title} {{Moat Regimes in QCD and
  their Signatures in Heavy-Ion Collisions}},\ }in\ \href@noop {} {\emph
  {\bibinfo {booktitle} {{International Conference on Critical Point and Onset
  of Deconfinement}}}}\ (\bibinfo {year} {2021})\ \Eprint
  {https://arxiv.org/abs/2110.02625} {arXiv:2110.02625 [hep-ph]} \BibitemShut
  {NoStop}%
\bibitem [{\citenamefont {Gross}\ and\ \citenamefont
  {Neveu}(1974)}]{Gross:1974jv}%
  \BibitemOpen
  \bibfield  {author} {\bibinfo {author} {\bibfnamefont {D.~J.}\ \bibnamefont
  {Gross}}\ and\ \bibinfo {author} {\bibfnamefont {A.}~\bibnamefont {Neveu}},\
  }\bibfield  {title} {\bibinfo {title} {{Dynamical Symmetry Breaking in
  Asymptotically Free Field Theories}},\ }\href
  {https://doi.org/10.1103/PhysRevD.10.3235} {\bibfield  {journal} {\bibinfo
  {journal} {Phys. Rev. D}\ }\textbf {\bibinfo {volume} {10}},\ \bibinfo
  {pages} {3235} (\bibinfo {year} {1974})}\BibitemShut {NoStop}%
\bibitem [{\citenamefont {Brzoska}\ and\ \citenamefont
  {Thies}(2002)}]{Brzoska:2001iq}%
  \BibitemOpen
  \bibfield  {author} {\bibinfo {author} {\bibfnamefont {A.}~\bibnamefont
  {Brzoska}}\ and\ \bibinfo {author} {\bibfnamefont {M.}~\bibnamefont
  {Thies}},\ }\bibfield  {title} {\bibinfo {title} {{No first order phase
  transition in the Gross-Neveu model?}},\ }\href
  {https://doi.org/10.1103/PhysRevD.65.125001} {\bibfield  {journal} {\bibinfo
  {journal} {Phys. Rev. D}\ }\textbf {\bibinfo {volume} {65}},\ \bibinfo
  {pages} {125001} (\bibinfo {year} {2002})},\ \Eprint
  {https://arxiv.org/abs/hep-th/0112105} {arXiv:hep-th/0112105} \BibitemShut
  {NoStop}%
\bibitem [{\citenamefont {Thies}(2004)}]{Thies:2003br}%
  \BibitemOpen
  \bibfield  {author} {\bibinfo {author} {\bibfnamefont {M.}~\bibnamefont
  {Thies}},\ }\bibfield  {title} {\bibinfo {title} {{Analytical solution of the
  Gross-Neveu model at finite density}},\ }\href
  {https://doi.org/10.1103/PhysRevD.69.067703} {\bibfield  {journal} {\bibinfo
  {journal} {Phys. Rev. D}\ }\textbf {\bibinfo {volume} {69}},\ \bibinfo
  {pages} {067703} (\bibinfo {year} {2004})},\ \Eprint
  {https://arxiv.org/abs/hep-th/0308164} {arXiv:hep-th/0308164} \BibitemShut
  {NoStop}%
\bibitem [{\citenamefont {Schnetz}\ \emph {et~al.}(2004)\citenamefont
  {Schnetz}, \citenamefont {Thies},\ and\ \citenamefont
  {Urlichs}}]{Schnetz:2004vr}%
  \BibitemOpen
  \bibfield  {author} {\bibinfo {author} {\bibfnamefont {O.}~\bibnamefont
  {Schnetz}}, \bibinfo {author} {\bibfnamefont {M.}~\bibnamefont {Thies}},\
  and\ \bibinfo {author} {\bibfnamefont {K.}~\bibnamefont {Urlichs}},\
  }\bibfield  {title} {\bibinfo {title} {{Phase diagram of the Gross-Neveu
  model: Exact results and condensed matter precursors}},\ }\href
  {https://doi.org/10.1016/j.aop.2004.06.009} {\bibfield  {journal} {\bibinfo
  {journal} {Annals Phys.}\ }\textbf {\bibinfo {volume} {314}},\ \bibinfo
  {pages} {425} (\bibinfo {year} {2004})},\ \Eprint
  {https://arxiv.org/abs/hep-th/0402014} {arXiv:hep-th/0402014} \BibitemShut
  {NoStop}%
\bibitem [{\citenamefont {Schnetz}\ \emph {et~al.}(2006)\citenamefont
  {Schnetz}, \citenamefont {Thies},\ and\ \citenamefont
  {Urlichs}}]{Schnetz:2005ih}%
  \BibitemOpen
  \bibfield  {author} {\bibinfo {author} {\bibfnamefont {O.}~\bibnamefont
  {Schnetz}}, \bibinfo {author} {\bibfnamefont {M.}~\bibnamefont {Thies}},\
  and\ \bibinfo {author} {\bibfnamefont {K.}~\bibnamefont {Urlichs}},\
  }\bibfield  {title} {\bibinfo {title} {{Full phase diagram of the massive
  Gross-Neveu model}},\ }\href {https://doi.org/10.1016/j.aop.2005.12.007}
  {\bibfield  {journal} {\bibinfo  {journal} {Annals Phys.}\ }\textbf {\bibinfo
  {volume} {321}},\ \bibinfo {pages} {2604} (\bibinfo {year} {2006})},\ \Eprint
  {https://arxiv.org/abs/hep-th/0511206} {arXiv:hep-th/0511206} \BibitemShut
  {NoStop}%
\bibitem [{\citenamefont {Schnetz}\ \emph {et~al.}(2005)\citenamefont
  {Schnetz}, \citenamefont {Thies},\ and\ \citenamefont
  {Urlichs}}]{Schnetz:2005vh}%
  \BibitemOpen
  \bibfield  {author} {\bibinfo {author} {\bibfnamefont {O.}~\bibnamefont
  {Schnetz}}, \bibinfo {author} {\bibfnamefont {M.}~\bibnamefont {Thies}},\
  and\ \bibinfo {author} {\bibfnamefont {K.}~\bibnamefont {Urlichs}},\
  }\href@noop {} {\bibinfo {title} {{The Phase diagram of the massive
  Gross-Neveu model, revisited}}} (\bibinfo {year} {2005}),\ \Eprint
  {https://arxiv.org/abs/hep-th/0507120} {arXiv:hep-th/0507120} \BibitemShut
  {NoStop}%
\bibitem [{\citenamefont {Thies}\ and\ \citenamefont
  {Urlichs}(2005{\natexlab{a}})}]{Thies:2005wv}%
  \BibitemOpen
  \bibfield  {author} {\bibinfo {author} {\bibfnamefont {M.}~\bibnamefont
  {Thies}}\ and\ \bibinfo {author} {\bibfnamefont {K.}~\bibnamefont
  {Urlichs}},\ }\bibfield  {title} {\bibinfo {title} {{From non-degenerate
  conducting polymers to dense matter in the massive Gross-Neveu model}},\
  }\href {https://doi.org/10.1103/PhysRevD.72.105008} {\bibfield  {journal}
  {\bibinfo  {journal} {Phys. Rev. D}\ }\textbf {\bibinfo {volume} {72}},\
  \bibinfo {pages} {105008} (\bibinfo {year} {2005}{\natexlab{a}})},\ \Eprint
  {https://arxiv.org/abs/hep-th/0505024} {arXiv:hep-th/0505024} \BibitemShut
  {NoStop}%
\bibitem [{\citenamefont {Thies}(2006)}]{Thies:2006ti}%
  \BibitemOpen
  \bibfield  {author} {\bibinfo {author} {\bibfnamefont {M.}~\bibnamefont
  {Thies}},\ }\bibfield  {title} {\bibinfo {title} {{From relativistic quantum
  fields to condensed matter and back again: Updating the Gross-Neveu phase
  diagram}},\ }\href {https://doi.org/10.1088/0305-4470/39/41/S04} {\bibfield
  {journal} {\bibinfo  {journal} {J. Phys. A}\ }\textbf {\bibinfo {volume}
  {39}},\ \bibinfo {pages} {12707} (\bibinfo {year} {2006})},\ \Eprint
  {https://arxiv.org/abs/hep-th/0601049} {arXiv:hep-th/0601049} \BibitemShut
  {NoStop}%
\bibitem [{\citenamefont {Dunne}\ and\ \citenamefont
  {Feinberg}(1998)}]{Dunne:1997ia}%
  \BibitemOpen
  \bibfield  {author} {\bibinfo {author} {\bibfnamefont {G.~V.}\ \bibnamefont
  {Dunne}}\ and\ \bibinfo {author} {\bibfnamefont {J.}~\bibnamefont
  {Feinberg}},\ }\bibfield  {title} {\bibinfo {title} {{Self isospectral
  periodic potentials and supersymmetric quantum mechanics}},\ }\href
  {https://doi.org/10.1103/PhysRevD.57.1271} {\bibfield  {journal} {\bibinfo
  {journal} {Phys. Rev. D}\ }\textbf {\bibinfo {volume} {57}},\ \bibinfo
  {pages} {1271} (\bibinfo {year} {1998})},\ \Eprint
  {https://arxiv.org/abs/hep-th/9706012} {arXiv:hep-th/9706012} \BibitemShut
  {NoStop}%
\bibitem [{\citenamefont {Cooper}\ \emph {et~al.}(2001)\citenamefont {Cooper},
  \citenamefont {Khare},\ and\ \citenamefont {Sukhatme}}]{Cooper:2001weo}%
  \BibitemOpen
  \bibfield  {author} {\bibinfo {author} {\bibfnamefont {F.}~\bibnamefont
  {Cooper}}, \bibinfo {author} {\bibfnamefont {A.}~\bibnamefont {Khare}},\ and\
  \bibinfo {author} {\bibfnamefont {U.}~\bibnamefont {Sukhatme}},\ }\href
  {https://doi.org/10.1142/4687} {\emph {\bibinfo {title} {{Supersymmetry and
  quantum mechanics}}}}\ (\bibinfo  {publisher} {World Scientific},\ \bibinfo
  {year} {2001})\BibitemShut {NoStop}%
\bibitem [{\citenamefont {Schön}\ and\ \citenamefont
  {Thies}(2000{\natexlab{a}})}]{Schon:2000he}%
  \BibitemOpen
  \bibfield  {author} {\bibinfo {author} {\bibfnamefont {V.}~\bibnamefont
  {Schön}}\ and\ \bibinfo {author} {\bibfnamefont {M.}~\bibnamefont {Thies}},\
  }\bibfield  {title} {\bibinfo {title} {{Emergence of Skyrme crystal in
  Gross-Neveu and 't Hooft models at finite density}},\ }\href
  {https://doi.org/10.1103/PhysRevD.62.096002} {\bibfield  {journal} {\bibinfo
  {journal} {Phys. Rev. D}\ }\textbf {\bibinfo {volume} {62}},\ \bibinfo
  {pages} {096002} (\bibinfo {year} {2000}{\natexlab{a}})},\ \Eprint
  {https://arxiv.org/abs/hep-th/0003195} {arXiv:hep-th/0003195} \BibitemShut
  {NoStop}%
\bibitem [{\citenamefont {Schön}\ and\ \citenamefont
  {Thies}(2000{\natexlab{b}})}]{Schon:2000qy}%
  \BibitemOpen
  \bibfield  {author} {\bibinfo {author} {\bibfnamefont {V.}~\bibnamefont
  {Schön}}\ and\ \bibinfo {author} {\bibfnamefont {M.}~\bibnamefont {Thies}},\
  }\bibinfo {title} {{2-D model field theories at finite temperature and
  density}},\ in\ \href {https://doi.org/10.1142/9789812810458_0041} {\emph
  {\bibinfo {booktitle} {At The Frontier of Particle Physics: Handbook of QCD,
  Boris Ioffe Festschrift}}},\ Vol.~\bibinfo {volume} {3},\ \bibinfo {editor}
  {edited by\ \bibinfo {editor} {\bibfnamefont {M.}~\bibnamefont {Shifman}}\
  and\ \bibinfo {editor} {\bibfnamefont {B.}~\bibnamefont {Ioffe}}}\ (\bibinfo
  {publisher} {World Scentific},\ \bibinfo {year} {2000})\ Chap.~\bibinfo
  {chapter} {33}, pp.\ \bibinfo {pages} {1945--2032},\ \Eprint
  {https://arxiv.org/abs/hep-th/0008175} {arXiv:hep-th/0008175} \BibitemShut
  {NoStop}%
\bibitem [{\citenamefont {Thies}(2020{\natexlab{b}})}]{Thies:2020ofv}%
  \BibitemOpen
  \bibfield  {author} {\bibinfo {author} {\bibfnamefont {M.}~\bibnamefont
  {Thies}},\ }\bibfield  {title} {\bibinfo {title} {{Duality study of the
  chiral Heisenberg-Gross-Neveu model in 1+1 dimensions}},\ }\href
  {https://doi.org/10.1103/PhysRevD.102.096006} {\bibfield  {journal} {\bibinfo
   {journal} {Phys. Rev. D}\ }\textbf {\bibinfo {volume} {102}},\ \bibinfo
  {pages} {096006} (\bibinfo {year} {2020}{\natexlab{b}})},\ \Eprint
  {https://arxiv.org/abs/2008.13119} {arXiv:2008.13119 [hep-th]} \BibitemShut
  {NoStop}%
\bibitem [{\citenamefont {Adhikari}\ \emph {et~al.}(2017)\citenamefont
  {Adhikari}, \citenamefont {Andersen},\ and\ \citenamefont
  {Kneschke}}]{Adhikari:2017ydi}%
  \BibitemOpen
  \bibfield  {author} {\bibinfo {author} {\bibfnamefont {P.}~\bibnamefont
  {Adhikari}}, \bibinfo {author} {\bibfnamefont {J.~O.}\ \bibnamefont
  {Andersen}},\ and\ \bibinfo {author} {\bibfnamefont {P.}~\bibnamefont
  {Kneschke}},\ }\bibfield  {title} {\bibinfo {title} {{Inhomogeneous chiral
  condensate in the quark-meson model}},\ }\href
  {https://doi.org/10.1103/PhysRevD.96.016013} {\bibfield  {journal} {\bibinfo
  {journal} {Phys. Rev. D}\ }\textbf {\bibinfo {volume} {96}},\ \bibinfo
  {pages} {016013} (\bibinfo {year} {2017})},\ \bibinfo {note} {[Erratum:
  Phys.Rev.D 98, 099902 (2018)]},\ \Eprint {https://arxiv.org/abs/1702.01324}
  {arXiv:1702.01324 [hep-ph]} \BibitemShut {NoStop}%
\bibitem [{\citenamefont {Steil}\ \emph {et~al.}(2021)\citenamefont {Steil},
  \citenamefont {Buballa},\ and\ \citenamefont {Schaefer}}]{Steil:2021RGMF}%
  \BibitemOpen
  \bibfield  {author} {\bibinfo {author} {\bibfnamefont {M.~J.}\ \bibnamefont
  {Steil}}, \bibinfo {author} {\bibfnamefont {M.}~\bibnamefont {Buballa}},\
  and\ \bibinfo {author} {\bibfnamefont {B.-J.}\ \bibnamefont {Schaefer}},\
  }\href@noop {} {\bibinfo {title} {{Inhomogeneous chiral condensates in the
  quark-meson model with the functional renormalization group}}},\ \bibinfo
  {howpublished} {in preparation} (\bibinfo {year} {2021})\BibitemShut
  {NoStop}%
\bibitem [{\citenamefont {Narayanan}(2020)}]{Narayanan:2020uqt}%
  \BibitemOpen
  \bibfield  {author} {\bibinfo {author} {\bibfnamefont {R.}~\bibnamefont
  {Narayanan}},\ }\bibfield  {title} {\bibinfo {title} {{Phase diagram of the
  large $N$ Gross-Neveu model in a finite periodic box}},\ }\href
  {https://doi.org/10.1103/PhysRevD.101.096001} {\bibfield  {journal} {\bibinfo
   {journal} {Phys. Rev. D}\ }\textbf {\bibinfo {volume} {101}},\ \bibinfo
  {pages} {096001} (\bibinfo {year} {2020})},\ \Eprint
  {https://arxiv.org/abs/2001.09200} {arXiv:2001.09200 [hep-th]} \BibitemShut
  {NoStop}%
\bibitem [{\citenamefont {Pannullo}\ \emph {et~al.}(2022)\citenamefont
  {Pannullo}, \citenamefont {Wagner},\ and\ \citenamefont
  {Winstel}}]{Pannullo:2021edr}%
  \BibitemOpen
  \bibfield  {author} {\bibinfo {author} {\bibfnamefont {L.}~\bibnamefont
  {Pannullo}}, \bibinfo {author} {\bibfnamefont {M.}~\bibnamefont {Wagner}},\
  and\ \bibinfo {author} {\bibfnamefont {M.}~\bibnamefont {Winstel}},\
  }\bibfield  {title} {\bibinfo {title} {{Inhomogeneous Phases in the Chirally
  Imbalanced 2 + 1-Dimensional Gross-Neveu Model and Their Absence in the
  Continuum Limit}},\ }\href {https://doi.org/10.3390/sym14020265} {\bibfield
  {journal} {\bibinfo  {journal} {Symmetry}\ }\textbf {\bibinfo {volume}
  {14}},\ \bibinfo {pages} {265} (\bibinfo {year} {2022})},\ \Eprint
  {https://arxiv.org/abs/2112.11183} {arXiv:2112.11183 [hep-lat]} \BibitemShut
  {NoStop}%
\bibitem [{\citenamefont {Abuki}(2014)}]{Abuki:2013pla}%
  \BibitemOpen
  \bibfield  {author} {\bibinfo {author} {\bibfnamefont {H.}~\bibnamefont
  {Abuki}},\ }\bibfield  {title} {\bibinfo {title} {{Ginzburg-Landau phase
  diagram of QCD near chiral critical point - chiral defect lattice and
  solitonic pion condensate}},\ }\href
  {https://doi.org/10.1016/j.physletb.2013.11.037} {\bibfield  {journal}
  {\bibinfo  {journal} {Phys. Lett. B}\ }\textbf {\bibinfo {volume} {728}},\
  \bibinfo {pages} {427} (\bibinfo {year} {2014})},\ \Eprint
  {https://arxiv.org/abs/1307.8173} {arXiv:1307.8173 [hep-ph]} \BibitemShut
  {NoStop}%
\bibitem [{\citenamefont {Zinn-Justin}(1991)}]{ZinnJustin:1991yn}%
  \BibitemOpen
  \bibfield  {author} {\bibinfo {author} {\bibfnamefont {J.}~\bibnamefont
  {Zinn-Justin}},\ }\bibfield  {title} {\bibinfo {title} {{Four fermion
  interaction near four-dimensions}},\ }\href
  {https://doi.org/10.1016/0550-3213(91)90043-W} {\bibfield  {journal}
  {\bibinfo  {journal} {Nucl. Phys. B}\ }\textbf {\bibinfo {volume} {367}},\
  \bibinfo {pages} {105} (\bibinfo {year} {1991})}\BibitemShut {NoStop}%
\bibitem [{\citenamefont {Zinn-Justin}(2002)}]{ZinnJustin:2002ru}%
  \BibitemOpen
  \bibfield  {author} {\bibinfo {author} {\bibfnamefont {J.}~\bibnamefont
  {Zinn-Justin}},\ }\href
  {https://doi.org/10.1093/acprof:oso/9780198509233.001.0001} {\emph {\bibinfo
  {title} {{Quantum field theory and critical phenomena}}}},\ \bibinfo
  {edition} {4th}\ ed.,\ \bibinfo {series} {Int. Ser. Monogr. Phys.}, Vol.\
  \bibinfo {volume} {113}\ (\bibinfo  {publisher} {Oxford University Press},\
  \bibinfo {year} {2002})\ pp.\ \bibinfo {pages} {1--1054},\ \bibinfo {note} {a
  Clarendon Press Publication}\BibitemShut {NoStop}%
\bibitem [{\citenamefont {Peskin}\ and\ \citenamefont
  {Schroeder}(1995)}]{Peskin:1995ev}%
  \BibitemOpen
  \bibfield  {author} {\bibinfo {author} {\bibfnamefont {M.~E.}\ \bibnamefont
  {Peskin}}\ and\ \bibinfo {author} {\bibfnamefont {D.~V.}\ \bibnamefont
  {Schroeder}},\ }\href@noop {} {\emph {\bibinfo {title} {{An introduction to
  quantum field theory}}}}\ (\bibinfo  {publisher} {Addison-Wesley},\ \bibinfo
  {address} {Reading, USA},\ \bibinfo {year} {1995})\BibitemShut {NoStop}%
\bibitem [{\citenamefont {Stoll}\ \emph {et~al.}(2021)\citenamefont {Stoll},
  \citenamefont {Zorbach}, \citenamefont {Koenigstein}, \citenamefont {Steil},\
  and\ \citenamefont {Rechenberger}}]{Stoll:2021ori}%
  \BibitemOpen
  \bibfield  {author} {\bibinfo {author} {\bibfnamefont {J.}~\bibnamefont
  {Stoll}}, \bibinfo {author} {\bibfnamefont {N.}~\bibnamefont {Zorbach}},
  \bibinfo {author} {\bibfnamefont {A.}~\bibnamefont {Koenigstein}}, \bibinfo
  {author} {\bibfnamefont {M.~J.}\ \bibnamefont {Steil}},\ and\ \bibinfo
  {author} {\bibfnamefont {S.}~\bibnamefont {Rechenberger}},\ }\href@noop {}
  {\bibinfo {title} {{Bosonic fluctuations in the $( 1 + 1 )$-dimensional
  Gross-Neveu(-Yukawa) model at varying $\mu$ and $T$ and finite $N$}}}
  (\bibinfo {year} {2021}),\ \Eprint {https://arxiv.org/abs/2108.10616}
  {arXiv:2108.10616 [hep-ph]} \BibitemShut {NoStop}%
\bibitem [{\citenamefont {Fitzner}\ and\ \citenamefont
  {Thies}(2011)}]{Fitzner:2010nv}%
  \BibitemOpen
  \bibfield  {author} {\bibinfo {author} {\bibfnamefont {C.}~\bibnamefont
  {Fitzner}}\ and\ \bibinfo {author} {\bibfnamefont {M.}~\bibnamefont
  {Thies}},\ }\bibfield  {title} {\bibinfo {title} {{Exact solution of N baryon
  problem in the Gross-Neveu model}},\ }\href
  {https://doi.org/10.1103/PhysRevD.83.085001} {\bibfield  {journal} {\bibinfo
  {journal} {Phys. Rev. D}\ }\textbf {\bibinfo {volume} {83}},\ \bibinfo
  {pages} {085001} (\bibinfo {year} {2011})},\ \Eprint
  {https://arxiv.org/abs/1010.5322} {arXiv:1010.5322 [hep-th]} \BibitemShut
  {NoStop}%
\bibitem [{\citenamefont {Dunne}\ \emph {et~al.}(2011)\citenamefont {Dunne},
  \citenamefont {Fitzner},\ and\ \citenamefont {Thies}}]{Dunne:2011wu}%
  \BibitemOpen
  \bibfield  {author} {\bibinfo {author} {\bibfnamefont {G.~V.}\ \bibnamefont
  {Dunne}}, \bibinfo {author} {\bibfnamefont {C.}~\bibnamefont {Fitzner}},\
  and\ \bibinfo {author} {\bibfnamefont {M.}~\bibnamefont {Thies}},\ }\bibfield
   {title} {\bibinfo {title} {{Baryon-baryon scattering in the Gross-Neveu
  model: the large $N$ solution}},\ }\href
  {https://doi.org/10.1103/PhysRevD.84.105014} {\bibfield  {journal} {\bibinfo
  {journal} {Phys. Rev. D}\ }\textbf {\bibinfo {volume} {84}},\ \bibinfo
  {pages} {105014} (\bibinfo {year} {2011})},\ \Eprint
  {https://arxiv.org/abs/1108.5888} {arXiv:1108.5888 [hep-th]} \BibitemShut
  {NoStop}%
\bibitem [{\citenamefont {Thies}(2017{\natexlab{a}})}]{Thies:2017fkr}%
  \BibitemOpen
  \bibfield  {author} {\bibinfo {author} {\bibfnamefont {M.}~\bibnamefont
  {Thies}},\ }\bibfield  {title} {\bibinfo {title} {{Beyond integrability:
  Baryon-baryon backward scattering in the massive Gross-Neveu model}},\ }\href
  {https://doi.org/10.1103/PhysRevD.96.076012} {\bibfield  {journal} {\bibinfo
  {journal} {Phys. Rev. D}\ }\textbf {\bibinfo {volume} {96}},\ \bibinfo
  {pages} {076012} (\bibinfo {year} {2017}{\natexlab{a}})},\ \Eprint
  {https://arxiv.org/abs/1706.06382} {arXiv:1706.06382 [hep-th]} \BibitemShut
  {NoStop}%
\bibitem [{\citenamefont {Lenz}\ \emph
  {et~al.}(2020{\natexlab{a}})\citenamefont {Lenz}, \citenamefont {Pannullo},
  \citenamefont {Wagner}, \citenamefont {Wellegehausen},\ and\ \citenamefont
  {Wipf}}]{Lenz:2020cuv}%
  \BibitemOpen
  \bibfield  {author} {\bibinfo {author} {\bibfnamefont {J.~J.}\ \bibnamefont
  {Lenz}}, \bibinfo {author} {\bibfnamefont {L.}~\bibnamefont {Pannullo}},
  \bibinfo {author} {\bibfnamefont {M.}~\bibnamefont {Wagner}}, \bibinfo
  {author} {\bibfnamefont {B.~H.}\ \bibnamefont {Wellegehausen}},\ and\
  \bibinfo {author} {\bibfnamefont {A.}~\bibnamefont {Wipf}},\ }\bibfield
  {title} {\bibinfo {title} {{Baryons in the Gross-Neveu model in 1+1
  dimensions at finite number of flavors}},\ }\href
  {https://doi.org/10.1103/PhysRevD.102.114501} {\bibfield  {journal} {\bibinfo
   {journal} {Phys. Rev. D}\ }\textbf {\bibinfo {volume} {102}},\ \bibinfo
  {pages} {114501} (\bibinfo {year} {2020}{\natexlab{a}})},\ \Eprint
  {https://arxiv.org/abs/2007.08382} {arXiv:2007.08382 [hep-lat]} \BibitemShut
  {NoStop}%
\bibitem [{\citenamefont {Harrington}\ and\ \citenamefont
  {Yildiz}(1975{\natexlab{a}})}]{Harrington:1974tf}%
  \BibitemOpen
  \bibfield  {author} {\bibinfo {author} {\bibfnamefont {B.~J.}\ \bibnamefont
  {Harrington}}\ and\ \bibinfo {author} {\bibfnamefont {A.}~\bibnamefont
  {Yildiz}},\ }\bibfield  {title} {\bibinfo {title} {{Restoration of
  Dynamically Broken Symmetries at Finite Temperature}},\ }\href
  {https://doi.org/10.1103/PhysRevD.11.779} {\bibfield  {journal} {\bibinfo
  {journal} {Phys. Rev. D}\ }\textbf {\bibinfo {volume} {11}},\ \bibinfo
  {pages} {779} (\bibinfo {year} {1975}{\natexlab{a}})}\BibitemShut {NoStop}%
\bibitem [{\citenamefont {Pannullo}(2020)}]{Pannullo:2019}%
  \BibitemOpen
  \bibfield  {author} {\bibinfo {author} {\bibfnamefont {L.}~\bibnamefont
  {Pannullo}},\ }\emph {\bibinfo {title} {{Inhomogeneous Phases in the $1 +
  1$-Dimensional Gross-Neveu Model at Finite Number of Fermion Flavors}}},\
  \href {https://itp.uni-frankfurt.de/~mwagner/theses/MA_Pannullo.pdf}
  {\bibinfo {type} {Master thesis}},\ \bibinfo  {school} {Goethe University
  Frankfurt} (\bibinfo {year} {2020}),\ \bibinfo {note} {updated version from
  June 30, 2020 with minor corrections.}\BibitemShut {Stop}%
\bibitem [{\citenamefont {Wolff}(1985)}]{Wolff:1985av}%
  \BibitemOpen
  \bibfield  {author} {\bibinfo {author} {\bibfnamefont {U.}~\bibnamefont
  {Wolff}},\ }\bibfield  {title} {\bibinfo {title} {{The phase diagram of the
  infinite-N Gross-Neveu model at finite temperature and chemical potential}},\
  }\href {https://doi.org/10.1016/0370-2693(85)90671-9} {\bibfield  {journal}
  {\bibinfo  {journal} {Phys. Lett. B}\ }\textbf {\bibinfo {volume} {157}},\
  \bibinfo {pages} {303} (\bibinfo {year} {1985})}\BibitemShut {NoStop}%
\bibitem [{\citenamefont {Jacobs}(1974)}]{Jacobs:1974ys}%
  \BibitemOpen
  \bibfield  {author} {\bibinfo {author} {\bibfnamefont {L.}~\bibnamefont
  {Jacobs}},\ }\bibfield  {title} {\bibinfo {title} {{Critical behavior in a
  class of $O(N)$-invariant field theories in two dimensions}},\ }\href
  {https://doi.org/10.1103/PhysRevD.10.3956} {\bibfield  {journal} {\bibinfo
  {journal} {Phys. Rev. D}\ }\textbf {\bibinfo {volume} {10}},\ \bibinfo
  {pages} {3956} (\bibinfo {year} {1974})}\BibitemShut {NoStop}%
\bibitem [{\citenamefont {Dashen}\ \emph
  {et~al.}(1975{\natexlab{a}})\citenamefont {Dashen}, \citenamefont {Ma},\ and\
  \citenamefont {Rajaraman}}]{Dashen:1974xz}%
  \BibitemOpen
  \bibfield  {author} {\bibinfo {author} {\bibfnamefont {R.~F.}\ \bibnamefont
  {Dashen}}, \bibinfo {author} {\bibfnamefont {S.-k.}\ \bibnamefont {Ma}},\
  and\ \bibinfo {author} {\bibfnamefont {R.}~\bibnamefont {Rajaraman}},\
  }\bibfield  {title} {\bibinfo {title} {{Finite temperature behavior of a
  relativistic field theory with dynamical symmetry breaking}},\ }\href
  {https://doi.org/10.1103/PhysRevD.11.1499} {\bibfield  {journal} {\bibinfo
  {journal} {Phys. Rev. D}\ }\textbf {\bibinfo {volume} {11}},\ \bibinfo
  {pages} {1499} (\bibinfo {year} {1975}{\natexlab{a}})}\BibitemShut {NoStop}%
\bibitem [{\citenamefont {Harrington}\ and\ \citenamefont
  {Yildiz}(1975{\natexlab{b}})}]{Harrington:1974te}%
  \BibitemOpen
  \bibfield  {author} {\bibinfo {author} {\bibfnamefont {B.~J.}\ \bibnamefont
  {Harrington}}\ and\ \bibinfo {author} {\bibfnamefont {A.}~\bibnamefont
  {Yildiz}},\ }\bibfield  {title} {\bibinfo {title} {{Chiral Symmetry Behavior
  at Large Densities}},\ }\href {https://doi.org/10.1103/PhysRevD.11.1705}
  {\bibfield  {journal} {\bibinfo  {journal} {Phys. Rev. D}\ }\textbf {\bibinfo
  {volume} {11}},\ \bibinfo {pages} {1705} (\bibinfo {year}
  {1975}{\natexlab{b}})}\BibitemShut {NoStop}%
\bibitem [{\citenamefont {Dashen}\ \emph
  {et~al.}(1975{\natexlab{b}})\citenamefont {Dashen}, \citenamefont
  {Hasslacher},\ and\ \citenamefont {Neveu}}]{Dashen:1975xh}%
  \BibitemOpen
  \bibfield  {author} {\bibinfo {author} {\bibfnamefont {R.~F.}\ \bibnamefont
  {Dashen}}, \bibinfo {author} {\bibfnamefont {B.}~\bibnamefont {Hasslacher}},\
  and\ \bibinfo {author} {\bibfnamefont {A.}~\bibnamefont {Neveu}},\ }\bibfield
   {title} {\bibinfo {title} {{Semiclassical Bound States in an Asymptotically
  Free Theory}},\ }\href {https://doi.org/10.1103/PhysRevD.12.2443} {\bibfield
  {journal} {\bibinfo  {journal} {Phys. Rev. D}\ }\textbf {\bibinfo {volume}
  {12}},\ \bibinfo {pages} {2443} (\bibinfo {year}
  {1975}{\natexlab{b}})}\BibitemShut {NoStop}%
\bibitem [{\citenamefont {Affleck}(1982)}]{Affleck:1981bn}%
  \BibitemOpen
  \bibfield  {author} {\bibinfo {author} {\bibfnamefont {I.~K.}\ \bibnamefont
  {Affleck}},\ }\bibfield  {title} {\bibinfo {title} {{Phase Transition in the
  Lattice {Gross-Neveu} Model}},\ }\href
  {https://doi.org/10.1016/0370-2693(82)90441-5} {\bibfield  {journal}
  {\bibinfo  {journal} {Phys. Lett. B}\ }\textbf {\bibinfo {volume} {109}},\
  \bibinfo {pages} {307} (\bibinfo {year} {1982})}\BibitemShut {NoStop}%
\bibitem [{\citenamefont {Cohen}\ \emph {et~al.}(1981)\citenamefont {Cohen},
  \citenamefont {Elitzur},\ and\ \citenamefont {Rabinovici}}]{Cohen:1981qz}%
  \BibitemOpen
  \bibfield  {author} {\bibinfo {author} {\bibfnamefont {Y.}~\bibnamefont
  {Cohen}}, \bibinfo {author} {\bibfnamefont {S.}~\bibnamefont {Elitzur}},\
  and\ \bibinfo {author} {\bibfnamefont {E.}~\bibnamefont {Rabinovici}},\
  }\bibfield  {title} {\bibinfo {title} {{Monte Carlo Study of Chiral
  Structure: The {Gross-Neveu} Model}},\ }\href
  {https://doi.org/10.1016/0370-2693(81)90128-3} {\bibfield  {journal}
  {\bibinfo  {journal} {Phys. Lett. B}\ }\textbf {\bibinfo {volume} {104}},\
  \bibinfo {pages} {289} (\bibinfo {year} {1981})}\BibitemShut {NoStop}%
\bibitem [{\citenamefont {Cohen}\ \emph {et~al.}(1983)\citenamefont {Cohen},
  \citenamefont {Elitzur},\ and\ \citenamefont {Rabinovici}}]{Cohen:1983nr}%
  \BibitemOpen
  \bibfield  {author} {\bibinfo {author} {\bibfnamefont {Y.}~\bibnamefont
  {Cohen}}, \bibinfo {author} {\bibfnamefont {S.}~\bibnamefont {Elitzur}},\
  and\ \bibinfo {author} {\bibfnamefont {E.}~\bibnamefont {Rabinovici}},\
  }\bibfield  {title} {\bibinfo {title} {{A Monte Carlo study of the
  Gross-Neveu model}},\ }\href {https://doi.org/10.1016/0550-3213(83)90136-0}
  {\bibfield  {journal} {\bibinfo  {journal} {Nucl. Phys. B}\ }\textbf
  {\bibinfo {volume} {220}},\ \bibinfo {pages} {102} (\bibinfo {year}
  {1983})}\BibitemShut {NoStop}%
\bibitem [{\citenamefont {Wetzel}(1985)}]{Wetzel:1984nw}%
  \BibitemOpen
  \bibfield  {author} {\bibinfo {author} {\bibfnamefont {W.}~\bibnamefont
  {Wetzel}},\ }\bibfield  {title} {\bibinfo {title} {{Two Loop Beta Function
  for the {Gross-Neveu} Model}},\ }\href
  {https://doi.org/10.1016/0370-2693(85)90551-9} {\bibfield  {journal}
  {\bibinfo  {journal} {Phys. Lett. B}\ }\textbf {\bibinfo {volume} {153}},\
  \bibinfo {pages} {297} (\bibinfo {year} {1985})}\BibitemShut {NoStop}%
\bibitem [{\citenamefont {Shankar}(1985)}]{Shankar:1985zc}%
  \BibitemOpen
  \bibfield  {author} {\bibinfo {author} {\bibfnamefont {R.}~\bibnamefont
  {Shankar}},\ }\bibfield  {title} {\bibinfo {title} {{Ashkin-Teller and
  Gross-Neveu models: New relations and results}},\ }\href
  {https://doi.org/10.1103/PhysRevLett.55.453} {\bibfield  {journal} {\bibinfo
  {journal} {Phys. Rev. Lett.}\ }\textbf {\bibinfo {volume} {55}},\ \bibinfo
  {pages} {453} (\bibinfo {year} {1985})}\BibitemShut {NoStop}%
\bibitem [{\citenamefont {Karsch}\ \emph {et~al.}(1987)\citenamefont {Karsch},
  \citenamefont {Kogut},\ and\ \citenamefont {Wyld}}]{Karsch:1986hm}%
  \BibitemOpen
  \bibfield  {author} {\bibinfo {author} {\bibfnamefont {F.}~\bibnamefont
  {Karsch}}, \bibinfo {author} {\bibfnamefont {J.~B.}\ \bibnamefont {Kogut}},\
  and\ \bibinfo {author} {\bibfnamefont {H.~W.}\ \bibnamefont {Wyld}},\
  }\bibfield  {title} {\bibinfo {title} {{The Gross-Neveu Model at finite
  temperature and density}},\ }\href
  {https://doi.org/10.1016/0550-3213(87)90149-0} {\bibfield  {journal}
  {\bibinfo  {journal} {Nucl. Phys. B}\ }\textbf {\bibinfo {volume} {280}},\
  \bibinfo {pages} {289} (\bibinfo {year} {1987})}\BibitemShut {NoStop}%
\bibitem [{\citenamefont {Treml}(1989)}]{Treml:1989}%
  \BibitemOpen
  \bibfield  {author} {\bibinfo {author} {\bibfnamefont {T.~F.}\ \bibnamefont
  {Treml}},\ }\bibfield  {title} {\bibinfo {title} {{Dynamical mass generation
  in the Gross-Neveu model at finite temperature and density}},\ }\href
  {https://doi.org/10.1103/PhysRevD.39.679} {\bibfield  {journal} {\bibinfo
  {journal} {Phys. Rev. D}\ }\textbf {\bibinfo {volume} {39}},\ \bibinfo
  {pages} {679} (\bibinfo {year} {1989})}\BibitemShut {NoStop}%
\bibitem [{\citenamefont {Rosenstein}\ \emph {et~al.}(1991)\citenamefont
  {Rosenstein}, \citenamefont {Warr},\ and\ \citenamefont
  {Park}}]{Rosenstein:1990nm}%
  \BibitemOpen
  \bibfield  {author} {\bibinfo {author} {\bibfnamefont {B.}~\bibnamefont
  {Rosenstein}}, \bibinfo {author} {\bibfnamefont {B.~J.}\ \bibnamefont
  {Warr}},\ and\ \bibinfo {author} {\bibfnamefont {S.~H.}\ \bibnamefont
  {Park}},\ }\bibfield  {title} {\bibinfo {title} {{Dynamical symmetry breaking
  in four Fermi interaction models}},\ }\href
  {https://doi.org/10.1016/0370-1573(91)90129-A} {\bibfield  {journal}
  {\bibinfo  {journal} {Phys. Rept.}\ }\textbf {\bibinfo {volume} {205}},\
  \bibinfo {pages} {59} (\bibinfo {year} {1991})}\BibitemShut {NoStop}%
\bibitem [{\citenamefont {Gracey}(1990)}]{Gracey:1990sx}%
  \BibitemOpen
  \bibfield  {author} {\bibinfo {author} {\bibfnamefont {J.~A.}\ \bibnamefont
  {Gracey}},\ }\bibfield  {title} {\bibinfo {title} {{Three loop calculations
  in the $O(N)$ Gross-Neveu model}},\ }\href
  {https://doi.org/10.1016/0550-3213(90)90186-H} {\bibfield  {journal}
  {\bibinfo  {journal} {Nucl. Phys. B}\ }\textbf {\bibinfo {volume} {341}},\
  \bibinfo {pages} {403} (\bibinfo {year} {1990})}\BibitemShut {NoStop}%
\bibitem [{\citenamefont {Gracey}(1991{\natexlab{a}})}]{Gracey:1990wi}%
  \BibitemOpen
  \bibfield  {author} {\bibinfo {author} {\bibfnamefont {J.~A.}\ \bibnamefont
  {Gracey}},\ }\bibfield  {title} {\bibinfo {title} {{Calculation of exponent
  eta to O(1/N**2) in the O(N) Gross-Neveu model}},\ }\href
  {https://doi.org/10.1142/S0217751X91000241} {\bibfield  {journal} {\bibinfo
  {journal} {Int. J. Mod. Phys. A}\ }\textbf {\bibinfo {volume} {6}},\ \bibinfo
  {pages} {395} (\bibinfo {year} {1991}{\natexlab{a}})},\ \bibinfo {note}
  {[Erratum: Int.J.Mod.Phys.A 6, 2755 (1991)]}\BibitemShut {NoStop}%
\bibitem [{\citenamefont {Gracey}(1991{\natexlab{b}})}]{Gracey:1991vy}%
  \BibitemOpen
  \bibfield  {author} {\bibinfo {author} {\bibfnamefont {J.~A.}\ \bibnamefont
  {Gracey}},\ }\bibfield  {title} {\bibinfo {title} {{Computation of the three
  loop Beta function of the $O(N)$ Gross-Neveu model in minimal subtraction}},\
  }\href {https://doi.org/10.1016/0550-3213(91)90012-M} {\bibfield  {journal}
  {\bibinfo  {journal} {Nucl. Phys. B}\ }\textbf {\bibinfo {volume} {367}},\
  \bibinfo {pages} {657} (\bibinfo {year} {1991}{\natexlab{b}})}\BibitemShut
  {NoStop}%
\bibitem [{\citenamefont {Pausch}\ \emph {et~al.}(1991)\citenamefont {Pausch},
  \citenamefont {Thies},\ and\ \citenamefont {Dolman}}]{Pausch:1991ee}%
  \BibitemOpen
  \bibfield  {author} {\bibinfo {author} {\bibfnamefont {R.}~\bibnamefont
  {Pausch}}, \bibinfo {author} {\bibfnamefont {M.}~\bibnamefont {Thies}},\ and\
  \bibinfo {author} {\bibfnamefont {V.~L.}\ \bibnamefont {Dolman}},\ }\bibfield
   {title} {\bibinfo {title} {{Solving the Gross-Neveu model with relativistic
  many body methods}},\ }\href {https://doi.org/10.1007/BF01295773} {\bibfield
  {journal} {\bibinfo  {journal} {Z. Phys. A}\ }\textbf {\bibinfo {volume}
  {338}},\ \bibinfo {pages} {441} (\bibinfo {year} {1991})}\BibitemShut
  {NoStop}%
\bibitem [{\citenamefont {Chodos}\ and\ \citenamefont
  {Minakata}(1994)}]{Chodos:1993mf}%
  \BibitemOpen
  \bibfield  {author} {\bibinfo {author} {\bibfnamefont {A.}~\bibnamefont
  {Chodos}}\ and\ \bibinfo {author} {\bibfnamefont {H.}~\bibnamefont
  {Minakata}},\ }\bibfield  {title} {\bibinfo {title} {{The Gross-Neveu model
  as an effective theory for polyacetylene}},\ }\href
  {https://doi.org/10.1016/0375-9601(94)90557-6} {\bibfield  {journal}
  {\bibinfo  {journal} {Phys. Lett. A}\ }\textbf {\bibinfo {volume} {191}},\
  \bibinfo {pages} {39} (\bibinfo {year} {1994})}\BibitemShut {NoStop}%
\bibitem [{\citenamefont {Barducci}\ \emph {et~al.}(1995)\citenamefont
  {Barducci}, \citenamefont {Casalbuoni}, \citenamefont {Modugno},
  \citenamefont {Pettini},\ and\ \citenamefont {Gatto}}]{Barducci:1994cb}%
  \BibitemOpen
  \bibfield  {author} {\bibinfo {author} {\bibfnamefont {A.}~\bibnamefont
  {Barducci}}, \bibinfo {author} {\bibfnamefont {R.}~\bibnamefont
  {Casalbuoni}}, \bibinfo {author} {\bibfnamefont {M.}~\bibnamefont {Modugno}},
  \bibinfo {author} {\bibfnamefont {G.}~\bibnamefont {Pettini}},\ and\ \bibinfo
  {author} {\bibfnamefont {R.}~\bibnamefont {Gatto}},\ }\bibfield  {title}
  {\bibinfo {title} {{Thermodynamics of the massive Gross-Neveu model}},\
  }\href {https://doi.org/10.1103/PhysRevD.51.3042} {\bibfield  {journal}
  {\bibinfo  {journal} {Phys. Rev. D}\ }\textbf {\bibinfo {volume} {51}},\
  \bibinfo {pages} {3042} (\bibinfo {year} {1995})},\ \Eprint
  {https://arxiv.org/abs/hep-th/9406117} {arXiv:hep-th/9406117} \BibitemShut
  {NoStop}%
\bibitem [{\citenamefont {Blaizot}\ \emph {et~al.}(2003)\citenamefont
  {Blaizot}, \citenamefont {Mendez-Galain},\ and\ \citenamefont
  {Wschebor}}]{Blaizot:2002nh}%
  \BibitemOpen
  \bibfield  {author} {\bibinfo {author} {\bibfnamefont {J.-P.}\ \bibnamefont
  {Blaizot}}, \bibinfo {author} {\bibfnamefont {R.}~\bibnamefont
  {Mendez-Galain}},\ and\ \bibinfo {author} {\bibfnamefont {N.}~\bibnamefont
  {Wschebor}},\ }\bibfield  {title} {\bibinfo {title} {{The Gross-Neveu model
  at finite temperature at next to leading order in the 1/N expansion}},\
  }\href {https://doi.org/10.1016/S0003-4916(03)00072-1} {\bibfield  {journal}
  {\bibinfo  {journal} {Annals Phys.}\ }\textbf {\bibinfo {volume} {307}},\
  \bibinfo {pages} {209} (\bibinfo {year} {2003})},\ \Eprint
  {https://arxiv.org/abs/hep-ph/0212084} {arXiv:hep-ph/0212084} \BibitemShut
  {NoStop}%
\bibitem [{\citenamefont {Thies}(2003)}]{Thies:2003zr}%
  \BibitemOpen
  \bibfield  {author} {\bibinfo {author} {\bibfnamefont {M.}~\bibnamefont
  {Thies}},\ }\bibfield  {title} {\bibinfo {title} {{Duality between quark
  quark and quark anti-quark pairing in 1+1 dimensional large N models}},\
  }\href {https://doi.org/10.1103/PhysRevD.68.047703} {\bibfield  {journal}
  {\bibinfo  {journal} {Phys. Rev. D}\ }\textbf {\bibinfo {volume} {68}},\
  \bibinfo {pages} {047703} (\bibinfo {year} {2003})},\ \Eprint
  {https://arxiv.org/abs/hep-th/0303026} {arXiv:hep-th/0303026} \BibitemShut
  {NoStop}%
\bibitem [{\citenamefont {Thies}\ and\ \citenamefont
  {Urlichs}(2005{\natexlab{b}})}]{Thies:2005vq}%
  \BibitemOpen
  \bibfield  {author} {\bibinfo {author} {\bibfnamefont {M.}~\bibnamefont
  {Thies}}\ and\ \bibinfo {author} {\bibfnamefont {K.}~\bibnamefont
  {Urlichs}},\ }\bibfield  {title} {\bibinfo {title} {{Baryons in massive
  Gross-Neveu models}},\ }\href {https://doi.org/10.1103/PhysRevD.71.105008}
  {\bibfield  {journal} {\bibinfo  {journal} {Phys. Rev. D}\ }\textbf {\bibinfo
  {volume} {71}},\ \bibinfo {pages} {105008} (\bibinfo {year}
  {2005}{\natexlab{b}})},\ \Eprint {https://arxiv.org/abs/hep-th/0502210}
  {arXiv:hep-th/0502210} \BibitemShut {NoStop}%
\bibitem [{\citenamefont {Karbstein}\ and\ \citenamefont
  {Thies}(2007{\natexlab{a}})}]{Karbstein:2006er}%
  \BibitemOpen
  \bibfield  {author} {\bibinfo {author} {\bibfnamefont {F.}~\bibnamefont
  {Karbstein}}\ and\ \bibinfo {author} {\bibfnamefont {M.}~\bibnamefont
  {Thies}},\ }\bibfield  {title} {\bibinfo {title} {{How to get from imaginary
  to real chemical potential}},\ }\href
  {https://doi.org/10.1103/PhysRevD.75.025003} {\bibfield  {journal} {\bibinfo
  {journal} {Phys. Rev. D}\ }\textbf {\bibinfo {volume} {75}},\ \bibinfo
  {pages} {025003} (\bibinfo {year} {2007}{\natexlab{a}})},\ \Eprint
  {https://arxiv.org/abs/hep-th/0610243} {arXiv:hep-th/0610243} \BibitemShut
  {NoStop}%
\bibitem [{\citenamefont {Karbstein}\ and\ \citenamefont
  {Thies}(2008)}]{Karbstein:2007be}%
  \BibitemOpen
  \bibfield  {author} {\bibinfo {author} {\bibfnamefont {F.}~\bibnamefont
  {Karbstein}}\ and\ \bibinfo {author} {\bibfnamefont {M.}~\bibnamefont
  {Thies}},\ }\bibfield  {title} {\bibinfo {title} {{Integrating out the Dirac
  sea: Effective field theory approach to exactly solvable four-fermion
  models}},\ }\href {https://doi.org/10.1103/PhysRevD.77.025008} {\bibfield
  {journal} {\bibinfo  {journal} {Phys. Rev. D}\ }\textbf {\bibinfo {volume}
  {77}},\ \bibinfo {pages} {025008} (\bibinfo {year} {2008})},\ \Eprint
  {https://arxiv.org/abs/0708.3176} {arXiv:0708.3176 [hep-th]} \BibitemShut
  {NoStop}%
\bibitem [{\citenamefont {Karbstein}\ and\ \citenamefont
  {Thies}(2007{\natexlab{b}})}]{Karbstein:2007bg}%
  \BibitemOpen
  \bibfield  {author} {\bibinfo {author} {\bibfnamefont {F.}~\bibnamefont
  {Karbstein}}\ and\ \bibinfo {author} {\bibfnamefont {M.}~\bibnamefont
  {Thies}},\ }\bibfield  {title} {\bibinfo {title} {{Divergence of the axial
  current and fermion density in Gross-Neveu models}},\ }\href
  {https://doi.org/10.1103/PhysRevD.76.085009} {\bibfield  {journal} {\bibinfo
  {journal} {Phys. Rev. D}\ }\textbf {\bibinfo {volume} {76}},\ \bibinfo
  {pages} {085009} (\bibinfo {year} {2007}{\natexlab{b}})},\ \Eprint
  {https://arxiv.org/abs/0706.0424} {arXiv:0706.0424 [hep-th]} \BibitemShut
  {NoStop}%
\bibitem [{\citenamefont {Basar}\ and\ \citenamefont
  {Dunne}(2008)}]{Basar:2008im}%
  \BibitemOpen
  \bibfield  {author} {\bibinfo {author} {\bibfnamefont {G.}~\bibnamefont
  {Basar}}\ and\ \bibinfo {author} {\bibfnamefont {G.~V.}\ \bibnamefont
  {Dunne}},\ }\bibfield  {title} {\bibinfo {title} {{Self-consistent
  crystalline condensate in chiral Gross-Neveu and Bogoliubov-de Gennes
  systems}},\ }\href {https://doi.org/10.1103/PhysRevLett.100.200404}
  {\bibfield  {journal} {\bibinfo  {journal} {Phys. Rev. Lett.}\ }\textbf
  {\bibinfo {volume} {100}},\ \bibinfo {pages} {200404} (\bibinfo {year}
  {2008})},\ \Eprint {https://arxiv.org/abs/0803.1501} {arXiv:0803.1501
  [hep-th]} \BibitemShut {NoStop}%
\bibitem [{\citenamefont {Brendel}\ and\ \citenamefont
  {Thies}(2010)}]{Brendel:2009pq}%
  \BibitemOpen
  \bibfield  {author} {\bibinfo {author} {\bibfnamefont {W.}~\bibnamefont
  {Brendel}}\ and\ \bibinfo {author} {\bibfnamefont {M.}~\bibnamefont
  {Thies}},\ }\bibfield  {title} {\bibinfo {title} {{Covariant boost and
  structure functions of baryons in Gross-Neveu models}},\ }\href
  {https://doi.org/10.1103/PhysRevD.81.085002} {\bibfield  {journal} {\bibinfo
  {journal} {Phys. Rev. D}\ }\textbf {\bibinfo {volume} {81}},\ \bibinfo
  {pages} {085002} (\bibinfo {year} {2010})},\ \Eprint
  {https://arxiv.org/abs/0910.5351} {arXiv:0910.5351 [hep-th]} \BibitemShut
  {NoStop}%
\bibitem [{\citenamefont {Zinn-Justin}(2010)}]{Zinn-Justin:2010}%
  \BibitemOpen
  \bibfield  {author} {\bibinfo {author} {\bibfnamefont {J.}~\bibnamefont
  {Zinn-Justin}},\ }\bibfield  {title} {\bibinfo {title} {{Critical Phenomena:
  field theoretical approach}},\ }\href
  {https://doi.org/10.4249/scholarpedia.8346} {\bibfield  {journal} {\bibinfo
  {journal} {Scholarpedia}\ }\textbf {\bibinfo {volume} {5}},\ \bibinfo {pages}
  {8346} (\bibinfo {year} {2010})},\ \bibinfo {note} {revision
  \#148508}\BibitemShut {NoStop}%
\bibitem [{\citenamefont {Fitzner}\ and\ \citenamefont
  {Thies}(2012)}]{Fitzner:2012gg}%
  \BibitemOpen
  \bibfield  {author} {\bibinfo {author} {\bibfnamefont {C.}~\bibnamefont
  {Fitzner}}\ and\ \bibinfo {author} {\bibfnamefont {M.}~\bibnamefont
  {Thies}},\ }\bibfield  {title} {\bibinfo {title} {{Evidence for factorized
  scattering of composite states in the Gross-Neveu model}},\ }\href
  {https://doi.org/10.1103/PhysRevD.85.105015} {\bibfield  {journal} {\bibinfo
  {journal} {Phys. Rev. D}\ }\textbf {\bibinfo {volume} {85}},\ \bibinfo
  {pages} {105015} (\bibinfo {year} {2012})},\ \Eprint
  {https://arxiv.org/abs/1202.0648} {arXiv:1202.0648 [hep-th]} \BibitemShut
  {NoStop}%
\bibitem [{\citenamefont {Fitzner}\ and\ \citenamefont
  {Thies}(2013)}]{Fitzner:2012kb}%
  \BibitemOpen
  \bibfield  {author} {\bibinfo {author} {\bibfnamefont {C.}~\bibnamefont
  {Fitzner}}\ and\ \bibinfo {author} {\bibfnamefont {M.}~\bibnamefont
  {Thies}},\ }\bibfield  {title} {\bibinfo {title} {{Breathers and their
  interaction in the massless Gross-Neveu model}},\ }\href
  {https://doi.org/10.1103/PhysRevD.87.025001} {\bibfield  {journal} {\bibinfo
  {journal} {Phys. Rev. D}\ }\textbf {\bibinfo {volume} {87}},\ \bibinfo
  {pages} {025001} (\bibinfo {year} {2013})},\ \Eprint
  {https://arxiv.org/abs/1210.4423} {arXiv:1210.4423 [hep-th]} \BibitemShut
  {NoStop}%
\bibitem [{\citenamefont {Dunne}\ and\ \citenamefont
  {Thies}(2014)}]{Dunne:2013rka}%
  \BibitemOpen
  \bibfield  {author} {\bibinfo {author} {\bibfnamefont {G.~V.}\ \bibnamefont
  {Dunne}}\ and\ \bibinfo {author} {\bibfnamefont {M.}~\bibnamefont {Thies}},\
  }\bibfield  {title} {\bibinfo {title} {{Full time-dependent Hartree-Fock
  solution of large N Gross-Neveu models}},\ }\href
  {https://doi.org/10.1103/PhysRevD.89.025008} {\bibfield  {journal} {\bibinfo
  {journal} {Phys. Rev. D}\ }\textbf {\bibinfo {volume} {89}},\ \bibinfo
  {pages} {025008} (\bibinfo {year} {2014})},\ \Eprint
  {https://arxiv.org/abs/1309.2443} {arXiv:1309.2443 [hep-th]} \BibitemShut
  {NoStop}%
\bibitem [{\citenamefont {Thies}(2014)}]{Thies:2014ida}%
  \BibitemOpen
  \bibfield  {author} {\bibinfo {author} {\bibfnamefont {M.}~\bibnamefont
  {Thies}},\ }\bibfield  {title} {\bibinfo {title} {{Integrable Gross-Neveu
  models with fermion-fermion and fermion-antifermion pairing}},\ }\href
  {https://doi.org/10.1103/PhysRevD.90.105017} {\bibfield  {journal} {\bibinfo
  {journal} {Phys. Rev. D}\ }\textbf {\bibinfo {volume} {90}},\ \bibinfo
  {pages} {105017} (\bibinfo {year} {2014})},\ \Eprint
  {https://arxiv.org/abs/1408.5506} {arXiv:1408.5506 [hep-th]} \BibitemShut
  {NoStop}%
\bibitem [{\citenamefont {Thies}(2017{\natexlab{b}})}]{Thies:2017mbl}%
  \BibitemOpen
  \bibfield  {author} {\bibinfo {author} {\bibfnamefont {M.}~\bibnamefont
  {Thies}},\ }\bibfield  {title} {\bibinfo {title} {{Untwisting twisted
  NJL$_2$-kinks by a bare fermion mass}},\ }\href
  {https://doi.org/10.1103/PhysRevD.96.116018} {\bibfield  {journal} {\bibinfo
  {journal} {Phys. Rev. D}\ }\textbf {\bibinfo {volume} {96}},\ \bibinfo
  {pages} {116018} (\bibinfo {year} {2017}{\natexlab{b}})},\ \Eprint
  {https://arxiv.org/abs/1709.01269} {arXiv:1709.01269 [hep-th]} \BibitemShut
  {NoStop}%
\bibitem [{\citenamefont {Ahmed}(2018)}]{Ahmed:2018tcs}%
  \BibitemOpen
  \bibfield  {author} {\bibinfo {author} {\bibfnamefont {A.}~\bibnamefont
  {Ahmed}},\ }\href@noop {} {\bibinfo {title} {{Ginzburg-Landau Type Approach
  to the 1+1 Gross Neveu Model - Beyond Lowest Non-Trivial Order}}} (\bibinfo
  {year} {2018}),\ \Eprint {https://arxiv.org/abs/1802.09095} {arXiv:1802.09095
  [hep-th]} \BibitemShut {NoStop}%
\bibitem [{\citenamefont {Bermudez}\ \emph {et~al.}(2018)\citenamefont
  {Bermudez}, \citenamefont {Tirrito}, \citenamefont {Rizzi}, \citenamefont
  {Lewenstein},\ and\ \citenamefont {Hands}}]{Bermudez:2018eyh}%
  \BibitemOpen
  \bibfield  {author} {\bibinfo {author} {\bibfnamefont {A.}~\bibnamefont
  {Bermudez}}, \bibinfo {author} {\bibfnamefont {E.}~\bibnamefont {Tirrito}},
  \bibinfo {author} {\bibfnamefont {M.}~\bibnamefont {Rizzi}}, \bibinfo
  {author} {\bibfnamefont {M.}~\bibnamefont {Lewenstein}},\ and\ \bibinfo
  {author} {\bibfnamefont {S.}~\bibnamefont {Hands}},\ }\bibfield  {title}
  {\bibinfo {title} {{Gross\textendash{}Neveu\textendash{}Wilson model and
  correlated symmetry-protected topological phases}},\ }\href
  {https://doi.org/10.1016/j.aop.2018.10.007} {\bibfield  {journal} {\bibinfo
  {journal} {Annals Phys.}\ }\textbf {\bibinfo {volume} {399}},\ \bibinfo
  {pages} {149} (\bibinfo {year} {2018})},\ \Eprint
  {https://arxiv.org/abs/1807.03202} {arXiv:1807.03202 [cond-mat.quant-gas]}
  \BibitemShut {NoStop}%
\bibitem [{\citenamefont {Roose}\ \emph {et~al.}(2020)\citenamefont {Roose},
  \citenamefont {Bultinck}, \citenamefont {Vanderstraeten}, \citenamefont
  {Verstraete}, \citenamefont {Van~Acoleyen},\ and\ \citenamefont
  {Haegeman}}]{Roose:2020znu}%
  \BibitemOpen
  \bibfield  {author} {\bibinfo {author} {\bibfnamefont {G.}~\bibnamefont
  {Roose}}, \bibinfo {author} {\bibfnamefont {N.}~\bibnamefont {Bultinck}},
  \bibinfo {author} {\bibfnamefont {L.}~\bibnamefont {Vanderstraeten}},
  \bibinfo {author} {\bibfnamefont {F.}~\bibnamefont {Verstraete}}, \bibinfo
  {author} {\bibfnamefont {K.}~\bibnamefont {Van~Acoleyen}},\ and\ \bibinfo
  {author} {\bibfnamefont {J.}~\bibnamefont {Haegeman}},\ }\bibfield  {title}
  {\bibinfo {title} {{Lattice regularisation and entanglement structure of the
  Gross-Neveu model}},\ }\href {https://doi.org/10.1007/JHEP07(2021)207}
  {\bibfield  {journal} {\bibinfo  {journal} {JHEP}\ }\textbf {\bibinfo
  {volume} {21}},\ \bibinfo {pages} {207}},\ \Eprint
  {https://arxiv.org/abs/2010.03441} {arXiv:2010.03441 [hep-lat]} \BibitemShut
  {NoStop}%
\bibitem [{\citenamefont {Quinto}\ \emph {et~al.}(2021)\citenamefont {Quinto},
  \citenamefont {Monroy},\ and\ \citenamefont {Ferrari}}]{Quinto:2021lqn}%
  \BibitemOpen
  \bibfield  {author} {\bibinfo {author} {\bibfnamefont {A.~G.}\ \bibnamefont
  {Quinto}}, \bibinfo {author} {\bibfnamefont {R.~V.}\ \bibnamefont {Monroy}},\
  and\ \bibinfo {author} {\bibfnamefont {A.~F.}\ \bibnamefont {Ferrari}},\
  }\href@noop {} {\bibinfo {title} {{Renormalization group improvement of the
  effective potential in a $( 1 + 1 )$ dimensional Gross-Neveu model}}}
  (\bibinfo {year} {2021}),\ \Eprint {https://arxiv.org/abs/2108.04079}
  {arXiv:2108.04079 [hep-th]} \BibitemShut {NoStop}%
\bibitem [{\citenamefont {Lopes}\ \emph {et~al.}(2021)\citenamefont {Lopes},
  \citenamefont {Continentino},\ and\ \citenamefont
  {Barci}}]{lopes2021excitonic}%
  \BibitemOpen
  \bibfield  {author} {\bibinfo {author} {\bibfnamefont {N.}~\bibnamefont
  {Lopes}}, \bibinfo {author} {\bibfnamefont {M.~A.}\ \bibnamefont
  {Continentino}},\ and\ \bibinfo {author} {\bibfnamefont {D.~G.}\ \bibnamefont
  {Barci}},\ }\href@noop {} {\bibinfo {title} {{Excitonic insulators and
  Gross-Neveu models}}} (\bibinfo {year} {2021}),\ \Eprint
  {https://arxiv.org/abs/2112.07362} {arXiv:2112.07362 [cond-mat.str-el]}
  \BibitemShut {NoStop}%
\bibitem [{\citenamefont {Olver}\ \emph {et~al.}(2021)\citenamefont {Olver},
  \citenamefont {{Olde Daalhuis}}, \citenamefont {Lozier}, \citenamefont
  {Schneider}, \citenamefont {Boisvert}, \citenamefont {Clark}, \citenamefont
  {Miller}, \citenamefont {Saunders}, \citenamefont {Cohl},\ and\ \citenamefont
  {McClain}}]{NIST:DLMF}%
  \BibitemOpen
  \bibfield  {author} {\bibinfo {author} {\bibfnamefont {F.~W.~J.}\
  \bibnamefont {Olver}}, \bibinfo {author} {\bibfnamefont {A.~B.}\ \bibnamefont
  {{Olde Daalhuis}}}, \bibinfo {author} {\bibfnamefont {D.~W.}\ \bibnamefont
  {Lozier}}, \bibinfo {author} {\bibfnamefont {B.~I.}\ \bibnamefont
  {Schneider}}, \bibinfo {author} {\bibfnamefont {R.~F.}\ \bibnamefont
  {Boisvert}}, \bibinfo {author} {\bibfnamefont {C.~W.}\ \bibnamefont {Clark}},
  \bibinfo {author} {\bibfnamefont {B.~R.}\ \bibnamefont {Miller}}, \bibinfo
  {author} {\bibfnamefont {B.~V.}\ \bibnamefont {Saunders}}, \bibinfo {author}
  {\bibfnamefont {H.~S.}\ \bibnamefont {Cohl}},\ and\ \bibinfo {author}
  {\bibfnamefont {M.~A.}\ \bibnamefont {McClain}},\ }\href
  {http://dlmf.nist.gov/} {\bibinfo {title} {{NIST Digital Library of
  Mathematical Function}}},\ \bibinfo {howpublished} {{Release 1.1.2 of
  2021-06-15}} (\bibinfo {year} {2021}),\ \bibinfo {note} {[Online; accessed
  2020.06.24]}\BibitemShut {NoStop}%
\bibitem [{\citenamefont {Pannullo}\ \emph {et~al.}(2020)\citenamefont
  {Pannullo}, \citenamefont {Lenz}, \citenamefont {Wagner}, \citenamefont
  {Wellegehausen},\ and\ \citenamefont {Wipf}}]{Pannullo:2019bfn}%
  \BibitemOpen
  \bibfield  {author} {\bibinfo {author} {\bibfnamefont {L.}~\bibnamefont
  {Pannullo}}, \bibinfo {author} {\bibfnamefont {J.}~\bibnamefont {Lenz}},
  \bibinfo {author} {\bibfnamefont {M.}~\bibnamefont {Wagner}}, \bibinfo
  {author} {\bibfnamefont {B.}~\bibnamefont {Wellegehausen}},\ and\ \bibinfo
  {author} {\bibfnamefont {A.}~\bibnamefont {Wipf}},\ }\bibfield  {title}
  {\bibinfo {title} {{Inhomogeneous phases in the 1+1 dimensional Gross-Neveu
  model at finite number of fermion flavors}},\ }\href
  {https://doi.org/10.5506/APhysPolBSupp.13.127} {\bibfield  {journal}
  {\bibinfo  {journal} {Acta Phys. Polon. Supp.}\ }\textbf {\bibinfo {volume}
  {13}},\ \bibinfo {pages} {127} (\bibinfo {year} {2020})},\ \Eprint
  {https://arxiv.org/abs/1902.11066} {arXiv:1902.11066 [hep-lat]} \BibitemShut
  {NoStop}%
\bibitem [{\citenamefont {Pannullo}\ \emph {et~al.}(2019)\citenamefont
  {Pannullo}, \citenamefont {Lenz}, \citenamefont {Wagner}, \citenamefont
  {Wellegehausen},\ and\ \citenamefont {Wipf}}]{Pannullo:2019prx}%
  \BibitemOpen
  \bibfield  {author} {\bibinfo {author} {\bibfnamefont {L.}~\bibnamefont
  {Pannullo}}, \bibinfo {author} {\bibfnamefont {J.}~\bibnamefont {Lenz}},
  \bibinfo {author} {\bibfnamefont {M.}~\bibnamefont {Wagner}}, \bibinfo
  {author} {\bibfnamefont {B.}~\bibnamefont {Wellegehausen}},\ and\ \bibinfo
  {author} {\bibfnamefont {A.}~\bibnamefont {Wipf}},\ }\bibfield  {title}
  {\bibinfo {title} {{Lattice investigation of the phase diagram of the 1+1
  dimensional Gross-Neveu model at finite number of fermion flavors}},\ }\href
  {https://doi.org/10.22323/1.363.0063} {\bibfield  {journal} {\bibinfo
  {journal} {PoS}\ }\textbf {\bibinfo {volume} {LATTICE2019}},\ \bibinfo
  {pages} {063} (\bibinfo {year} {2019})},\ \Eprint
  {https://arxiv.org/abs/1909.11513} {arXiv:1909.11513 [hep-lat]} \BibitemShut
  {NoStop}%
\bibitem [{\citenamefont {Lenz}\ \emph
  {et~al.}(2020{\natexlab{b}})\citenamefont {Lenz}, \citenamefont {Pannullo},
  \citenamefont {Wagner}, \citenamefont {Wellegehausen},\ and\ \citenamefont
  {Wipf}}]{Lenz:2020bxk}%
  \BibitemOpen
  \bibfield  {author} {\bibinfo {author} {\bibfnamefont {J.}~\bibnamefont
  {Lenz}}, \bibinfo {author} {\bibfnamefont {L.}~\bibnamefont {Pannullo}},
  \bibinfo {author} {\bibfnamefont {M.}~\bibnamefont {Wagner}}, \bibinfo
  {author} {\bibfnamefont {B.}~\bibnamefont {Wellegehausen}},\ and\ \bibinfo
  {author} {\bibfnamefont {A.}~\bibnamefont {Wipf}},\ }\bibfield  {title}
  {\bibinfo {title} {{Inhomogeneous phases in the Gross-Neveu model in 1+1
  dimensions at finite number of flavors}},\ }\href
  {https://doi.org/10.1103/PhysRevD.101.094512} {\bibfield  {journal} {\bibinfo
   {journal} {Phys. Rev. D}\ }\textbf {\bibinfo {volume} {101}},\ \bibinfo
  {pages} {094512} (\bibinfo {year} {2020}{\natexlab{b}})},\ \Eprint
  {https://arxiv.org/abs/2004.00295} {arXiv:2004.00295 [hep-lat]} \BibitemShut
  {NoStop}%
\bibitem [{\citenamefont {Lenz}\ \emph {et~al.}(2022)\citenamefont {Lenz},
  \citenamefont {Mandl},\ and\ \citenamefont {Wipf}}]{Lenz:2021kzo}%
  \BibitemOpen
  \bibfield  {author} {\bibinfo {author} {\bibfnamefont {J.~J.}\ \bibnamefont
  {Lenz}}, \bibinfo {author} {\bibfnamefont {M.}~\bibnamefont {Mandl}},\ and\
  \bibinfo {author} {\bibfnamefont {A.}~\bibnamefont {Wipf}},\ }\bibfield
  {title} {\bibinfo {title} {{Inhomogeneities in the two-flavor chiral
  Gross-Neveu model}},\ }\href {https://doi.org/10.1103/PhysRevD.105.034512}
  {\bibfield  {journal} {\bibinfo  {journal} {Phys. Rev. D}\ }\textbf {\bibinfo
  {volume} {105}},\ \bibinfo {pages} {034512} (\bibinfo {year} {2022})},\
  \Eprint {https://arxiv.org/abs/2109.05525} {arXiv:2109.05525 [hep-lat]}
  \BibitemShut {NoStop}%
\bibitem [{\citenamefont {Horie}\ and\ \citenamefont
  {Nonaka}(2021)}]{Horie:2021wnn}%
  \BibitemOpen
  \bibfield  {author} {\bibinfo {author} {\bibfnamefont {K.}~\bibnamefont
  {Horie}}\ and\ \bibinfo {author} {\bibfnamefont {C.}~\bibnamefont {Nonaka}},\
  }\bibfield  {title} {\bibinfo {title} {{Inhomogeneous Phases in the Chiral
  Gross-Neveu Model on the Lattice}},\ }in\ \href@noop {} {\emph {\bibinfo
  {booktitle} {{38th International Symposium on Lattice Field Theory}}}}\
  (\bibinfo {year} {2021})\ \Eprint {https://arxiv.org/abs/2112.02261}
  {arXiv:2112.02261 [hep-lat]} \BibitemShut {NoStop}%
\bibitem [{\citenamefont {Matsubara}(1955)}]{Matsubara:1955ws}%
  \BibitemOpen
  \bibfield  {author} {\bibinfo {author} {\bibfnamefont {T.}~\bibnamefont
  {Matsubara}},\ }\bibfield  {title} {\bibinfo {title} {{A new approach to
  quantum statistical mechanics}},\ }\href {https://doi.org/10.1143/PTP.14.351}
  {\bibfield  {journal} {\bibinfo  {journal} {Prog. Theor. Phys.}\ }\textbf
  {\bibinfo {volume} {14}},\ \bibinfo {pages} {351} (\bibinfo {year}
  {1955})}\BibitemShut {NoStop}%
\bibitem [{\citenamefont {Fermi}(1926)}]{Fermi:1926}%
  \BibitemOpen
  \bibfield  {author} {\bibinfo {author} {\bibfnamefont {E.}~\bibnamefont
  {Fermi}},\ }\bibfield  {title} {\bibinfo {title} {{Sulla quantizzazione del
  gas perfetto monoatomico}},\ }\href@noop {} {\bibfield  {journal} {\bibinfo
  {journal} {Rend. Lincei}\ }\textbf {\bibinfo {volume} {3}},\ \bibinfo {pages}
  {145} (\bibinfo {year} {1926})}\BibitemShut {NoStop}%
\bibitem [{\citenamefont {Fermi}(1999)}]{Fermi:1999ncp}%
  \BibitemOpen
  \bibfield  {author} {\bibinfo {author} {\bibfnamefont {E.}~\bibnamefont
  {Fermi}},\ }\href@noop {} {\bibinfo {title} {{On the Quantization of the
  Monoatomic Ideal Gas}}} (\bibinfo {year} {1999}),\ \Eprint
  {https://arxiv.org/abs/cond-mat/9912229} {arXiv:cond-mat/9912229}
  \BibitemShut {NoStop}%
\bibitem [{\citenamefont {Dirac}(1926)}]{Dirac:1926jz}%
  \BibitemOpen
  \bibfield  {author} {\bibinfo {author} {\bibfnamefont {P.~A.~M.}\
  \bibnamefont {Dirac}},\ }\bibfield  {title} {\bibinfo {title} {{On the theory
  of quantum mechanics}},\ }\href {https://doi.org/10.1098/rspa.1926.0133}
  {\bibfield  {journal} {\bibinfo  {journal} {Proc. Roy. Soc. Lond. A}\
  }\textbf {\bibinfo {volume} {112}},\ \bibinfo {pages} {661} (\bibinfo {year}
  {1926})}\BibitemShut {NoStop}%
\bibitem [{\citenamefont {Steil}(2022)}]{Steil:2022phd}%
  \BibitemOpen
  \bibfield  {author} {\bibinfo {author} {\bibfnamefont {M.~J.}\ \bibnamefont
  {Steil}},\ }\emph {\bibinfo {title} {{From zero-dimensional theories to
  inhomogeneous phases with the functional renormalization group}}},\
  \href@noop {} {\bibinfo {type} {Phd thesis}},\ \bibinfo  {school} {Technische
  Universität Darmstadt} (\bibinfo {year} {2022}),\ \bibinfo {note} {in
  preparation}\BibitemShut {NoStop}%
\bibitem [{\citenamefont {Koenigstein}(2022)}]{Koenigstein:2022phd}%
  \BibitemOpen
  \bibfield  {author} {\bibinfo {author} {\bibfnamefont {A.}~\bibnamefont
  {Koenigstein}},\ }\emph {\bibinfo {title} {{Non-perturbative aspects of
  low-dimensional quantum-field theories}}},\ \href@noop {} {\bibinfo {type}
  {Phd thesis}},\ \bibinfo  {school} {Goethe University Frankfurt} (\bibinfo
  {year} {2022}),\ \bibinfo {note} {in preparation}\BibitemShut {NoStop}%
\bibitem [{\citenamefont {Roscher}\ \emph {et~al.}(2015)\citenamefont
  {Roscher}, \citenamefont {Braun},\ and\ \citenamefont
  {Drut}}]{Roscher:2015xha}%
  \BibitemOpen
  \bibfield  {author} {\bibinfo {author} {\bibfnamefont {D.}~\bibnamefont
  {Roscher}}, \bibinfo {author} {\bibfnamefont {J.}~\bibnamefont {Braun}},\
  and\ \bibinfo {author} {\bibfnamefont {J.~E.}\ \bibnamefont {Drut}},\
  }\bibfield  {title} {\bibinfo {title} {{Phase structure of mass- and
  spin-imbalanced unitary Fermi gases}},\ }\href
  {https://doi.org/10.1103/PhysRevA.91.053611} {\bibfield  {journal} {\bibinfo
  {journal} {Phys. Rev. A}\ }\textbf {\bibinfo {volume} {91}},\ \bibinfo
  {pages} {053611} (\bibinfo {year} {2015})},\ \Eprint
  {https://arxiv.org/abs/1501.05544} {arXiv:1501.05544 [cond-mat.quant-gas]}
  \BibitemShut {NoStop}%
\bibitem [{\citenamefont {Dolan}\ and\ \citenamefont
  {Jackiw}(1974)}]{Dolan:1973qd}%
  \BibitemOpen
  \bibfield  {author} {\bibinfo {author} {\bibfnamefont {L.~A.}\ \bibnamefont
  {Dolan}}\ and\ \bibinfo {author} {\bibfnamefont {R.~W.}\ \bibnamefont
  {Jackiw}},\ }\bibfield  {title} {\bibinfo {title} {{Symmetry behavior at
  finite temperature}},\ }\href {https://doi.org/10.1103/PhysRevD.9.3320}
  {\bibfield  {journal} {\bibinfo  {journal} {Phys. Rev. D}\ }\textbf {\bibinfo
  {volume} {9}},\ \bibinfo {pages} {3320} (\bibinfo {year} {1974})}\BibitemShut
  {NoStop}%
\bibitem [{\citenamefont {Weinberg}(1974)}]{Weinberg:1974hy}%
  \BibitemOpen
  \bibfield  {author} {\bibinfo {author} {\bibfnamefont {S.}~\bibnamefont
  {Weinberg}},\ }\bibfield  {title} {\bibinfo {title} {{Gauge and Global
  Symmetries at High Temperature}},\ }\href
  {https://doi.org/10.1103/PhysRevD.9.3357} {\bibfield  {journal} {\bibinfo
  {journal} {Phys. Rev. D}\ }\textbf {\bibinfo {volume} {9}},\ \bibinfo {pages}
  {3357} (\bibinfo {year} {1974})}\BibitemShut {NoStop}%
\bibitem [{\citenamefont {Eser}\ \emph {et~al.}(2018)\citenamefont {Eser},
  \citenamefont {Divotgey}, \citenamefont {Mitter},\ and\ \citenamefont
  {Rischke}}]{Eser:2018jqo}%
  \BibitemOpen
  \bibfield  {author} {\bibinfo {author} {\bibfnamefont {J.}~\bibnamefont
  {Eser}}, \bibinfo {author} {\bibfnamefont {F.}~\bibnamefont {Divotgey}},
  \bibinfo {author} {\bibfnamefont {M.}~\bibnamefont {Mitter}},\ and\ \bibinfo
  {author} {\bibfnamefont {D.~H.}\ \bibnamefont {Rischke}},\ }\bibfield
  {title} {\bibinfo {title} {{Low-energy limit of the $O(4)$ quark-meson model
  from the functional renormalization group approach}},\ }\href
  {https://doi.org/10.1103/PhysRevD.98.014024} {\bibfield  {journal} {\bibinfo
  {journal} {Phys. Rev. D}\ }\textbf {\bibinfo {volume} {98}},\ \bibinfo
  {pages} {014024} (\bibinfo {year} {2018})},\ \Eprint
  {https://arxiv.org/abs/1804.01787} {arXiv:1804.01787 [hep-ph]} \BibitemShut
  {NoStop}%
\bibitem [{\citenamefont {Eser}\ \emph {et~al.}(2019)\citenamefont {Eser},
  \citenamefont {Divotgey},\ and\ \citenamefont {Mitter}}]{Eser:2019pvd}%
  \BibitemOpen
  \bibfield  {author} {\bibinfo {author} {\bibfnamefont {J.}~\bibnamefont
  {Eser}}, \bibinfo {author} {\bibfnamefont {F.}~\bibnamefont {Divotgey}},\
  and\ \bibinfo {author} {\bibfnamefont {M.}~\bibnamefont {Mitter}},\
  }\bibfield  {title} {\bibinfo {title} {{Low-energy limit of the $O(4)$
  quark-meson model}},\ }\href {https://doi.org/10.22323/1.317.0060} {\bibfield
   {journal} {\bibinfo  {journal} {PoS}\ }\textbf {\bibinfo {volume}
  {CD2018}},\ \bibinfo {pages} {060} (\bibinfo {year} {2019})},\ \Eprint
  {https://arxiv.org/abs/1902.04804} {arXiv:1902.04804 [hep-ph]} \BibitemShut
  {NoStop}%
\bibitem [{\citenamefont {Cichutek}\ \emph {et~al.}(2020)\citenamefont
  {Cichutek}, \citenamefont {Divotgey},\ and\ \citenamefont
  {Eser}}]{Cichutek:2020bli}%
  \BibitemOpen
  \bibfield  {author} {\bibinfo {author} {\bibfnamefont {N.}~\bibnamefont
  {Cichutek}}, \bibinfo {author} {\bibfnamefont {F.}~\bibnamefont {Divotgey}},\
  and\ \bibinfo {author} {\bibfnamefont {J.}~\bibnamefont {Eser}},\ }\bibfield
  {title} {\bibinfo {title} {{Fluctuation-induced higher-derivative couplings
  and infrared dynamics of the quark-meson-diquark model}},\ }\href
  {https://doi.org/10.1103/PhysRevD.102.034030} {\bibfield  {journal} {\bibinfo
   {journal} {Phys. Rev. D}\ }\textbf {\bibinfo {volume} {102}},\ \bibinfo
  {pages} {034030} (\bibinfo {year} {2020})},\ \Eprint
  {https://arxiv.org/abs/2006.12473} {arXiv:2006.12473 [hep-ph]} \BibitemShut
  {NoStop}%
\bibitem [{\citenamefont {Divotgey}\ \emph {et~al.}(2019)\citenamefont
  {Divotgey}, \citenamefont {Eser},\ and\ \citenamefont
  {Mitter}}]{Divotgey:2019xea}%
  \BibitemOpen
  \bibfield  {author} {\bibinfo {author} {\bibfnamefont {F.}~\bibnamefont
  {Divotgey}}, \bibinfo {author} {\bibfnamefont {J.}~\bibnamefont {Eser}},\
  and\ \bibinfo {author} {\bibfnamefont {M.}~\bibnamefont {Mitter}},\
  }\bibfield  {title} {\bibinfo {title} {{Dynamical generation of low-energy
  couplings from quark-meson fluctuations}},\ }\href
  {https://doi.org/10.1103/PhysRevD.99.054023} {\bibfield  {journal} {\bibinfo
  {journal} {Phys. Rev. D}\ }\textbf {\bibinfo {volume} {99}},\ \bibinfo
  {pages} {054023} (\bibinfo {year} {2019})},\ \Eprint
  {https://arxiv.org/abs/1901.02472} {arXiv:1901.02472 [hep-ph]} \BibitemShut
  {NoStop}%
\bibitem [{\citenamefont {Pawlowski}\ and\ \citenamefont
  {Rennecke}(2014)}]{Pawlowski:2014zaa}%
  \BibitemOpen
  \bibfield  {author} {\bibinfo {author} {\bibfnamefont {J.~M.}\ \bibnamefont
  {Pawlowski}}\ and\ \bibinfo {author} {\bibfnamefont {F.}~\bibnamefont
  {Rennecke}},\ }\bibfield  {title} {\bibinfo {title} {{Higher order
  quark-mesonic scattering processes and the phase structure of QCD}},\ }\href
  {https://doi.org/10.1103/PhysRevD.90.076002} {\bibfield  {journal} {\bibinfo
  {journal} {Phys. Rev. D}\ }\textbf {\bibinfo {volume} {90}},\ \bibinfo
  {pages} {076002} (\bibinfo {year} {2014})},\ \Eprint
  {https://arxiv.org/abs/1403.1179} {arXiv:1403.1179 [hep-ph]} \BibitemShut
  {NoStop}%
\bibitem [{\citenamefont {Grossi}\ \emph {et~al.}(2021)\citenamefont {Grossi},
  \citenamefont {Ihssen}, \citenamefont {Pawlowski},\ and\ \citenamefont
  {Wink}}]{Grossi:2021ksl}%
  \BibitemOpen
  \bibfield  {author} {\bibinfo {author} {\bibfnamefont {E.}~\bibnamefont
  {Grossi}}, \bibinfo {author} {\bibfnamefont {F.~J.}\ \bibnamefont {Ihssen}},
  \bibinfo {author} {\bibfnamefont {J.~M.}\ \bibnamefont {Pawlowski}},\ and\
  \bibinfo {author} {\bibfnamefont {N.}~\bibnamefont {Wink}},\ }\bibfield
  {title} {\bibinfo {title} {{Shocks and quark-meson scatterings at large
  density}},\ }\href {https://doi.org/10.1103/PhysRevD.104.016028} {\bibfield
  {journal} {\bibinfo  {journal} {Phys. Rev. D}\ }\textbf {\bibinfo {volume}
  {104}},\ \bibinfo {pages} {016028} (\bibinfo {year} {2021})},\ \Eprint
  {https://arxiv.org/abs/2102.01602} {arXiv:2102.01602 [hep-ph]} \BibitemShut
  {NoStop}%
\bibitem [{\citenamefont {Otto}\ \emph
  {et~al.}(2020{\natexlab{a}})\citenamefont {Otto}, \citenamefont {Oertel},\
  and\ \citenamefont {Schaefer}}]{Otto:2019zjy}%
  \BibitemOpen
  \bibfield  {author} {\bibinfo {author} {\bibfnamefont {K.}~\bibnamefont
  {Otto}}, \bibinfo {author} {\bibfnamefont {M.}~\bibnamefont {Oertel}},\ and\
  \bibinfo {author} {\bibfnamefont {B.-J.}\ \bibnamefont {Schaefer}},\
  }\bibfield  {title} {\bibinfo {title} {{Hybrid and quark star matter based on
  a nonperturbative equation of state}},\ }\href
  {https://doi.org/10.1103/PhysRevD.101.103021} {\bibfield  {journal} {\bibinfo
   {journal} {Phys. Rev. D}\ }\textbf {\bibinfo {volume} {101}},\ \bibinfo
  {pages} {103021} (\bibinfo {year} {2020}{\natexlab{a}})},\ \Eprint
  {https://arxiv.org/abs/1910.11929} {arXiv:1910.11929 [hep-ph]} \BibitemShut
  {NoStop}%
\bibitem [{\citenamefont {Otto}\ \emph
  {et~al.}(2020{\natexlab{b}})\citenamefont {Otto}, \citenamefont {Oertel},\
  and\ \citenamefont {Schaefer}}]{Otto:2020hoz}%
  \BibitemOpen
  \bibfield  {author} {\bibinfo {author} {\bibfnamefont {K.}~\bibnamefont
  {Otto}}, \bibinfo {author} {\bibfnamefont {M.}~\bibnamefont {Oertel}},\ and\
  \bibinfo {author} {\bibfnamefont {B.-J.}\ \bibnamefont {Schaefer}},\
  }\bibfield  {title} {\bibinfo {title} {{Nonperturbative quark matter
  equations of state with vector interactions}},\ }\href
  {https://doi.org/10.1140/epjst/e2020-000155-y} {\bibfield  {journal}
  {\bibinfo  {journal} {Eur. Phys. J. ST}\ }\textbf {\bibinfo {volume} {229}},\
  \bibinfo {pages} {3629} (\bibinfo {year} {2020}{\natexlab{b}})},\ \Eprint
  {https://arxiv.org/abs/2007.07394} {arXiv:2007.07394 [hep-ph]} \BibitemShut
  {NoStop}%
\bibitem [{\citenamefont {Dupuis}\ \emph {et~al.}(2021)\citenamefont {Dupuis},
  \citenamefont {Canet}, \citenamefont {Eichhorn}, \citenamefont {Metzner},
  \citenamefont {Pawlowski}, \citenamefont {Tissier},\ and\ \citenamefont
  {Wschebor}}]{Dupuis:2020fhh}%
  \BibitemOpen
  \bibfield  {author} {\bibinfo {author} {\bibfnamefont {N.}~\bibnamefont
  {Dupuis}}, \bibinfo {author} {\bibfnamefont {L.}~\bibnamefont {Canet}},
  \bibinfo {author} {\bibfnamefont {A.}~\bibnamefont {Eichhorn}}, \bibinfo
  {author} {\bibfnamefont {W.}~\bibnamefont {Metzner}}, \bibinfo {author}
  {\bibfnamefont {J.~M.}\ \bibnamefont {Pawlowski}}, \bibinfo {author}
  {\bibfnamefont {M.}~\bibnamefont {Tissier}},\ and\ \bibinfo {author}
  {\bibfnamefont {N.}~\bibnamefont {Wschebor}},\ }\bibfield  {title} {\bibinfo
  {title} {{The nonperturbative functional renormalization group and its
  applications}},\ }\href {https://doi.org/10.1016/j.physrep.2021.01.001}
  {\bibfield  {journal} {\bibinfo  {journal} {Phys. Rept.}\ }\textbf {\bibinfo
  {volume} {910}},\ \bibinfo {pages} {1} (\bibinfo {year} {2021})},\ \Eprint
  {https://arxiv.org/abs/2006.04853} {arXiv:2006.04853 [cond-mat.stat-mech]}
  \BibitemShut {NoStop}%
\bibitem [{\citenamefont {Eser}\ and\ \citenamefont
  {Blaizot}(2021)}]{Eser:2021ivo}%
  \BibitemOpen
  \bibfield  {author} {\bibinfo {author} {\bibfnamefont {J.}~\bibnamefont
  {Eser}}\ and\ \bibinfo {author} {\bibfnamefont {J.-P.}\ \bibnamefont
  {Blaizot}},\ }\href@noop {} {\bibinfo {title} {{S-wave pion-pion scattering
  lengths from nucleon-meson fluctuations}}} (\bibinfo {year} {2021}),\ \Eprint
  {https://arxiv.org/abs/2112.14579} {arXiv:2112.14579 [hep-ph]} \BibitemShut
  {NoStop}%
\bibitem [{\citenamefont {{The pandas development
  team}}(2020)}]{reback2020pandas}%
  \BibitemOpen
  \bibfield  {author} {\bibinfo {author} {\bibnamefont {{The pandas development
  team}}},\ }\href {https://doi.org/10.5281/zenodo.3509134} {\bibinfo {title}
  {pandas-dev/pandas: Pandas}},\ \bibinfo {howpublished}
  {\url{https://doi.org/10.5281/zenodo.3509134}} (\bibinfo {year}
  {2020})\BibitemShut {NoStop}%
\bibitem [{\citenamefont {{W}es
  {M}c{K}inney}(2010)}]{mckinney-proc-scipy-2010}%
  \BibitemOpen
  \bibfield  {author} {\bibinfo {author} {\bibnamefont {{W}es {M}c{K}inney}},\
  }\bibfield  {title} {\bibinfo {title} {{D}ata {S}tructures for {S}tatistical
  {C}omputing in {P}ython},\ }in\ \href
  {https://doi.org/10.25080/Majora-92bf1922-00a} {\emph {\bibinfo {booktitle}
  {{P}roceedings of the 9th {P}ython in {S}cience {C}onference}}},\ \bibinfo
  {editor} {edited by\ \bibinfo {editor} {\bibnamefont {{S}t\'efan van~der
  {W}alt}}\ and\ \bibinfo {editor} {\bibnamefont {{J}arrod {M}illman}}}\
  (\bibinfo {year} {2010})\ pp.\ \bibinfo {pages} {56 -- 61}\BibitemShut
  {NoStop}%
\bibitem [{\citenamefont {Virtanen}\ \emph {et~al.}(2020)\citenamefont
  {Virtanen}, \citenamefont {Gommers}, \citenamefont {Oliphant}, \citenamefont
  {Haberland}, \citenamefont {Reddy}, \citenamefont {Cournapeau}, \citenamefont
  {Burovski}, \citenamefont {Peterson}, \citenamefont {Weckesser},
  \citenamefont {Bright}, \citenamefont {{van der Walt}}, \citenamefont
  {Brett}, \citenamefont {Wilson}, \citenamefont {Millman}, \citenamefont
  {Mayorov}, \citenamefont {Nelson}, \citenamefont {Jones}, \citenamefont
  {Kern}, \citenamefont {Larson}, \citenamefont {Carey}, \citenamefont {Polat},
  \citenamefont {Feng}, \citenamefont {Moore}, \citenamefont {{VanderPlas}},
  \citenamefont {Laxalde}, \citenamefont {Perktold}, \citenamefont {Cimrman},
  \citenamefont {Henriksen}, \citenamefont {Quintero}, \citenamefont {Harris},
  \citenamefont {Archibald}, \citenamefont {Ribeiro}, \citenamefont
  {Pedregosa}, \citenamefont {{van Mulbregt}},\ and\ \citenamefont {{SciPy 1.0
  Contributors}}}]{2020SciPy-NMeth}%
  \BibitemOpen
  \bibfield  {author} {\bibinfo {author} {\bibfnamefont {P.}~\bibnamefont
  {Virtanen}}, \bibinfo {author} {\bibfnamefont {R.}~\bibnamefont {Gommers}},
  \bibinfo {author} {\bibfnamefont {T.~E.}\ \bibnamefont {Oliphant}}, \bibinfo
  {author} {\bibfnamefont {M.}~\bibnamefont {Haberland}}, \bibinfo {author}
  {\bibfnamefont {T.}~\bibnamefont {Reddy}}, \bibinfo {author} {\bibfnamefont
  {D.}~\bibnamefont {Cournapeau}}, \bibinfo {author} {\bibfnamefont
  {E.}~\bibnamefont {Burovski}}, \bibinfo {author} {\bibfnamefont
  {P.}~\bibnamefont {Peterson}}, \bibinfo {author} {\bibfnamefont
  {W.}~\bibnamefont {Weckesser}}, \bibinfo {author} {\bibfnamefont
  {J.}~\bibnamefont {Bright}}, \bibinfo {author} {\bibfnamefont {S.~J.}\
  \bibnamefont {{van der Walt}}}, \bibinfo {author} {\bibfnamefont
  {M.}~\bibnamefont {Brett}}, \bibinfo {author} {\bibfnamefont
  {J.}~\bibnamefont {Wilson}}, \bibinfo {author} {\bibfnamefont {K.~J.}\
  \bibnamefont {Millman}}, \bibinfo {author} {\bibfnamefont {N.}~\bibnamefont
  {Mayorov}}, \bibinfo {author} {\bibfnamefont {A.~R.~J.}\ \bibnamefont
  {Nelson}}, \bibinfo {author} {\bibfnamefont {E.}~\bibnamefont {Jones}},
  \bibinfo {author} {\bibfnamefont {R.}~\bibnamefont {Kern}}, \bibinfo {author}
  {\bibfnamefont {E.}~\bibnamefont {Larson}}, \bibinfo {author} {\bibfnamefont
  {C.~J.}\ \bibnamefont {Carey}}, \bibinfo {author} {\bibfnamefont
  {{\.I}.}~\bibnamefont {Polat}}, \bibinfo {author} {\bibfnamefont
  {Y.}~\bibnamefont {Feng}}, \bibinfo {author} {\bibfnamefont {E.~W.}\
  \bibnamefont {Moore}}, \bibinfo {author} {\bibfnamefont {J.}~\bibnamefont
  {{VanderPlas}}}, \bibinfo {author} {\bibfnamefont {D.}~\bibnamefont
  {Laxalde}}, \bibinfo {author} {\bibfnamefont {J.}~\bibnamefont {Perktold}},
  \bibinfo {author} {\bibfnamefont {R.}~\bibnamefont {Cimrman}}, \bibinfo
  {author} {\bibfnamefont {I.}~\bibnamefont {Henriksen}}, \bibinfo {author}
  {\bibfnamefont {E.~A.}\ \bibnamefont {Quintero}}, \bibinfo {author}
  {\bibfnamefont {C.~R.}\ \bibnamefont {Harris}}, \bibinfo {author}
  {\bibfnamefont {A.~M.}\ \bibnamefont {Archibald}}, \bibinfo {author}
  {\bibfnamefont {A.~H.}\ \bibnamefont {Ribeiro}}, \bibinfo {author}
  {\bibfnamefont {F.}~\bibnamefont {Pedregosa}}, \bibinfo {author}
  {\bibfnamefont {P.}~\bibnamefont {{van Mulbregt}}},\ and\ \bibinfo {author}
  {\bibnamefont {{SciPy 1.0 Contributors}}},\ }\bibfield  {title} {\bibinfo
  {title} {{SciPy 1.0: Fundamental Algorithms for Scientific Computing in
  Python}},\ }\href {https://doi.org/10.1038/s41592-019-0686-2} {\bibfield
  {journal} {\bibinfo  {journal} {Nature Methods}\ }\textbf {\bibinfo {volume}
  {17}},\ \bibinfo {pages} {261} (\bibinfo {year} {2020})}\BibitemShut
  {NoStop}%
\bibitem [{\citenamefont {Harris}\ \emph {et~al.}(2020)\citenamefont {Harris},
  \citenamefont {Millman}, \citenamefont {van~der Walt}, \citenamefont
  {Gommers}, \citenamefont {Virtanen}, \citenamefont {Cournapeau},
  \citenamefont {Wieser}, \citenamefont {Taylor}, \citenamefont {Berg},
  \citenamefont {Smith}, \citenamefont {Kern}, \citenamefont {Picus},
  \citenamefont {Hoyer}, \citenamefont {van Kerkwijk}, \citenamefont {Brett},
  \citenamefont {Haldane}, \citenamefont {Fernández~del Río}, \citenamefont
  {Wiebe}, \citenamefont {Peterson}, \citenamefont {Gérard-Marchant},
  \citenamefont {Sheppard}, \citenamefont {Reddy}, \citenamefont {Weckesser},
  \citenamefont {Abbasi}, \citenamefont {Gohlke},\ and\ \citenamefont
  {Oliphant}}]{2020NumPy-Array}%
  \BibitemOpen
  \bibfield  {author} {\bibinfo {author} {\bibfnamefont {C.~R.}\ \bibnamefont
  {Harris}}, \bibinfo {author} {\bibfnamefont {K.~J.}\ \bibnamefont {Millman}},
  \bibinfo {author} {\bibfnamefont {S.~J.}\ \bibnamefont {van~der Walt}},
  \bibinfo {author} {\bibfnamefont {R.}~\bibnamefont {Gommers}}, \bibinfo
  {author} {\bibfnamefont {P.}~\bibnamefont {Virtanen}}, \bibinfo {author}
  {\bibfnamefont {D.}~\bibnamefont {Cournapeau}}, \bibinfo {author}
  {\bibfnamefont {E.}~\bibnamefont {Wieser}}, \bibinfo {author} {\bibfnamefont
  {J.}~\bibnamefont {Taylor}}, \bibinfo {author} {\bibfnamefont
  {S.}~\bibnamefont {Berg}}, \bibinfo {author} {\bibfnamefont {N.~J.}\
  \bibnamefont {Smith}}, \bibinfo {author} {\bibfnamefont {R.}~\bibnamefont
  {Kern}}, \bibinfo {author} {\bibfnamefont {M.}~\bibnamefont {Picus}},
  \bibinfo {author} {\bibfnamefont {S.}~\bibnamefont {Hoyer}}, \bibinfo
  {author} {\bibfnamefont {M.~H.}\ \bibnamefont {van Kerkwijk}}, \bibinfo
  {author} {\bibfnamefont {M.}~\bibnamefont {Brett}}, \bibinfo {author}
  {\bibfnamefont {A.}~\bibnamefont {Haldane}}, \bibinfo {author} {\bibfnamefont
  {J.}~\bibnamefont {Fernández~del Río}}, \bibinfo {author} {\bibfnamefont
  {M.}~\bibnamefont {Wiebe}}, \bibinfo {author} {\bibfnamefont
  {P.}~\bibnamefont {Peterson}}, \bibinfo {author} {\bibfnamefont
  {P.}~\bibnamefont {Gérard-Marchant}}, \bibinfo {author} {\bibfnamefont
  {K.}~\bibnamefont {Sheppard}}, \bibinfo {author} {\bibfnamefont
  {T.}~\bibnamefont {Reddy}}, \bibinfo {author} {\bibfnamefont
  {W.}~\bibnamefont {Weckesser}}, \bibinfo {author} {\bibfnamefont
  {H.}~\bibnamefont {Abbasi}}, \bibinfo {author} {\bibfnamefont
  {C.}~\bibnamefont {Gohlke}},\ and\ \bibinfo {author} {\bibfnamefont {T.~E.}\
  \bibnamefont {Oliphant}},\ }\bibfield  {title} {\bibinfo {title} {{Array
  programming with {NumPy}}},\ }\href
  {https://doi.org/10.1038/s41586-020-2649-2} {\bibfield  {journal} {\bibinfo
  {journal} {Nature}\ }\textbf {\bibinfo {volume} {585}},\ \bibinfo {pages}
  {357–362} (\bibinfo {year} {2020})}\BibitemShut {NoStop}%
\bibitem [{\citenamefont {Johnson}(2020)}]{cubature:2020}%
  \BibitemOpen
  \bibfield  {author} {\bibinfo {author} {\bibfnamefont {S.~G.}\ \bibnamefont
  {Johnson}},\ }\href@noop {} {\bibinfo {title} {{Cubature v1.0.4}}},\ \bibinfo
  {howpublished} {\url{https://github.com/stevengj/cubature}} (\bibinfo {year}
  {2020})\BibitemShut {NoStop}%
\bibitem [{\citenamefont {Nelder}\ and\ \citenamefont
  {Mead}(1965)}]{Nelder:1965}%
  \BibitemOpen
  \bibfield  {author} {\bibinfo {author} {\bibfnamefont {J.~A.}\ \bibnamefont
  {Nelder}}\ and\ \bibinfo {author} {\bibfnamefont {R.}~\bibnamefont {Mead}},\
  }\bibfield  {title} {\bibinfo {title} {{A Simplex Method for Function
  Minimization}},\ }\href {https://doi.org/10.1093/comjnl/7.4.308} {\bibfield
  {journal} {\bibinfo  {journal} {The Computer Journal}\ }\textbf {\bibinfo
  {volume} {7}},\ \bibinfo {pages} {308} (\bibinfo {year} {1965})}\BibitemShut
  {NoStop}%
\bibitem [{\citenamefont {{Wolfram Research{,}
  Inc.}}(2020)}]{Mathematica:12.2}%
  \BibitemOpen
  \bibfield  {author} {\bibinfo {author} {\bibnamefont {{Wolfram Research{,}
  Inc.}}},\ }\href {https://www.wolfram.com/mathematica} {\bibinfo {title}
  {{Mathematica, {V}ersion 12.2}}} (\bibinfo {year} {2020}),\ \bibinfo {note}
  {{Champaign, IL}}\BibitemShut {NoStop}%
\bibitem [{\citenamefont {Collins}\ and\ \citenamefont
  {Vermaseren}(2016)}]{Collins:2016aya}%
  \BibitemOpen
  \bibfield  {author} {\bibinfo {author} {\bibfnamefont {J.~C.}\ \bibnamefont
  {Collins}}\ and\ \bibinfo {author} {\bibfnamefont {J.~A.}\ \bibnamefont
  {Vermaseren}},\ }\href@noop {} {\bibinfo {title} {{Axodraw version 2}}}
  (\bibinfo {year} {2016}),\ \Eprint {https://arxiv.org/abs/1606.01177}
  {arXiv:1606.01177 [cs.OH]} \BibitemShut {NoStop}%
\bibitem [{\citenamefont {Hunter}(2007)}]{Hunter:2007}%
  \BibitemOpen
  \bibfield  {author} {\bibinfo {author} {\bibfnamefont {J.~D.}\ \bibnamefont
  {Hunter}},\ }\bibfield  {title} {\bibinfo {title} {{Matplotlib: A 2D graphics
  environment}},\ }\href {https://doi.org/10.1109/MCSE.2007.55} {\bibfield
  {journal} {\bibinfo  {journal} {Computing in Science \& Engineering}\
  }\textbf {\bibinfo {volume} {9}},\ \bibinfo {pages} {90} (\bibinfo {year}
  {2007})}\BibitemShut {NoStop}%
\end{thebibliography}%

\end{document}